\documentclass{ws-ijgmmp}

\begin{document}

\markboth{Zi-Hua Weng}
{Four interactions in the sedenion curved spaces}

%
\catchline{}{}{}{}{}
%

\title{Four interactions in the sedenion curved spaces
}

\author{Zi-Hua Weng
}
\address{School of Aerospace Engineering, Xiamen University, Xiamen, China
\\
College of Physical Science and Technology, Xiamen University, Xiamen, China
\\
\email{xmuwzh@xmu.edu.cn
}
}


\maketitle

\begin{history}
\received{(Day Month Year)}
\revised{(Day Month Year)}
\end{history}

\begin{abstract}
  The paper aims to apply the complex-sedenions to explore the field equations of four fundamental interactions, which are relevant to the classical mechanics and quantum mechanics, in the curved spaces. J. C. Maxwell was the first to utilize the quaternions to describe the property of electromagnetic fields. Nowadays the scholars introduce the complex-octonions to depict the electromagnetic and gravitational fields. And the complex-sedenions can be applied to study the field equations of the four interactions in the classical mechanics and quantum mechanics. Further, it is able to extend the field equations from the flat space into the curved space described with the complex-sedenions, by means of the tangent-frames and tensors. The research states that a few physical quantities will make a contribution to certain spatial parameters of the curved spaces. These spatial parameters may exert an influence on some operators (such as, divergence, gradient, and curl), impacting the field equations in the curved spaces, especially the field equations of the four quantum-fields in the quantum mechanics. Apparently the paper and General Relativity both confirm and succeed to the Cartesian academic thought of `the space is the extension of substance'.
\end{abstract}


\section{\label{sec:level1}Introduction}

How to describe the existing four fundamental interactions simultaneously is one bewildering and intriguing puzzle. Since a long time, this puzzle and correlative topics are attracting the attention of many scholars. Are there merely four sorts of fundamental interactions? May we find one method to describe simultaneously the physical properties of so-called four fundamental interactions? In the curved space, how do we write out the field equations of four fundamental interactions? Until recently, the appearance of four fundamental interactions, in the curved space described with the complex-sedenions (in Section 2), replies to some of these puzzles. According to this field theory, it is capable of deducing not only the field equations of the classical mechanics on the macroscopic scale, but also the field equations of the quantum mechanics on the microscopic scale, in the complex-sedenion curved space.

A. Einstein was the first to propose the unified field theory, attempting to unify the gravitational field and electromagnetic field. As a further complication, two new sorts of fields, weak-nuclear field and strong-nuclear field, emerged suddenly from the experiments, although the unified field theory has not been achieved yet. By the twentieth century, the scholars had found four sorts of fields, including the gravitational field, electromagnetic field, weak-nuclear field, and strong-nuclear field. Consequently, the unified field theory should be redeveloped and expanded.

Through protracted and unremitting efforts, the electromagnetic field and weak-nuclear field were successfully unified into the electroweak field. The amazing achievements acquired in the electroweak field theory inspirit some scholars to endeavor to propose the Quantum Chromodynamics. Subsequently a few scholars put forward the Standard Model (SM for short) for the elementary particle \cite{GUT1, GUT2, GUT3}, attempting to unify the Quantum Chromodynamics and electroweak theory \cite{GUT4, GUT5, GUT6}, although this theoretical model excludes the gravitational field. For the purpose of containing the gravitational field, certain scholars advanced several new theoretical assumptions \cite{GUT7, GUT8, GUT9}, including the Grand Unified Theory \cite{GUT10, GUT11}, `Beyond the SM' \cite{GUT12, GUT13}, superstring theory \cite{GUT14, GUT15}, and `Beyond the General Relativity' and so forth \cite{GUT16, GUT17}. Although it may be encountered a large number of difficulties, these ambitious theoretical assumptions \cite{GUT18, GUT19} do offer promising hope for the future to unify the existing four fields. By means of these ambitious plans, scholars expect the researches to be able to resolve some remaining problems \cite{GUT20, GUT21} in the existing field theories, especially the field equations of the four fields \cite{GUT22, GUT23, GUT24}, which are relevant to the classical mechanics and quantum mechanics, in the curved spaces.

Making an analysis and comparison of preceding studies, it is able to find several primal problems associate with the four fields as follows:

1) Four fields. The existing physical theories are unable to describe simultaneously the four fields, especially the gravitational field. In the four fields, the gravitational field was the first to be found in the history. However, a part of recent unified field theories exclude the gravitational field, such as SM and so forth. It reveals that some unified field theories may be seized of a few unacceptable defects essentially. Moreover, these unified field theories should possess the predictive power and expandability, predicting and accommodating new unknown fields, which may emerge from the unification process of fields.

2) Curved space. In the General Relativity (GR for short), it is able to achieve the inference of `the physical quantities dominate the spatial parameters', according to the Cartesian academic thought of `the space is the extension of substance'. This viewpoint belongs to the groundbreaking insight, nevertheless the GR is unable to express this extraordinary viewpoint adequately. As a result, the applicable scope of GR is restricted severely, and the GR can merely explain a small quantity of physical phenomena. In the classical mechanics, the universally applicable field equations have not yet been derived from the curved spaces. It means that the Cartesian academic thought has not yet been spread out extensively in the classical mechanics on the macroscopic scale.

3) Wave equation. In the recent physical theories, the unification of quantum mechanics and GR is not successful adequately. It is assumed that this is because both of two theories are not perfect enough. For instance, it is still able to establish the field equations in the curved spaces, even though we take no account of the inferences derived from the E\"{o}tv\"{o}s experiment in the recent studies. And it is unable to explain clearly the spin angular momentum \cite{weng1}. Consequently, it is not yet able to establish the applicable and effective wave equations in the curved spaces, for the quantum mechanics on the microscopic scale. At present, we still have a long way to attempt to spread out the Cartesian academic thought in the quantum mechanics.

In 1843, W. R. Hamilton invented the quaternions. Subsequently, others introduced the octonions and sedenions. J. C. Maxwell was the first to apply the quaternions to explore the properties of electromagnetic fields. Presently, some scholars utilize the quaternions \cite{quaternion2, quaternion5}, octonions \cite{quaternion6, quaternion7, quaternion8}, and sedenions \cite{quaternion11, quaternion12, quaternion14} to research the gravitational field equations \cite{quaternion13, quaternion4}, electromagnetic field equations \cite{quaternion3}, Dirac wave equations, Yang-Mills equations, curved space \cite{quaternion9}, astrophysical jet, and dark matter \cite{weng2, quaternion1, quaternion10} and so forth.

In the complex-sedenion space, it is able to describe simultaneously the physical properties of four fundamental interactions. In terms of the four fundamental interactions, the paper is capable of deducing not only the physical quantities and field equations for the classical mechanics on the macroscopic scale, but also the wavefunctions and field equations for the quantum mechanics on the microscopic scale. Subsequently, these arguments and field equations, in the classical mechanics and quantum mechanics, can be extended from the flat space into the curved space.

1) Sedenion space. As is well known, the complex-quaternion space can be utilized to describe the electromagnetic field equations. By a logical extension of this point, each of complex-quaternion spaces is able to study one field. As a result, the complex-octonion space is capable of depicting simultaneously the electromagnetic and gravitational fields. Further the complex-sedenion space can be applied to explore simultaneously the physical properties of four fundamental-fields and twelve adjoint-fields.

2) Classical mechanics. In the curved fundamental-space (in Section 2) described with the complex-sedenions, it is able to explore the field equations in the classical mechanics, including the conventional electromagnetic field equations and gravitational field equations and so forth. Further, in the curved composite-space (in Section 6) described with the complex-sedenions, it is capable of concluding the inference of `the physical quantities dominate the spatial parameters'. This inference is concordant with that derived from the GR.

3) Quantum mechanics. In the curved quantum-space (in Section 5) described with the complex-sedenions, it is capable of researching the wave functions and wave equations in the quantum mechanics, including the conventional Dirac equation, Schrodinger equation, and Yang-Mills equation and so forth. Meanwhile, in the curved product-space (in Section 7) described with the complex-sedenions, it is capable of achieving the deduction of `the physical quantities dominate the spatial parameters' also. This deduction inherits and extends the Cartesian academic thought of `the space is the extension of substance' apparently.

In the paper, by means of the composite vector \cite{weng3} , it is able to achieve the complex-sedenion composite-space from the complex-sedenion fundamental-space, for the classical mechanics on the macroscopic scale. Making use of the auxiliary quantity, it is capable of deducing the complex-sedenion quantum-space from the complex-sedenion fundamental-space, in the quantum mechanics on the microscopic scale. Subsequently, in virtue of the composite vector and auxiliary quantity, the complex-sedenion product-space can be derived from the fundamental-space and/or quantum-space. Meanwhile, one can infer some arguments and field equations in these flat spaces. For these different flat spaces, a majority of arguments and field equations in one flat space may be different from that of other flat spaces.

In the complex-sedenion curved space, it is able to extend these arguments and field equations from the complex-sedenion flat space into the complex-sedenion curved space, by means of the properties of tangent-frames and tensors. Furthermore, in the curved composite-space, some physical quantities dominate several spatial parameters. The latter exerts an influence on the arguments and field equations relevant to four classical-fields (in Section 4), in the curved fundamental-space. Next, in the curved product-space, some physical quantities will dominate several spatial parameters also. The latter has an influence on the arguments and field equations relevant to four fundamental quantum-fields (in Section 5), in the curved quantum-space. In the curved composite-space or product-space, it is able to deduce a few conclusions accordant with the GR, under some approximate circumstances (in Section 8). In other words, the paper and GR both succeed to the Cartesian academic thought of `the space is the extension of substance'.

It is worth noting that several physics quantities, in the gravitational and electromagnetic fields, may exert an influence on the spatial parameters of the complex-octonion curved space, and then the complex-octonion curved space will impact on the forces. These are the researches we have discussed in Refs. [25] and [41], they are very noticeably different from the paper, especially the sorts of interactions and the dimensions of space. In the paper, we place great emphasis on some physics properties of four interactions in the complex-sedenion curved space, including the contradistinctions between the complex-sedenions with complex-octonions, the contributions of physics quantities regarding the four interactions on the complex-sedenion curved space, and the influence of complex-sedenion curved space on the forces.

\section{\label{sec:level1}Flat space}

It is well known that two perpendicular quaternion spaces compose one octonion space. And four quaternion spaces, which are perpendicular to each other, will combine together to become one sedenion space. When a part of coordinate values are complex numbers, the quaternion, octonion, and sedenion are the complex-quaternion, complex-octonion, and complex-sedenion respectively. Meanwhile, the standard (rather than split- and others) quaternion space, octonion space, and sedenion space are called as the fundamental-spaces temporarily. From these fundamental-spaces, it is able to evolve further some function spaces, including the composite-space, quantum-space, and product-space. The complex-octonion space can be utilized to research the physical property of electromagnetic and gravitational fields, including their properties in the flat space and curved space. On the basis of the studies relevant to the complex-octonion spaces, the complex-sedenion spaces can be applied to explore several theories associated with the four fundamental interactions, including the flat space, curved space, classical mechanics on the macroscopic scale, and quantum mechanics on the microscopic scale.

\subsection{\label{sec:level1}Four fields}

According to the basic postulates (see Ref.[25]), one quaternion space is able to describe a type of field. As a result, two perpendicular complex-quaternion spaces (or one complex-octonion space) can be utilized to express the physical properties of two fields (such as, electromagnetic and gravitational fields). Four complex-quaternion spaces, which are perpendicular to each other, can be combined together to become one complex-sedenion space, exploring the physical properties of four fundamental interactions.

It is noteworthy that the four fields, in the paper, comprise the gravitational field, electromagnetic field, and strong-nuclear field, except for the weak-nuclear field. In the existing electroweak theory, the weak-nuclear field and electromagnetic field can be unified into the `electroweak field'. Each of weak-nuclear field and electromagnetic field is merely a constituent of the electroweak field. In the quantum mechanics (in Section 5) described with the complex-sedenions, the complex-sedenion electromagnetic field is able to be degenerated into the existing electroweak field, while the weak-nuclear field can be regarded as the adjoint-field of the electromagnetic field. Therefore, the weak-nuclear field is no longer considered as an independent field. In other words, as one crucial constituent, the weak-nuclear field is merged into the electromagnetic field described with the complex-sedenions.

Nevertheless, the multiplicative closure of sedenions affirms that there must be four types of fields in the complex-sedenion space simultaneously. So there will be a new species of unknown field. It is called as W-nuclear field temporarily, marking with the initial of `weak' traditionally. In the following context, the four fields designate the gravitational field, electromagnetic field, W-nuclear field, and strong-nuclear field.

\subsection{\label{sec:level1}Sedenion space}

In the complex-quaternion space $\mathbb{H}_g$ for the gravitational field, the coordinate values are $i R^{g0}$ and $R^{gq}$ , the basis vector is $\textbf{\emph{I}}_{gj}$ , and the complex-quaternion radius vector is, $\mathbb{R}_g = i \textbf{\emph{I}}_{g0} R^{g0} + \textbf{\emph{I}}_{gq} R^{gq}$ . Similarly, in the complex 2-quaternion (short for the second quaternion) space $\mathbb{H}_e$ for the electromagnetic field, the coordinate values are $i R^{e0}$ and $R^{eq}$ , the basis vector is $\textbf{\emph{I}}_{ej}$ , and the complex 2-quaternion radius vector is, $\mathbb{R}_e = i \textbf{\emph{I}}_{e0} R^{e0} + \textbf{\emph{I}}_{eq} R^{eq}$ . In the complex 3-quaternion (short for the third quaternion) space $\mathbb{H}_w$ for the W-nuclear field, the coordinate values are $i R^{w0}$ and $R^{wq}$ , the basis vector is $\textbf{\emph{I}}_{wj}$ , and the complex 3-quaternion radius vector is, $\mathbb{R}_w = i \textbf{\emph{I}}_{w0} R^{w0} + \textbf{\emph{I}}_{wq} R^{wq}$ . In the complex 4-quaternion (short for the fourth quaternion) space $\mathbb{H}_s$ for the strong-nuclear field, the coordinate values are $i R^{s0}$ and $R^{sq}$ , the basis vector is $\textbf{\emph{I}}_{sj}$ , and the complex 4-quaternion radius vector is, $\mathbb{R}_s = i \textbf{\emph{I}}_{s0} R^{s0} + \textbf{\emph{I}}_{sq} R^{sq}$ . Herein, the superscripts or subscripts, $g$, $e$, $w$, and $s$, are applied to mark respectively the physical quantities in the gravitational field, electromagnetic field, W-nuclear field, and strong-nuclear field. $R^{gj}$, $R^{ej}$, $R^{wj}$, and $R^{sj}$ are all real. $R^{g0} = v^0 t$. $v^0$ is the speed of light, and $t$ is the time. $\textbf{\emph{I}}_{g0} = 1$. $( \textbf{\emph{I}}_{g0} )^2 = 1$ . $( \textbf{\emph{I}}_{gq} )^2 = - 1$. $\textbf{\emph{I}}_{ej} = \textbf{\emph{I}}_{gj} \circ \textbf{\emph{I}}_{e0}$ . $( \textbf{\emph{I}}_{ej} )^2 = - 1$ . $\textbf{\emph{I}}_{wj} = \textbf{\emph{I}}_{gj} \circ \textbf{\emph{I}}_{w0}$. $( \textbf{\emph{I}}_{wj} )^2 = - 1$ . $\textbf{\emph{I}}_{s0} = \textbf{\emph{I}}_{g0} \circ \textbf{\emph{I}}_{s0}$. $\textbf{\emph{I}}_{sq} = - \textbf{\emph{I}}_{gq} \circ \textbf{\emph{I}}_{s0}$ . $( \textbf{\emph{I}}_{sj} )^2 = - 1$ . The symbol $\circ$ denotes the sedenion multiplication. $i$ is the imaginary unit. $j, k = 0, 1, 2, 3$. $q = 1, 2, 3$.

In the above, four complex-quaternion spaces, $\mathbb{H}_g$ , $\mathbb{H}_e$ , $\mathbb{H}_w$ , and $\mathbb{H}_s$ , are perpendicular to each other. They can be combined together to become one complex-sedenion space, $\mathbb{K}$ . In the complex-sedenion space, $\mathbb{K}$ , the basis vectors are, $\textbf{\emph{I}}_{gj}$ ,
$\textbf{\emph{I}}_{ej}$, $\textbf{\emph{I}}_{wj}$ , and $\textbf{\emph{I}}_{sj}$ . The complex-sedenion radius vector is, $\mathbb{R} = \mathbb{R}_g + k_{eg} \mathbb{R}_e + k_{wg} \mathbb{R}_w + k_{sg} \mathbb{R}_s$. Herein, $k_{eg}$ , $k_{wg}$ , and $k_{sg}$ are coefficients, to meet the requirement of the dimensional homogeneity.

In the complex-quaternion space $\mathbb{H}_g$ for the gravitational field, the complex-quaternion operator is, $\lozenge_g = i \textbf{\emph{I}}_{g0} \partial_{g0} + \textbf{\emph{I}}_{gq} \partial_{gq}$ , and $\partial_{gj} = \partial / \partial R^{gj}$ . In the complex 2-quaternion space $\mathbb{H}_e$ for the electromagnetic field, the complex 2-quaternion operator is, $\lozenge_e = i \textbf{\emph{I}}_{e0} \partial_{e0} + \textbf{\emph{I}}_{eq} \partial_{eq}$ , and $\partial_{ej} = \partial / \partial R^{ej}$ . In the complex 3-quaternion space $\mathbb{H}_w$ for the W-nuclear field, the complex 3-quaternion operator is, $\lozenge_w = i \textbf{\emph{I}}_{w0} \partial_{w0} + \textbf{\emph{I}}_{wq} \partial_{wq}$, and $\partial_{wj} = \partial / \partial R^{wj}$ . In the complex 4-quaternion space $\mathbb{H}_s$ for the strong-nuclear field, the complex 4-quaternion operator is, $\lozenge_s = i \textbf{\emph{I}}_{s0} \partial_{s0} + \textbf{\emph{I}}_{sq} \partial_{sq}$ , and $\partial_{sj} = \partial / \partial R^{sj}$. Consequently, the complex-sedenion operator is, $\lozenge = \lozenge_g + k_{eg}^{~~-1} \lozenge_e + k_{wg}^{~~-1} \lozenge_w + k_{sg}^{~~-1} \lozenge_s$, in the complex-sedenion space, $\mathbb{K}$ . Herein, $\lozenge_g = i \textbf{\emph{I}}_{g0} \partial_{g0} + \nabla_g$ . $\lozenge_e = i \textbf{\emph{I}}_{e0} \partial_{e0} + \nabla_e$. $\lozenge_w = i \textbf{\emph{I}}_{w0} \partial_{w0} + \nabla_w$ . $\lozenge_s = i \textbf{\emph{I}}_{s0} \partial_{s0} + \nabla_s$. $\nabla_g = \textbf{\emph{I}}_{gq} \partial_{gq} $ . $\nabla_e = \textbf{\emph{I}}_{eq} \partial_{eq} $ . $\nabla_w = \textbf{\emph{I}}_{wq} \partial_{wq} $ . $\nabla_s = \textbf{\emph{I}}_{sq} \partial_{sq} $ .

\section{\label{sec:level1}Curved space}

For the sake of explaining some comparative complicated physical phenomena, it is necessary to expand the field theory, associated with the four fundamental interactions, from the flat space into the curved space. In virtue of the properties of tangent space and tensor, it is capable of exploring the curved spaces described with the quaternions, octonions, and sedenions respectively. In the complex-octonion space, without the help of the ¡®equivalence principle¡¯ relevant to the E\"{o}tv\"{o}s experiment, it is still able to introduce the curved spaces into the field theory, researching the physical properties of electromagnetic and gravitational fields. Similarly, in the complex-sedenion space, we can introduce the curved spaces also, investigating the physical properties of four fundamental interactions. And these theories are able to deduce some conclusions, which are accordant with the GR, under the approximate circumstances. In the complex-sedenion curved space, the paper is able to achieve not only the arguments and field equations of the classical mechanics on the macroscopic scale, but also the arguments and field equations of the quantum mechanics on the macroscopic scale.

In the complex-sedenion curved space, by means of the properties of manifold, affine connection space, parallel translation, and metric space, it is able to deduce some geometric parameters in the orthogonal and unequal-length tangent-frame, including the metric coefficients, connection coefficients, covariant derivative, and curvature tensor and so forth.

\subsection{\label{sec:level1}Metric space}

The manifold is a type of indefinite geometric object. The coordinate systems are allowed to establish in the manifold. The manifold possesses neither the metric tensor nor the affine connection. Later, one can append properly a few geometric properties in the manifolds. Even though there is no metric property in the manifold, it is still able to define the connection among different tensors in contiguous points. The manifold with the connection is the affine connection space, which is seized of the connection coefficient and curvature tensor, and even the torsion tensor.

In some manifolds, when a tensor transfers from one point to another to meet the need of parallel translation, it means that the differential of this tensor is equal to zero. According to the definition of parallel translation, we can achieve the relations between the tensor component and the connection coefficient. After defining the affine connection among the tensors, it is able to append subsequently the metric property to the affine connection space. The manifold, with the parallel translation and metric property, is called as the metric space, including the Riemannian space and pseudo-Riemannian space.

In the metric space, there are several interrelationships of the connection coefficient and metric coefficient. The connection coefficient can be derived from the metric coefficient, enabling these two coefficients to be consistent with each other. Later the curvature tensor is inferred from the connection coefficient. In the complex-sedenion curved space, it is able to define the tangent-frame component, by means of the partial derivative of complex-sedenion tensor with respect to the coordinate value. According to the definition of sedenion norm, the metric coefficient can be derived from the tangent-frame component, and it is sedenion-Hermitian.

\subsection{\label{sec:level1}Tangent space}

In the curved space described with the complex-sedenions, not all tangent-frames are suitable to be chosen as the curved coordinate systems, in terms of the four fundamental interactions. As a result, it is essential to demarcate and filter out the appropriate tangent-frames further. When we study the physics quantities and relevant properties in the complex-sedenion curved space, it is proper to stipulate to choose the orthogonal and unequal-length tangent-frame (or the orthogonal curved coordinate system), reducing the mathematical difficulty related with the nonassociativity of octonions and sedenions in the following context. Especially, it will be beneficial to lower down the difficulty of mathematics encountered in the paper, as the metric coefficient and connection coefficient both are scalars.

Apparently, the physical quantities of the four fundamental interactions, described with the complex-sedenions, can be extended from the flat space into the curved space. Firstly, convert the physical quantities of four fundamental interactions in the flat space into that under the orthogonal and equal-length tangent-frame of the tangent space in the curved space. Secondly, in terms of the tangent space, convert the physical quantities of four fundamental interactions in the orthogonal and equal-length tangent-frame into that in the orthogonal and unequal-length tangent-frame. Thirdly, convert the physical quantities of four fundamental interactions in the orthogonal and unequal-length tangent-frame into that in the unorthogonal and unequal-length tangent-frame, and even others.

In the complex-sedenion curved space, the covariant derivative of any argument may comprise several parameters of the curved space, enabling a physical quantity with the covariant derivative in the curved space to departure from that in the flat space to a certain extent. Firstly, it is able to convert the physical quantities of four fundamental interactions in the unorthogonal and unequal-length tangent-frame into that in the orthogonal and unequal-length tangent-frame of the tangent space. Secondly, the physical quantities of four fundamental interactions in the orthogonal and unequal-length tangent-frame are converted into that in the orthogonal and equal-length tangent-frame. Lastly, it is possible to distinguish the bending degrees of a curved space, to contrast the physics quantities in the tangent space with that in the flat space. That is, by means of contrasting the measurement result in the tangent space with the theoretical prediction in the flat space, one can estimate the bending degree of curved space.

\subsection{\label{sec:level1}Connection coefficient}

In the complex-sedenion flat space, it is appropriate to choose an orthogonal and equal-length tangent-frame as the coordinate system. In the coordinate system, the complex-sedenion radius vector can be written as,
\begin{eqnarray}
&& \mathbb{R} ( h^\alpha ) = i h^0 \textbf{\emph{i}}_0 + h^q \textbf{\emph{i}}_q + i h^4 \textbf{\emph{i}}_4 + h^{4+q} \textbf{\emph{i}}_{4+q}
\nonumber
\\
&&
~~~~~~~~~~~~
+ i h^8 \textbf{\emph{i}}_8 + h^{8+q} \textbf{\emph{i}}_{8+q} + i h^{12} \textbf{\emph{i}}_{12} + h^{12+q} \textbf{\emph{i}}_{12+q}   ~ ,
\end{eqnarray}
where $h^\alpha$ is the coordinate value. $h^j = R^{gj}$ , $h^{j+4} = k_{eg} R^{ej}$ , $h^{j+8} = k_{wg} R^{wj}$, and $h^{j+12} = k_{sg} R^{sj}$ . In the paper, $(k_{eg})^2 < 0$ , furthermore it is supposed that, $(k_{wg})^2 < 0$, and $(k_{sg})^2 < 0$ . According to the dimensional homogeneity, $h^j$ is real, while $h^{j+4}$ , $h^{j+8}$ , and $h^{j+12}$ are the imaginary numbers. $\textbf{\emph{i}}_\alpha$ is the basis vector. $\textbf{\emph{i}}_j = \textbf{\emph{I}}_{gj}$ . $\textbf{\emph{i}}_{j+4} = \textbf{\emph{I}}_{ej}$. $\textbf{\emph{i}}_{j+8} = \textbf{\emph{I}}_{wj}$ . $\textbf{\emph{i}}^{j+12} = \textbf{\emph{I}}_{sj}$ . $\xi = 1, 2, 3, 4, 5, 6, 7, 8, 9, 10, 11, 12, 13, 14, 15$ . $\textbf{\emph{i}}_0 = 1$. $(\textbf{\emph{i}}_0)^2 = 1$. $(\textbf{\emph{i}}_\xi)^2 = -1$ . $\alpha, \beta, \gamma, \lambda, \nu = 0, 1, 2, 3, 4, 5, 6, 7, 8, 9, 10, 11, 12, 13, 14, 15$ .

In the complex-sedenion curved space, a tangent space may be considered as a locally flat space. As a result, the physical laws and research methods can be extended from the complex-sedenion flat space into the tangent spaces. The physical quantities and field equations in the rectangular coordinate system of the flat space are converted into that in the orthogonal and equal-length tangent-frame of the tangent space. Further they are converted into that in the orthogonal and unequal-length tangent-frame, and even others.

In the tangent space of the complex-sedenion curved space, one orthogonal and unequal-length tangent-frame is chosen as the coordinate system. In the coordinate system, the complex-sedenion radius vector is expanded in terms of the tangent-frame component $\textbf{\emph{e}}_\alpha$ ,
\begin{eqnarray}
&& \mathbb{R} ( c^\alpha ) = i c^0 \textbf{\emph{e}}_0 + c^q \textbf{\emph{e}}_q + i c^4 \textbf{\emph{e}}_4 + c^{4+q} \textbf{\emph{e}}_{4+q}
\nonumber
\\
&&
~~~~~~~~~~~~
+ i c^8 \textbf{\emph{e}}_8 + c^{8+q} \textbf{\emph{e}}_{8+q} + i c^{12} \textbf{\emph{e}}_{12} + c^{12+q} \textbf{\emph{e}}_{12+q}   ~ ,
\end{eqnarray}
where $c^\alpha$ is the coordinate value. According to the dimensional homogeneity, $c^j$ is real, while $c^{j+4}$ , $c^{j+8}$ , and $c^{j+12}$ are imaginary numbers. $\textbf{\emph{e}}_\alpha$ is unequal-length. $(\textbf{\emph{e}}_0)^2 > 0$ , and $(\textbf{\emph{e}}_\xi)^2 < 0$ . In the space $\mathbb{H}_g$ , choosing an appropriate coordinate system to satisfy the conditions that, $\textbf{\emph{e}}_0$ is the scalar part (corresponding to $\textbf{\emph{i}}_0$), $\textbf{\emph{e}}_q$ is the component of vector part (corresponding to $\textbf{\emph{i}}_q$). Similarly, $\textbf{\emph{e}}_4$ , $\textbf{\emph{e}}_8$ , and $\textbf{\emph{e}}_{12}$ are the `scalar' parts in the spaces, $\mathbb{H}_e$ , $\mathbb{H}_w$ , $\mathbb{H}_s$ , respectively, and correspond to $\textbf{\emph{i}}_4$, $\textbf{\emph{i}}_8$ , and $\textbf{\emph{i}}_{12}$ respectively. Meanwhile $\textbf{\emph{e}}_{4+q}$ , $\textbf{\emph{e}}_{8+q}$ , and $\textbf{\emph{e}}_{12+q}$ are the components of `vector' parts, and correspond to $\textbf{\emph{i}}_{4+q}$, $\textbf{\emph{i}}_{8+q}$ , and $\textbf{\emph{i}}_{12+q}$ .

In the orthogonal and unequal-length tangent-frame, making use of the tangent-frame component and the norm of complex-sedenion radius vector, one can define the metric of complex-sedenion curved space $\mathbb{K}$ as follows,
\begin{eqnarray}
d S^2 = g_{\overline{\alpha}\beta} d \overline{u^\alpha} d u^\beta  ~ ,
\end{eqnarray}
where the metric coefficient, $g_{\overline{\alpha}\beta} = \textbf{\emph{e}}_\alpha^\ast \circ \textbf{\emph{e}}_\beta$ , is sedenion-Hermitian. $\textbf{\emph{e}}_\alpha$ is the component of tangent-frame, with $\textbf{\emph{e}}_\alpha = \partial \mathbb{R} / \partial u^\alpha$ . $\ast$ denotes the sedenion conjugate. $u^0 = i c^0$ , $u^q = c^q$ . $u^4 = i c^4$ , $u^{q+4} = c^{q+4}$ . $u^8 = i c^8$ , $u^{q+8} = c^{q+8}$. $u^{12} = i c^{12}$, $u^{q+12} = c^{q+12}$. $( u^\alpha )^\ast = \overline{u^\alpha}$ , and it indicates that the correlated tangent-frame component, $\textbf{\emph{e}}_\alpha$ , is sedenion conjugate. $ u^\alpha $ is a complex number. $g_{\overline{\alpha}\beta}$ is one real number and even complex number, due to the orthogonal tangent-frames.

\subsection{\label{sec:level1}Curvature tensor}

In the complex-sedenion curved space, by means of the metric coefficient, it is able to deduce the connection coefficient, covariant derivative, and curvature tensor. From the metric coefficient, one can infer the connection coefficient (Appendix A) as follows,
\begin{eqnarray}
\Gamma_{\overline{\lambda} , \beta \gamma } = (1/2) ( \partial g_{\overline{\gamma} \lambda} / \partial u^\beta + \partial g_{\overline{\lambda} \beta} / \partial u^\gamma - \partial g_{\overline{\gamma} \beta} /\partial u^\lambda ) ~ ,
\\
\Gamma_{\overline{\lambda} , \overline{\beta} \gamma } = (1/2) ( \partial g_{\overline{\gamma} \lambda} / \partial \overline{u^\beta} + \partial g_{\overline{\lambda} \beta} / \partial \overline{u^\gamma} - \partial g_{\overline{\gamma} \beta} /\partial \overline{u^\lambda} ) ~ ,
\end{eqnarray}
where $\Gamma_{\overline{\lambda} , \beta \gamma }$ and $\Gamma_{\overline{\lambda} , \overline{\beta} \gamma }$ both are connection coefficients, and are all scalar. $\Gamma_{\overline{\lambda} , \beta \gamma } = g_{\overline{\lambda} \alpha} \Gamma_{\beta \gamma}^\alpha$. $ \Gamma_{\beta \gamma}^\alpha = g^{\alpha \overline{\lambda}} \Gamma_{\overline{\lambda} , \beta \gamma } $ . $\Gamma_{\overline{\lambda} , \overline{\beta} \gamma } = g_{\overline{\lambda} \alpha} \Gamma_{\overline{\beta} \gamma}^\alpha$ . $ \Gamma_{\overline{\beta} \gamma}^\alpha =  g^{\alpha \overline{\lambda}} \Gamma_{\overline{\lambda} , \overline{\beta} \gamma } $ . $\Gamma_{\beta \gamma}^\alpha = \Gamma_{\gamma \beta}^\alpha $ . $\Gamma_{\overline{\beta} \gamma}^\alpha = \Gamma_{\gamma \overline{\beta}}^\alpha $ . $ g^{\alpha \overline{\lambda}}  g_{\overline{\lambda} \beta} = \delta^\alpha_\beta $. $[ ( \Gamma_{\overline{\beta} \gamma}^\alpha )^* ]^T = \Gamma_{\gamma \overline{\beta}}^\alpha $ . The superscript T denotes the transpose of matrix.

When a complex-sedenion quantity, $\mathbb{Y} = Y^\beta \textbf{\emph{e}}_\beta $ , is transferred from a point $M_1$ to the next point $M_2$ , to meet the requirement of parallel translation, it means that the differential of quantity $\mathbb{Y}$ equals to zero. And the condition of parallel translation, $d \mathbb{Y} = 0$ , will yield,
\begin{equation}
d Y^\beta = - \Gamma_{\alpha \gamma}^\beta Y^\alpha d u^\gamma  ~,
\end{equation}
with
\begin{equation}
\partial^2 \mathbb{R} / \partial u^\beta \partial u^\gamma = \Gamma_{\beta \gamma}^\alpha \textbf{\emph{e}}_\alpha  ~ .
\end{equation}

In the complex-sedenion curved space, for the first-rank contravariant tensor $Y^\beta$ of a point $M_2$ , the component of covariant derivative with respect to the coordinate $u^\gamma$ is written as,
\begin{eqnarray}
\nabla_\gamma Y^\beta = \partial ( \delta_\alpha^\beta Y^\alpha ) / \partial u^\gamma + \Gamma_{\alpha \gamma}^\beta Y^\alpha   ~ ,
\\
\nabla_{\overline{\gamma}} Y^\beta = \partial ( \delta_\alpha^\beta Y^\alpha ) / \partial \overline{u^\gamma} + \Gamma_{\alpha \overline{\gamma}}^\beta Y^\alpha   ~ ,
\end{eqnarray}
where $Y^\beta$ and $\Gamma_{\alpha \overline{\gamma}}^\beta$ both are scalar.

Further, from the covariant derivative of tensor, it is able to infer the curvature tensor, in the complex-sedenion curved space,
\begin{equation}
\nabla_{\overline{\alpha}} ( \nabla_\beta Y^\gamma ) - \nabla_\beta ( \nabla_{\overline{\alpha}} Y^\gamma ) = R_{\beta \overline{\alpha} \nu}^{~~~~\gamma} Y^\nu
+ T_{\beta \overline{\alpha}}^\lambda ( \nabla_\lambda Y^\gamma )  ~   ,
\end{equation}
with
\begin{equation}
R_{\beta \overline{\alpha} \nu}^{~~~~\gamma} = \partial \Gamma_{\nu \beta}^\gamma / \partial \overline{u^\alpha} - \partial \Gamma_{\nu \overline{\alpha}}^\gamma  / \partial u^\beta
+ \Gamma_{\lambda \overline{\alpha}}^\gamma \Gamma_{\nu \beta}^\lambda - \Gamma_{\lambda \beta}^\gamma \Gamma_{\nu \overline{\alpha}}^\lambda   ~~ ,
\end{equation}
where $T_{\beta \overline{\alpha}}^\lambda = \Gamma_{\overline{\alpha} \beta }^\lambda - \Gamma_{\beta \overline{\alpha}}^\lambda $ . $R_{\beta \overline{\alpha} \nu}^{~~~~\gamma}$ is the curvature tensor, while $T_{\beta \overline{\alpha}}^\lambda$ is the torsion tensor. $R_{\beta \overline{\alpha} \nu}^{~~~~\gamma}$ and $T_{\beta \overline{\alpha}}^\lambda$ both are scalar. In the paper, we merely discuss the case, $T_{\beta \overline{\alpha}}^\lambda = 0$, that is, $\Gamma_{\overline{\alpha} \beta }^\lambda = \Gamma_{\beta \overline{\alpha}}^\lambda $ .

In the complex-octonion space, the physical quantities and field equations can be extended from the flat space into the tangent space of the curved space, exploring the influences of covariant derivative and curvature tensor on the physical quantities and field equations of the gravitational and electromagnetic fields. Similarly, in the complex-sedenion space, a part of physical quantities and field equations of the four fundamental interactions can be extended from the flat space into the tangent space of the curved space. And the covariant derivative and curvature tensor will make a contribution to these physical quantities and field equations to a certain extent.

\begin{table}[h]
\tbl{The physical quantities and definitions of the four classical-fields, which are relevant to the classical mechanics on the macroscopic scale, in the complex-sedenion curved space.}
{\begin{tabular}{@{}ll@{}} \toprule
octonion~physics~quantity ~~~~~~~~    &  definition                                                                                           \\
\colrule
integrating~function                  &  $\mathbb{X} = \mathbb{X}_g + k_{eg} \mathbb{X}_e + k_{wg} \mathbb{X}_w + k_{sg} \mathbb{X}_s$        \\
field~potential                       &  $\mathbb{A} = i \lozenge^\star \circ \mathbb{X}  $                                                   \\
field~strength                        &  $\mathbb{F} = \lozenge \circ \mathbb{A}  $                                                           \\
field~source                          &  $\mu \mathbb{S} = - ( i \mathbb{F} / v^0 + \lozenge )^* \circ \mathbb{F} $                           \\
linear~momentum                       &  $\mathbb{P} = \mu \mathbb{S} / \mu_g^g $                                                             \\
angular~momentum                      &  $\mathbb{L} = ( \mathbb{R} + k_{rx} \mathbb{X} )^\star \circ \mathbb{P} $                            \\
octonion~torque                       &  $\mathbb{W} = - v^0 ( i \mathbb{F} / v^0 + \lozenge ) \circ \mathbb{L} $                             \\
octonion~force                        &  $\mathbb{N} = - ( i \mathbb{F} / v^0 + \lozenge ) \circ \mathbb{W} $                                 \\
\botrule
\end{tabular}}
\end{table}

\section{\label{sec:level1}Classical field equations}

In virtue of the properties of complex-sedenions, it is able to deduce the physical quantities and field equations, for the classical mechanics on the macroscopic scale. They are relevant to the physical quantities of four classical-fields (gravitational field, electromagnetic field, W-nuclear field, and strong-nuclear field), including the complex-sedenion field potential, field strength, field source, linear momentum, angular momentum, torque, and force (Table 1).

\subsection{\label{sec:level1}Integrating function}

In the tangent space of the complex-sedenion curved space $\mathbb{K}$ , one orthogonal and equal-length tangent-frame $\Pi$ is chosen as the coordinate system. In this tangent space, the complex-sedenion integrating function of field potential, $\mathbb{X} = \mathbb{X}_g + k_{eg} \mathbb{X}_e + k_{wg} \mathbb{X}_w + k_{sg} \mathbb{X}_s$, is written as,
\begin{eqnarray}
&& \mathbb{X} ( x^\alpha ) = i x^0 \textbf{\emph{i}}_0 + x^q \textbf{\emph{i}}_q + i x^4 \textbf{\emph{i}}_4 + x^{4+q} \textbf{\emph{i}}_{4+q}
\nonumber
\\
&&
~~~~~~~~~~~~
+ i x^8 \textbf{\emph{i}}_8 + x^{8+q} \textbf{\emph{i}}_{8+q} + i x^{12} \textbf{\emph{i}}_{12} + x^{12+q} \textbf{\emph{i}}_{12+q}   ~ ,
\end{eqnarray}
where $\mathbb{X}_g$ , $\mathbb{X}_e$ , $\mathbb{X}_w$ , and $\mathbb{X}_s$ are respectively the components of the integrating function of field potential $\mathbb{X}$ in four complex-quaternion spaces, $\mathbb{H}_g$ , $\mathbb{H}_e$ , $\mathbb{H}_w$, and $\mathbb{H}_s$ . $\mathbb{X}_g = i \textbf{\emph{I}}_{g0} X^{g0} + \textbf{\emph{I}}_{gq} X^{gq}$ . $\mathbb{X}_e = i \textbf{\emph{I}}_{e0} X^{e0} + \textbf{\emph{I}}_{eq} X^{eq}$ . $\mathbb{X}_w = i \textbf{\emph{I}}_{w0} X^{w0} + \textbf{\emph{I}}_{wq} X^{wq}$. $\mathbb{X}_s = i \textbf{\emph{I}}_{s0} X^{s0} + \textbf{\emph{I}}_{sq} X^{sq}$ . $X^{gj}$ , $X^{ej}$ , $X^{wj}$ , and $X^{sj}$ are all real. The components, $X^{gj}$, $k_{eg} X^{ej}$ , $k_{wg} X^{wj}$ , and $k_{sg} X^{sj}$ , are respectively written as, $x^j$, $x^{j+4}$ , $x^{j+8}$ , and $x^{j+12}$ .

After the tangent-frame $\Pi$ is converted into the orthogonal and unequal-length tangent-frame $\Theta$ , the complex-sedenion integrating function of field potential, $\mathbb{X}$ , is expanded in terms of the tangent-frame component $\textbf{\emph{e}}_\alpha$ ,
\begin{eqnarray}
&& \mathbb{X} ( y^\alpha ) = i y^0 \textbf{\emph{e}}_0 + y^q \textbf{\emph{e}}_q + i y^4 \textbf{\emph{e}}_4 + y^{4+q} \textbf{\emph{e}}_{4+q}
\nonumber
\\
&&
~~~~~~~~~~~~
+ i y^8 \textbf{\emph{e}}_8 + y^{8+q} \textbf{\emph{e}}_{8+q} + i y^{12} \textbf{\emph{e}}_{12} + y^{12+q} \textbf{\emph{e}}_{12+q}   ~ ,
\end{eqnarray}
where $y^a$ corresponds to $x^a$ , while $\textbf{\emph{e}}_\alpha$ corresponds to $\textbf{\emph{i}}_\alpha$ .

Further the above can be written as,
\begin{eqnarray}
\mathbb{X} ( z^\alpha ) = z^\alpha \textbf{\emph{e}}_\alpha ~,
\end{eqnarray}
where $z^0 = i y^0$ , $z^q = y^q$ . $z^4 = i y^4$ , $z^{q+4} = y^{q+4}$ . $z^8 = i y^8$ , $z^{q+8} = y^{q+8}$ . $z^{12} = i y^{12}$ , $z^{q+12} = y^{q+12}$ .

\subsection{\label{sec:level1}Field potential}

In the orthogonal and equal-length tangent-frame $\Pi$ , the complex-sedenion field potential can be written as,
\begin{eqnarray}
\mathbb{A} = i \lozenge^\star \circ \mathbb{X}   ~,
\end{eqnarray}
where $\mathbb{A} = \mathbb{A}_g + k_{eg} \mathbb{A}_e + k_{wg} \mathbb{A}_w + k_{sg} \mathbb{A}_s$ . $\mathbb{A}_g$ , $\mathbb{A}_e$ , $\mathbb{A}_w$ , and $\mathbb{A}_s$ are respectively the components of field potential $\mathbb{A}$ in four complex-quaternion spaces, $\mathbb{H}_g$, $\mathbb{H}_e$ , $\mathbb{H}_w$ , and $\mathbb{H}_s$ . $\mathbb{A}_g = i \textbf{\emph{I}}_{g0} A^{g0} + \textbf{\emph{I}}_{gq} A^{gq}$ . $\mathbb{A}_e = i \textbf{\emph{I}}_{e0} A^{e0} + \textbf{\emph{I}}_{eq} A^{eq}$ . $\mathbb{A}_w = i \textbf{\emph{I}}_{w0} A^{w0} + \textbf{\emph{I}}_{wq} A^{wq}$. $\mathbb{A}_s = i \textbf{\emph{I}}_{s0} A^{s0} + \textbf{\emph{I}}_{sq} A^{sq}$ . $A^{gj}$ , $A^{ej}$ , $A^{wj}$ , and $A^{sj}$ are all real. The symbol $\star$ stands for the complex conjugate. $ \lozenge^\star = \lozenge_g^\star + k_{eg}^{~~-1} \lozenge_e^\star + k_{wg}^{~~-1} \lozenge_w^\star + k_{sg}^{~~-1} \lozenge_s^\star $ . Especially, in the context, the operation of complex conjugate will be applied to some physical quantities, except for the coefficients, $k_{eg}$ , $k_{wg}$ , and $k_{sg}$ .

In the orthogonal and unequal-length tangent-frame $\Theta$ , the complex-sedenion field potential $\mathbb{A}$ can be expanded in terms of the tangent-frame component $\textbf{\emph{e}}_\alpha$,
\begin{eqnarray}
\mathbb{A} ( A^\alpha ) = A^\alpha \textbf{\emph{e}}_\alpha ~,
\end{eqnarray}
where $A^0 = i a^0$ , $A^q = a^q$ . $A^4 = i a^4$ , $A^{q+4} = a^{q+4}$ . $A^8 = i a^8$ , $A^{q+8} = a^{q+8}$. $A^{12} = i a^{12}$, $A^{q+12} = a^{q+12}$ . The  components, $a^j$ , $a^{j+4}$ , $a^{j+8}$ , and $a^{j+12}$ , correspond to $A^{gj}$ , $k_{eg} A^{ej}$, $k_{wg} A^{wj}$ , and $k_{sg} A^{sj}$ respectively.

In the tangent-frame $\Theta$ , by means of the property of the integrating function of field potential $\mathbb{X}$ , the complex-sedenion field potential, $\mathbb{A} = i \lozenge^\star \circ \mathbb{X}$ , can be separated into,
\begin{eqnarray}
\mathbb{A} = i \lozenge^\star \circledcirc \mathbb{X} + i \lozenge^\star \circledast \mathbb{X}  ~,
\end{eqnarray}
where $i \lozenge^\star \circledcirc \mathbb{X}$ and $i \lozenge^\star \circledast \mathbb{X}$ are respectively the scalar part and vector part of field potential $\mathbb{A}$ . Apparently, in the tangent-frame $\Theta$ , the field potential $\mathbb{A}$ consists of the derivative of the integrating function of field potential $\mathbb{X}$ , and the spatial parameter of complex-sedenion curved space.

\subsection{\label{sec:level1}Field strength}

In the orthogonal and equal-length tangent-frame $\Pi$ , the complex-sedenion field strength can be written as,
\begin{eqnarray}
\mathbb{F} = \lozenge \circ \mathbb{A}   ~,
\end{eqnarray}
where $\mathbb{F} = \mathbb{F}_g + k_{eg} \mathbb{F}_e + k_{wg} \mathbb{F}_w + k_{sg} \mathbb{F}_s$ . $\mathbb{F}_g$ , $\mathbb{F}_e$ , $\mathbb{F}_w$ , and $\mathbb{F}_s$ are respectively the components of field strength $\mathbb{F}$ in four complex-quaternion spaces, $\mathbb{H}_g$, $\mathbb{H}_e$ , $\mathbb{H}_w$ , and $\mathbb{H}_s$ . $\mathbb{F}_g = i \textbf{\emph{I}}_{g0} F^{g0} + \textbf{\emph{I}}_{gq} F^{gq}$ . $\mathbb{F}_e = i \textbf{\emph{I}}_{e0} F^{e0} + \textbf{\emph{I}}_{eq} F^{eq}$ . $\mathbb{F}_w = i \textbf{\emph{I}}_{w0} F^{w0} + \textbf{\emph{I}}_{wq} F^{wq}$. $\mathbb{F}_s = i \textbf{\emph{I}}_{s0} F^{s0} + \textbf{\emph{I}}_{sq} F^{sq}$ . $F^{gj}$ , $F^{ej}$ , $F^{wj}$ , and $F^{sj}$ are the complex numbers.

In the orthogonal and unequal-length tangent-frame $\Theta$ , the complex-sedenion field strength $\mathbb{F}$ can be expanded in terms of the tangent-frame component $\textbf{\emph{e}}_\alpha$ ,
\begin{eqnarray}
\mathbb{F} ( F^\alpha ) = F^\alpha \textbf{\emph{e}}_\alpha ~,
\end{eqnarray}
where $F^0 = i f^0$ , $F^q = f^q$ . $F^4 = i f^4$ , $F^{q+4} = f^{q+4}$ . $F^8 = i f^8$ , $F^{q+8} = f^{q+8}$. $F^{12} = i f^{12}$ , $F^{q+12} = f^{q+12}$ . The  components, $f^j$ , $f^{j+4}$ , $f^{j+8}$ , and $f^{j+12}$ , correspond to $F^{gj}$ , $k_{eg} F^{ej}$ , $k_{wg} F^{wj}$ , and $k_{sg} F^{sj}$ respectively.

In the tangent-frame $\Theta$ , by means of the property of the field potential $\mathbb{A}$ , the complex-sedenion field strength, $\mathbb{F} = \lozenge \circ \mathbb{A}$ , can be separated into,
\begin{eqnarray}
\mathbb{F} = \lozenge \circledcirc \mathbb{A} + \lozenge \circledast \mathbb{A}  ~,
\end{eqnarray}
where $\lozenge \circledcirc \mathbb{A}$ and $\lozenge \circledast \mathbb{A}$ are respectively the scalar part and vector part of field strength $\mathbb{F}$ . Apparently, in the tangent-frame $\Theta$ , the field strength $\mathbb{F}$ contains the derivative of the field potential $\mathbb{A}$ , and the spatial parameter of complex-sedenion curved space.

\subsection{\label{sec:level1}Field source}

In the orthogonal and equal-length tangent-frame $\Pi$ , the complex-sedenion field source can be written as,
\begin{eqnarray}
\mu \mathbb{S} = - ( i \mathbb{F} / v^0 + \lozenge )^\ast \circ \mathbb{F}  ~,
\end{eqnarray}
where $\mu \mathbb{S} = \mu_g \mathbb{S}_g + k_{eg} \mu_e \mathbb{S}_e + k_{wg} \mu_w \mathbb{S}_w + k_{sg} \mu_s \mathbb{S}_s - i \mathbb{F}^\ast \circ \mathbb{F} / v^0 $ . $\mathbb{S}_g$ , $\mathbb{S}_e$ , $\mathbb{S}_w$ , and $\mathbb{S}_s$ are respectively the components of field source $\mathbb{S}$ in four complex-quaternion spaces, $\mathbb{H}_g$ , $\mathbb{H}_e$, $\mathbb{H}_w$ , and $\mathbb{H}_s$ . $\mathbb{S}_g = i \textbf{\emph{I}}_{g0} S^{g0} + \textbf{\emph{I}}_{gq} S^{gq}$ . $\mathbb{S}_e = i \textbf{\emph{I}}_{e0} S^{e0} + \textbf{\emph{I}}_{eq} S^{eq}$. $\mathbb{S}_w = i \textbf{\emph{I}}_{w0} S^{w0} + \textbf{\emph{I}}_{wq} S^{wq}$. $\mathbb{S}_s = i \textbf{\emph{I}}_{s0} S^{s0} + \textbf{\emph{I}}_{sq} S^{sq}$ . $S^{gj}$ , $S^{ej}$ , $S^{wj}$ , and $S^{sj}$ are all real. $\mu$ , $\mu_g$ , $\mu_e$ , $\mu_w$, and $\mu_s$ are coefficients. From the above, it is able to deduce the electromagnetic field equations and gravitational field equations, in the orthogonal and equal-length tangent-frame $\Pi$ of the complex-sedenion flat space.

In the orthogonal and unequal-length tangent-frame $\Theta$ , the complex-sedenion field source $\mathbb{S}$ can be expanded in terms of the tangent-frame component $\textbf{\emph{e}}_\alpha$ ,
\begin{eqnarray}
\mathbb{S} ( S^\alpha ) = S^\alpha \textbf{\emph{e}}_\alpha ~,
\end{eqnarray}
where $S^0 = i s^0$ , $S^q = s^q$ . $S^4 = i s^4$ , $S^{q+4} = s^{q+4}$ . $S^8 = i s^8$ , $S^{q+8} = s^{q+8}$. $S^{12} = i s^{12}$, $S^{q+12} = s^{q+12}$ . The components, $s^j$ , $s^{j+4}$ , $s^{j+8}$ , and $s^{j+12}$ , correspond to $\mu_g S^{gj}$ , $k_{eg} \mu_e S^{ej}$ , $k_{wg} \mu_w S^{wj}$ , and $k_{sg} \mu_s S^{sj}$ respectively.

In the tangent-frame $\Theta$ , by means of the property of the field strength $\mathbb{F}$ , the complex-sedenion field source, $\mu \mathbb{S} = - ( i \mathbb{F} / v^0 + \lozenge )^\ast \circ \mathbb{F} $ , can be separated into,
\begin{eqnarray}
\mu \mathbb{S} = - ( i \mathbb{F} / v^0 + \lozenge )^\ast \circledcirc \mathbb{F} - ( i \mathbb{F} / v^0 + \lozenge )^\ast \circledast \mathbb{F}  ~,
\end{eqnarray}
where $- ( i \mathbb{F} / v^0 + \lozenge )^\ast \circledcirc \mathbb{F}$ and $- ( i \mathbb{F} / v^0 + \lozenge )^\ast \circledast \mathbb{F}$ are respectively the scalar part and vector part of field source $\mu \mathbb{S}$ . Apparently, in the tangent-frame $\Theta$, the field source $\mathbb{S}$ comprises the derivative of the field strength $\mathbb{F}$ , and the spatial parameter of complex-sedenion curved space. From the above, it is able to deduce the electromagnetic field equations and gravitational field equations, in the orthogonal and unequal-length tangent-frame $\Theta$ of the complex-sedenion curved space.

\subsection{\label{sec:level1}Linear momentum}

In the orthogonal and equal-length tangent-frame $\Pi$ , the complex-sedenion linear momentum can be written as,
\begin{eqnarray}
\mathbb{P} = \mu \mathbb{S} / \mu_g^g  ~,
\end{eqnarray}
where $\mathbb{P} = \mathbb{P}_g + k_{eg} \mathbb{P}_e + k_{wg} \mathbb{P}_w + k_{sg} \mathbb{P}_s$ . $\mathbb{P}_g$ , $\mathbb{P}_e$ , $\mathbb{P}_w$ , and $\mathbb{P}_s$ are respectively the components of linear momentum $\mathbb{P}$ in four complex-quaternion spaces, $\mathbb{H}_g$, $\mathbb{H}_e$, $\mathbb{H}_w$, and $\mathbb{H}_s$ . $\mathbb{P}_g = \{ \mu_g \mathbb{S}_g - i \mathbb{F}^\ast \circ \mathbb{F} / v^0 \} / \mu_g^g $ . $\mathbb{P}_e = \mu_e \mathbb{S}_e / \mu_g^g $ . $\mathbb{P}_w = \mu_w \mathbb{S}_w / \mu_g^g $. $\mathbb{P}_s = \mu_s \mathbb{S}_s / \mu_g^g $. $\mathbb{P}_g = i \textbf{\emph{I}}_{g0} P^{g0} + \textbf{\emph{I}}_{gq} P^{gq}$ . $\mathbb{P}_e = i \textbf{\emph{I}}_{e0} P^{e0} + \textbf{\emph{I}}_{eq} P^{eq}$ . $\mathbb{P}_w = i \textbf{\emph{I}}_{w0} P^{w0} + \textbf{\emph{I}}_{wq} P^{wq}$. $\mathbb{P}_s = i \textbf{\emph{I}}_{s0} P^{s0} + \textbf{\emph{I}}_{sq} P^{sq}$. $P^{gj}$ , $P^{ej}$ , $P^{wj}$ , and $P^{sj}$ are all real. $\mu_g^g$ is the gravitational constant.

In the orthogonal and unequal-length tangent-frame $\Theta$ , the complex-sedenion linear momentum $\mathbb{P}$ is expanded in terms of the tangent-frame component $\textbf{\emph{e}}_\alpha$ ,
\begin{eqnarray}
\mathbb{P} ( P^\alpha ) = P^\alpha \textbf{\emph{e}}_\alpha ~,
\end{eqnarray}
where $P^0 = i p^0$ , $P^q = p^q$ . $P^4 = i p^4$ , $P^{q+4} = p^{q+4}$ . $P^8 = i p^8$ , $P^{q+8} = p^{q+8}$. $P^{12} = i p^{12}$, $P^{q+12} = p^{q+12}$ . The components, $p^j$ , $p^{j+4}$ , $p^{j+8}$ , and $p^{j+12}$ , correspond to $P^{gj}$ , $k_{eg} P^{ej}$, $k_{wg} P^{wj}$ , and $k_{sg} P^{sj}$ respectively.

\subsection{\label{sec:level1}Angular momentum}

In the orthogonal and equal-length tangent-frame $\Pi$ , the complex-sedenion angular momentum can be written as,
\begin{eqnarray}
\mathbb{L} = \mathbb{U}^\star \circ \mathbb{P} ~,
\end{eqnarray}
where $\mathbb{L} = \mathbb{L}_g + k_{eg} \mathbb{L}_e + k_{wg} \mathbb{L}_w + k_{sg} \mathbb{L}_s$ . $\mathbb{L}_g$ , $\mathbb{L}_e$ , $\mathbb{L}_w$ , and $\mathbb{L}_s$ are respectively the components of linear momentum $\mathbb{L}$ in four complex-quaternion spaces, $\mathbb{H}_g$, $\mathbb{H}_e$, $\mathbb{H}_w$, and $\mathbb{H}_s$ . $\mathbb{R}^\star = \mathbb{R}_g^\star + k_{eg} \mathbb{R}_e^\star + k_{wg} \mathbb{R}_w^\star + k_{sg} \mathbb{R}_s^\star$ . $\mathbb{X}^\star = \mathbb{X}_g^\star + k_{eg} \mathbb{X}_e^\star + k_{wg} \mathbb{X}_w^\star + k_{sg} \mathbb{X}_s^\star$. The radius vector, $\mathbb{R}$ , and the integrating function of field potential, $\mathbb{X}$, can be combined together to become one new radius vector, $\mathbb{U} = \mathbb{R} + k_{rx} \mathbb{X}$ , which is called as the composite radius vector temporarily. $\mathbb{U}_g = \mathbb{R}_g + k_{rx} \mathbb{X}_g$ . $\mathbb{U}_e = \mathbb{R}_e + k_{rx} \mathbb{X}_e$ . $\mathbb{U}_w = \mathbb{R}_w + k_{rx} \mathbb{X}_w$. $\mathbb{U}_s = \mathbb{R}_s + k_{rx} \mathbb{X}_s$. $k_{rx}$ is a coefficient, to meet the requirement of dimensional homogeneity. $\mathbb{L}_g = i \textbf{\emph{I}}_{g0} L^{g0} + \textbf{\emph{I}}_{gq} L^{gq}$ . $\mathbb{L}_e = i \textbf{\emph{I}}_{e0} L^{e0} + \textbf{\emph{I}}_{eq} L^{eq}$ . $\mathbb{L}_w = i \textbf{\emph{I}}_{w0} L^{w0} + \textbf{\emph{I}}_{wq} L^{wq}$. $\mathbb{L}_s = i \textbf{\emph{I}}_{s0} L^{s0} + \textbf{\emph{I}}_{sq} L^{sq}$ . $L^{gj}$ , $L^{ej}$ , $L^{wj}$ , and $L^{sj}$ are complex numbers. The angular momentum $\mathbb{L}$ includes the orbital angular momentum, electric dipole moment, and magnetic dipole moment.

In the orthogonal and unequal-length tangent-frame $\Theta$ , the complex-sedenion angular momentum $\mathbb{L}$ is expanded in terms of the tangent-frame component $\textbf{\emph{e}}_\alpha$ ,
\begin{eqnarray}
\mathbb{L} ( L^\alpha ) = L^\alpha \textbf{\emph{e}}_\alpha ~,
\end{eqnarray}
where $L^0 = i l^0$ , $L^q = l^q$ . $L^4 = i l^4$ , $L^{q+4} = l^{q+4}$ . $L^8 = i l^8$ , $L^{q+8} = l^{q+8}$. $L^{12} = i l^{12}$, $L^{q+12} = l^{q+12}$ . The components, $l^j$ , $l^{j+4}$ , $l^{j+8}$ , and $l^{j+12}$ , correspond to $L^{gj}$ , $k_{eg} L^{ej}$, $k_{wg} L^{wj}$ , and $k_{sg} L^{sj}$ respectively.

In the tangent-frame $\Theta$ , making use of the properties of the composite radius vector $\mathbb{U}$ and linear momentum $\mathbb{P}$ , the complex-sedenion angular momentum, $\mathbb{L} = \mathbb{U}^\star \circ \mathbb{P}$ , can be separated into,
\begin{eqnarray}
\mathbb{L} = \mathbb{U}^\star \circledcirc \mathbb{P} + \mathbb{U}^\star \circledast \mathbb{P}  ~,
\end{eqnarray}
where $\mathbb{U}^\star \circledcirc \mathbb{P}$ and $\mathbb{U}^\star \circledast \mathbb{P}$ are respectively the scalar part and vector part of angular momentum $\mathbb{L}$ . Apparently, in the tangent-frame $\Theta$ , the angular momentum $\mathbb{L}$ consists of the derivative of the composite radius vector $\mathbb{U}$ and linear momentum $\mathbb{P}$, and the spatial parameter of complex-sedenion curved space.

\subsection{\label{sec:level1}Torque}

In the orthogonal and equal-length tangent-frame $\Pi$ , the complex-sedenion torque can be written as,
\begin{eqnarray}
\mathbb{W} = - v^0 ( i \mathbb{F} / v^0 + \lozenge ) \circ \mathbb{L}   ~,
\end{eqnarray}
where $\mathbb{W} = \mathbb{W}_g + k_{eg} \mathbb{W}_e + k_{wg} \mathbb{W}_w + k_{sg} \mathbb{W}_s$ . $\mathbb{W}_g$ , $\mathbb{W}_e$ , $\mathbb{W}_w$ , and $\mathbb{W}_s$ are respectively the components of torque $\mathbb{W}$ in four complex-quaternion spaces, $\mathbb{H}_g$ , $\mathbb{H}_e$ , $\mathbb{H}_w$ , and $\mathbb{H}_s$. $\mathbb{W}_g = i \textbf{\emph{I}}_{g0} W^{g0} + \textbf{\emph{I}}_{gq} W^{gq}$ . $\mathbb{W}_e = i \textbf{\emph{I}}_{e0} W^{e0} + \textbf{\emph{I}}_{eq} W^{eq}$. $\mathbb{W}_w = i \textbf{\emph{I}}_{w0} W^{w0} + \textbf{\emph{I}}_{wq} W^{wq}$. $\mathbb{W}_s = i \textbf{\emph{I}}_{s0} W^{s0} + \textbf{\emph{I}}_{sq} W^{sq}$ . $W^{gj}$, $W^{ej}$ , $W^{wj}$ , and $W^{sj}$ are complex numbers. The torque $\mathbb{W}$ includes the conventional torque and energy and so forth. Especially, the energy consists of the proper energy, kinetic energy, potential energy, work, and the interacting energy between the electric field intensity with the electric dipole moment, and the interacting energy between the magnetic flux density with the magnetic dipole moment, and so forth.

In the orthogonal and unequal-length tangent-frame $\Theta$ , the complex-sedenion torque $\mathbb{W}$ can be expanded in terms of the tangent-frame component $\textbf{\emph{e}}_\alpha$ ,
\begin{eqnarray}
\mathbb{W} ( W^\alpha ) = W^\alpha \textbf{\emph{e}}_\alpha ~,
\end{eqnarray}
where $W^0 = i w^0$ , $W^q = w^q$ . $W^4 = i w^4$ , $W^{q+4} = w^{q+4}$ . $W^8 = i w^8$, $W^{q+8} = w^{q+8}$. $W^{12} = i w^{12}$ , $W^{q+12} = w^{q+12}$. The components, $w^j$ , $w^{j+4}$, $w^{j+8}$ , and $w^{j+12}$ , correspond to $W^{gj}$ , $k_{eg} W^{ej}$ , $k_{wg} W^{wj}$ , and $k_{sg} W^{sj}$ respectively.

In the tangent-frame $\Theta$ , in virtue of the property of angular momentum $\mathbb{L}$ , the complex-sedenion torque, $\mathbb{W} = - v^0 ( i \mathbb{F} / v^0 + \lozenge ) \circ \mathbb{L}$ , can be separated into,
\begin{eqnarray}
\mathbb{W} = - v^0 ( i \mathbb{F} / v^0 + \lozenge ) \circledcirc \mathbb{L} - v^0 ( i \mathbb{F} / v^0 + \lozenge ) \circledast \mathbb{L}  ~,
\end{eqnarray}
where $- v^0 ( i \mathbb{F} / v^0 + \lozenge ) \circledcirc \mathbb{L}$ and $- v^0 ( i \mathbb{F} / v^0 + \lozenge ) \circledast \mathbb{L}$ are respectively the scalar part and vector part of torque $\mathbb{W}$ . Apparently, in the tangent-frame $\Theta$ , the torque $\mathbb{W}$ contains the derivative of the angular momentum $\mathbb{L}$ , and the spatial parameter of complex-sedenion curved space.

\subsection{\label{sec:level1}Force}

In the orthogonal and equal-length tangent-frame $\Pi$ , the complex-sedenion force can be written as,
\begin{eqnarray}
\mathbb{N} = - ( i \mathbb{F} / v^0 + \lozenge ) \circ \mathbb{W}   ~,
\end{eqnarray}
where $\mathbb{N} = \mathbb{N}_g + k_{eg} \mathbb{N}_e + k_{wg} \mathbb{N}_w + k_{sg} \mathbb{N}_s$ . $\mathbb{N}_g$ , $\mathbb{N}_e$ , $\mathbb{N}_w$ , and $\mathbb{N}_s$ are respectively the components of force $\mathbb{N}$ in four complex-quaternion spaces, $\mathbb{H}_g$, $\mathbb{H}_e$ , $\mathbb{H}_w$ , and $\mathbb{H}_s$. $\mathbb{N}_g = i \textbf{\emph{I}}_{g0} N^{g0} + \textbf{\emph{I}}_{gq} N^{gq}$ . $\mathbb{N}_e = i \textbf{\emph{I}}_{e0} N^{e0} + \textbf{\emph{I}}_{eq} N^{eq}$. $\mathbb{N}_w = i \textbf{\emph{I}}_{w0} N^{w0} + \textbf{\emph{I}}_{wq} N^{wq}$. $\mathbb{N}_s = i \textbf{\emph{I}}_{s0} N^{s0} + \textbf{\emph{I}}_{sq} N^{sq}$ . $N^{gj}$ , $N^{ej}$ , $N^{wj}$ , and $N^{sj}$ are complex numbers. When $\mathbb{N} = 0$, it is able to deduce the force equilibrium equation, precessional equilibrium equation, mass continuity equation, and current continuity equation and so forth, from the above.

In the orthogonal and unequal-length tangent-frame $\Theta$ , the complex-sedenion force $\mathbb{N}$ can be expanded in terms of the tangent-frame component $\textbf{\emph{e}}_\alpha$ ,
\begin{eqnarray}
\mathbb{N} ( N^\alpha ) = N^\alpha \textbf{\emph{e}}_\alpha ~,
\end{eqnarray}
where $N^0 = i n^0$ , $N^q = n^q$ . $N^4 = i n^4$ , $N^{q+4} = n^{q+4}$ . $N^8 = i n^8$ , $N^{q+8} = n^{q+8}$. $N^{12} = i n^{12}$ , $N^{q+12} = n^{q+12}$. The components, $n^j$ , $n^{j+4}$ , $n^{j+8}$ , and $n^{j+12}$ , correspond to $N^{gj}$ , $k_{eg} N^{ej}$ , $k_{wg} N^{wj}$ , and $k_{sg} N^{sj}$ respectively.

In the tangent-frame $\Theta$ , in virtue of the property of torque momentum $\mathbb{W}$, the complex-sedenion force, $\mathbb{N} = - ( i \mathbb{F} / v^0 + \lozenge ) \circ \mathbb{W}$ , can be separated into,
\begin{eqnarray}
\mathbb{N} = - ( i \mathbb{F} / v^0 + \lozenge ) \circledcirc \mathbb{W} - ( i \mathbb{F} / v^0 + \lozenge ) \circledast \mathbb{W}  ~,
\end{eqnarray}
where $- ( i \mathbb{F} / v^0 + \lozenge ) \circledcirc \mathbb{W}$ and $- ( i \mathbb{F} / v^0 + \lozenge ) \circledast \mathbb{W}$ are respectively the scalar part and vector part of force $\mathbb{N}$ . Apparently, in the tangent-frame $\Theta$ , the force $\mathbb{N}$ contains not only the derivative of torque $\mathbb{W}$ , but also the spatial parameter of complex-sedenion curved space.

In terms of the vector part, $- ( i \mathbb{F} / v^0 + \lozenge ) \circledast \mathbb{W}$ , its component in the complex-quaternion space $\mathbb{H}_g$ can be written as, $\textbf{N}^f = N^q \textbf{\emph{e}}_q$ . The imaginary part of $\textbf{N}^f$ is the force term $\textbf{N}^i$ , while the real part of $\textbf{N}^f$ is the precessional term $\textbf{N}^r$ . When $\textbf{N}^f = 0$, the force equilibrium equation is $\textbf{N}^i = 0$, while the precessional equilibrium equation is $\textbf{N}^r = 0$ . Further, it is able to infer the linear acceleration and angular velocity of revolution from $\textbf{N}^i = 0$ , and the angular velocity of precession from $\textbf{N}^r = 0$ . The force is seized of several terms, including the inertial force, gravity, electromagnetic force, energy gradient, and additional force term, $\textbf{N}^i_{(curved)}$ , caused by the curved space. The term, $\textbf{N}^i_{(curved)}$ , is relevant to the connection coefficient and curvature tensor and so forth of the complex-sedenion curved space. In terms of the force term, $\textbf{N}^i$, the contribution of the additional force term, $\textbf{N}^i_{(curved)}$ , is quite tiny, under most circumstances.

The above states that the complex-sedenion angular momentum, in the complex-sedenion curved space, comprises the orbital angular momentum, magnetic moment, and electric moment and so forth. The complex-sedenion torque consists of the conventional torque, energy, and power and so on. When the complex-sedenion force equals to zero, it is able to deduce the force equilibrium equation, precessional equilibrium equation, mass continuity equation, and current continuity equation and so forth. The force contains the inertial force, gravity, electromagnetic force, and energy gradient and so forth. The energy gradient can be applied to explore the gravitational mass, astrophysical jet, driving force, condensed dark matter, new accelerator and decelerator and so on. The precessional equilibrium equation can be utilized to explain the reason of the precessional phenomena, inferring the angular frequency of Larmor precession and so forth.

\begin{table}[h]
\tbl{Some quantum physical quantities, auxiliary quantities, and wavefunctions, in the complex-sedenion quantum space.}
{\begin{tabular}{@{}lcll@{}}
\toprule
classical~physical        & auxiliary           & wave~function                                        & quantum~physical                                      \\
quantity                  & quantity            &                                                      & quantity                                              \\
\colrule
composite~radius~vector   & $\mathbb{Z}_U$      & $\Psi_{ZU} = \mathbb{Z}_U \circ \mathbb{U} / \hbar$  & $\mathbb{U}_{(\Psi)} = \mathbb{Z}_U \circ \mathbb{U}$ \\
integrating~function      & $\mathbb{Z}_X$      & $\Psi_{ZX} = \mathbb{Z}_X \circ \mathbb{X} / \hbar$  & $\mathbb{X}_{(\Psi)} = \mathbb{Z}_X \circ \mathbb{X}$ \\
field~potential           & $\mathbb{Z}_A$      & $\Psi_{ZA} = \mathbb{Z}_A \circ \mathbb{A} / \hbar$  & $\mathbb{A}_{(\Psi)} = \mathbb{Z}_A \circ \mathbb{A}$ \\
field~strength            & $\mathbb{Z}_F$      & $\Psi_{ZF} = \mathbb{Z}_F \circ \mathbb{F} / \hbar$  & $\mathbb{F}_{(\Psi)} = \mathbb{Z}_F \circ \mathbb{F}$ \\
field~source              & $\mathbb{Z}_S$      & $\Psi_{ZS} = \mathbb{Z}_S \circ \mathbb{S} / \hbar$  & $\mathbb{S}_{(\Psi)} = \mathbb{Z}_S \circ \mathbb{S}$ \\
linear~momentum           & $\mathbb{Z}_P$      & $\Psi_{ZP} = \mathbb{Z}_P \circ \mathbb{P} / \hbar$  & $\mathbb{P}_{(\Psi)} = \mathbb{Z}_P \circ \mathbb{P}$ \\
angular~momentum          & $\mathbb{Z}_L$      & $\Psi_{ZL} = \mathbb{Z}_L \circ \mathbb{L} / \hbar$  & $\mathbb{L}_{(\Psi)} = \mathbb{Z}_L \circ \mathbb{L}$ \\
torque                    & $\mathbb{Z}_W$      & $\Psi_{ZW} = \mathbb{Z}_W \circ \mathbb{W} / \hbar$  & $\mathbb{W}_{(\Psi)} = \mathbb{Z}_W \circ \mathbb{W}$ \\
force                     & $\mathbb{Z}_N$      & $\Psi_{ZN} = \mathbb{Z}_N \circ \mathbb{N} / \hbar$  & $\mathbb{N}_{(\Psi)} = \mathbb{Z}_N \circ \mathbb{N}$ \\
\botrule
\end{tabular}}
\end{table}

\section{\label{sec:level1}Quantum-field equations}

In the complex-quaternion space, one complex-quaternion physical quantity can be rewritten as the exponential form. In virtue of the properties of auxiliary quantity and function, the product of the complex-quaternion physical quantity with the auxiliary quantity is able to be expressed as one complex-quaternion wavefunction. And it can be reduced into the complex-number wavefunction in the conventional quantum mechanics. Similarly, in the complex-octonion space, the product of the complex-octonion physical quantity with its auxiliary quantity can be written as one complex-octonion wavefunction. And the product of the complex-sedenion physical quantity with its auxiliary quantity is able to be written as one complex-sedenion wavefunction, in the complex-sedenion space.

By means of the introduction of auxiliary quantities (Table 2), from the complex-sedenion space $\mathbb{K}$ , it is capable of achieving one complex-sedenion function space, which is called as the quantum space, $\mathbb{K}_Z$ , temporarily. And it is able to deduce the quantum-field equations, for the quantum mechanics on the microscopic scale, from the classical field equations on the macroscopic scale, in the complex-sedenion curved space. In the complex-sedenion quantum space, $\mathbb{K}_Z$ , two of quantum-field equations can be respectively degenerated into the complex-number Dirac wave equation, and the Yang-Mills equation of the non-Abelian gauge fields \cite{weng4} , under some circumstances.

In the complex-sedenion quantum space, $\mathbb{K}_Z$ , the quantum-field, in the quantum mechanics on the microscopic scale, can be regarded as the function or transformation of the conventional classical-field, in the classical mechanics on the macroscopic scale. By analogy with the deduction method of the field equations for the classical mechanics on the macroscopic scale, it is able to achieve the field equations for the quantum mechanics on the microscopic scale (Table 3). In other words, in virtue of the complex-sedenion exponential form and wavefunction, the paper is capable of inferring the field equations for the quantum mechanics on the microscopic scale. They are relevant to the physical quantities of the four quantum-fields (gravitational quantum-field, electromagnetic quantum-field, W-nuclear quantum-field, and strong-nuclear quantum-field), including the quantum-field potential, quantum-field strength, quantum-field source, quantum linear momentum, quantum angular momentum, quantum torque, and quantum force.

In terms of different quantum physical quantities, their relevant auxiliary quantities will be different from each other. In the practical problems, we can choose some appropriate auxiliary quantities, to satisfy fully the specific requirements. Sometimes, there may be several types of selections, for one auxiliary quantity. Therefore, a majority of physical quantities in the quantum space, $\mathbb{K}_Z$, are different from that in the complex-sedenion space, $\mathbb{K}$ .

\subsection{\label{sec:level1}Quantum composite radius vector}

In the orthogonal and equal-length tangent-frame $\Pi$ of the complex-sedenion curved space, the quantum composite radius vector can be written as,
\begin{eqnarray}
\mathbb{U}_{(\Psi)} = \mathbb{Z}_U \circ \mathbb{U}   ~,
\end{eqnarray}
where $\mathbb{U}_{(\Psi)} = \mathbb{U}_{(\Psi)g} + k_{eg} \mathbb{U}_{(\Psi)e} + k_{wg} \mathbb{U}_{(\Psi)w} + k_{sg} \mathbb{U}_{(\Psi)s}$ .
$\mathbb{Z}_U$ is a dimensionless auxiliary quantity. $\mathbb{U}_{(\Psi)g}$ , $\mathbb{U}_{(\Psi)e}$, $\mathbb{U}_{(\Psi)w}$, and $\mathbb{U}_{(\Psi)s}$ are respectively the components of quantum composite radius vector $\mathbb{U}_{(\Psi)}$ in four complex-quaternion spaces, $\mathbb{H}_g$, $\mathbb{H}_e$, $\mathbb{H}_w$ , and $\mathbb{H}_s$. $\mathbb{U}_{(\Psi)g} = i \textbf{\emph{I}}_{g0} U^{g0}_{(\Psi)} + \textbf{\emph{I}}_{gq} U^{gq}_{(\Psi)}$ . $\mathbb{U}_{(\Psi)e} = i \textbf{\emph{I}}_{e0} U^{e0}_{(\Psi)} + \textbf{\emph{I}}_{eq} U^{eq}_{(\Psi)}$. $\mathbb{U}_{(\Psi)w} = i \textbf{\emph{I}}_{w0} U^{w0}_{(\Psi)} + \textbf{\emph{I}}_{wq} U^{wq}_{(\Psi)}$. $\mathbb{U}_{(\Psi)s} = i \textbf{\emph{I}}_{s0} U^{s0}_{(\Psi)} + \textbf{\emph{I}}_{sq} U^{sq}_{(\Psi)}$. $U^{gj}_{(\Psi)}$ , $U^{ej}_{(\Psi)}$ , $U^{wj}_{(\Psi)}$ , and $U^{sj}_{(\Psi)}$ are complex numbers.

In the orthogonal and unequal-length tangent-frame $\Theta$ , the complex-sedenion quantum composite radius vector, $\mathbb{U}_{(\Psi)}$ , can be expanded in terms of the tangent-frame component $\textbf{\emph{e}}_\alpha$ ,
\begin{eqnarray}
\mathbb{U}_{(\Psi)} ( U^\alpha_{(\Psi)} ) = U^\alpha_{(\Psi)} \textbf{\emph{e}}_\alpha ~,
\end{eqnarray}
where $U_{(\Psi)}^0 = i C_{(\Psi)}^0$ , $U_{(\Psi)}^q = C_{(\Psi)}^q$ . $U_{(\Psi)}^4 = i C_{(\Psi)}^4$ , $U_{(\Psi)}^{q+4} = C_{(\Psi)}^{q+4}$ . $U_{(\Psi)}^8 = i C_{(\Psi)}^8$, $U_{(\Psi)}^{q+8} = C_{(\Psi)}^{q+8}$ . $U_{(\Psi)}^{12} = i C_{(\Psi)}^{12}$ , $U_{(\Psi)}^{q+12} = C_{(\Psi)}^{q+12}$. And the components, $C_{(\Psi)}^j$ , $C_{(\Psi)}^{j+4}$ , $C_{(\Psi)}^{j+8}$, and $C_{(\Psi)}^{j+12}$ , correspond to $U_{(\Psi)}^{gj}$ , $k_{eg} U_{(\Psi)}^{ej}$ , $k_{wg} U_{(\Psi)}^{wj}$ , and $k_{sg} U_{(\Psi)}^{sj}$ respectively.

\subsection{\label{sec:level1}Composite operator}

In the orthogonal and equal-length tangent-frame $\Pi$ , one new composite operator can be written as,
\begin{eqnarray}
\mathbb{D}_Z = i \mathbb{Z}_W \circ \mathbb{W}^\star / ( \hbar v^0 ) + \lozenge  ~,
\end{eqnarray}
where $\mathbb{Z}_W$ is a dimensionless auxiliary quantity. $\mathbb{D}_Z = \mathbb{D}_{Zg} + k_{eg} \mathbb{D}_{Ze} + k_{wg} \mathbb{D}_{Zw} + k_{sg} \mathbb{D}_{Zs}$. $\mathbb{D}_{Zg}$ , $\mathbb{D}_{Ze}$ , $\mathbb{D}_{Zw}$ , and $\mathbb{D}_{Zs}$ are the components of composite operator $\mathbb{D}_Z$ in four complex-quaternion spaces, $\mathbb{H}_g$ , $\mathbb{H}_e$ , $\mathbb{H}_w$ , and $\mathbb{H}_s$ , respectively. $\mathbb{D}_{Zg} = i \mathbb{D}_{(ZW)g} / ( \hbar v^0 ) + \lozenge_g$.
$\mathbb{D}_{Ze} = i \mathbb{D}_{(ZW)e} / ( \hbar v^0 ) + k_{eg}^{~~-2} \lozenge_e$ . $\mathbb{D}_{Zw} = i \mathbb{D}_{(ZW)w} / ( \hbar v^0 ) + k_{wg}^{~~-2} \lozenge_w$. $\mathbb{D}_{Zs} = i \mathbb{D}_{(ZW)s} / ( \hbar v^0 ) + k_{sg}^{~~-2} \lozenge_s$ . Moreover, $\mathbb{D}_{(ZW)} = \mathbb{Z}_W \circ \mathbb{W}^\star$ . $\mathbb{D}_{(ZW)} = \mathbb{D}_{(ZW)g} + k_{eg} \mathbb{D}_{(ZW)e} + k_{wg} \mathbb{D}_{(ZW)w} + k_{sg} \mathbb{D}_{(ZW)s}$. $\mathbb{D}_{(ZW)g}$ , $\mathbb{D}_{(ZW)e}$ , $\mathbb{D}_{(ZW)w}$ , and $\mathbb{D}_{(ZW)s}$ are the components of operator $\mathbb{D}_{(ZW)}$ in four complex-quaternion spaces, $\mathbb{H}_g$ , $\mathbb{H}_e$ , $\mathbb{H}_w$ , and $\mathbb{H}_s$ , respectively.

Apparently, by means of the appropriate transformation, the composite operator, $\mathbb{D}_Z = i \mathbb{Z}_W \circ \mathbb{W}^\star / ( \hbar v^0 ) + \lozenge$ , can be converted from the tangent-frame $\Pi$, into the tangent-frame $\Theta$ .

\subsection{\label{sec:level1}Quantum-field potential}

In the orthogonal and equal-length tangent-frame $\Pi$ of the complex-sedenion curved space, the quantum-field potential can be written as,
\begin{eqnarray}
\mathbb{A}_{(\Psi)} = i \mathbb{D}_Z^\star \circ \mathbb{X}_{(\Psi)}   ~,
\end{eqnarray}
where $\mathbb{A}_{(\Psi)} = \mathbb{Z}_A \circ \mathbb{A}$ , $\mathbb{X}_{(\Psi)} = \mathbb{Z}_X \circ \mathbb{X}$ . $\mathbb{Z}_A$ and $\mathbb{Z}_X$ are dimensionless auxiliary quantities. $\mathbb{A}_{(\Psi)} = \mathbb{A}_{(\Psi)g} + k_{eg} \mathbb{A}_{(\Psi)e} + k_{wg} \mathbb{A}_{(\Psi)w} + k_{sg} \mathbb{A}_{(\Psi)s}$ . $\mathbb{A}_{(\Psi)g}$ , $\mathbb{A}_{(\Psi)e}$, $\mathbb{A}_{(\Psi)w}$, and $\mathbb{A}_{(\Psi)s}$ are the components of quantum-field potential $\mathbb{A}_{(\Psi)}$ in four complex-quaternion spaces, $\mathbb{H}_g$ , $\mathbb{H}_e$ , $\mathbb{H}_w$ , and $\mathbb{H}_s$ , respectively. $\mathbb{D}_Z^\star = \mathbb{D}_{Zg}^\star + k_{eg} \mathbb{D}_{Ze}^\star + k_{wg} \mathbb{D}_{Zw}^\star + k_{sg} \mathbb{D}_{Zs}^\star$ . $\mathbb{A}_{(\Psi)g} = i \textbf{\emph{I}}_{g0} A^{g0}_{(\Psi)} + \textbf{\emph{I}}_{gq} A^{gq}_{(\Psi)}$ . $\mathbb{A}_{(\Psi)e} = i \textbf{\emph{I}}_{e0} A^{e0}_{(\Psi)} + \textbf{\emph{I}}_{eq} A^{eq}_{(\Psi)}$ . $\mathbb{A}_{(\Psi)w} = i \textbf{\emph{I}}_{w0} A^{w0}_{(\Psi)} + \textbf{\emph{I}}_{wq} A^{wq}_{(\Psi)}$. $\mathbb{A}_{(\Psi)s} = i \textbf{\emph{I}}_{s0} A^{s0}_{(\Psi)} + \textbf{\emph{I}}_{sq} A^{sq}_{(\Psi)}$. $A^{gj}_{(\Psi)}$ , $A^{ej}_{(\Psi)}$ , $A^{wj}_{(\Psi)}$ , and $A^{sj}_{(\Psi)}$ are complex numbers. The case of $\mathbb{X}_{(\Psi)} $ is similar to that of $\mathbb{A}_{(\Psi)}$ .

In the orthogonal and unequal-length tangent-frame $\Theta$ , the complex-sedenion quantum-field potential, $\mathbb{A}_{(\Psi)}$ , can be expanded in terms of the tangent-frame component $\textbf{\emph{e}}_\alpha$ ,
\begin{eqnarray}
\mathbb{A}_{(\Psi)} ( A^\alpha_{(\Psi)} ) = A^\alpha_{(\Psi)} \textbf{\emph{e}}_\alpha ~,
\end{eqnarray}
where $A_{(\Psi)}^0 = i a_{(\Psi)}^0$ , $A_{(\Psi)}^q = a_{(\Psi)}^q$ . $A_{(\Psi)}^4 = i a_{(\Psi)}^4$ , $A_{(\Psi)}^{q+4} = a_{(\Psi)}^{q+4}$ . $A_{(\Psi)}^8 = i a_{(\Psi)}^8$, $A_{(\Psi)}^{q+8} = a_{(\Psi)}^{q+8}$ . $A_{(\Psi)}^{12} = i a_{(\Psi)}^{12}$ , $A_{(\Psi)}^{q+12} = a_{(\Psi)}^{q+12}$. And the components, $a_{(\Psi)}^j$ , $a_{(\Psi)}^{j+4}$ , $a_{(\Psi)}^{j+8}$ , and $a_{(\Psi)}^{j+12}$ , correspond to $A_{(\Psi)}^{gj}$ , $k_{eg} A_{(\Psi)}^{ej}$ , $k_{wg} A_{(\Psi)}^{wj}$ , and $k_{sg} A_{(\Psi)}^{sj}$ respectively.

In the tangent-frame $\Theta$ , in virtue of the property of the quantum integrating function of field potential, $\mathbb{X}_{(\Psi)}$ , the complex-sedenion quantum-field potential, $\mathbb{A}_{(\Psi)} = i \mathbb{D}_Z^\star \circ \mathbb{X}_{(\Psi)}$ , can be separated into,
\begin{eqnarray}
\mathbb{A}_{(\Psi)} = i \mathbb{D}_Z^\star \circledcirc \mathbb{X}_{(\Psi)} + i \mathbb{D}_Z^\star \circledast \mathbb{X}_{(\Psi)}  ~,
\end{eqnarray}
where $i \mathbb{D}_Z^\star \circledcirc \mathbb{X}_{(\Psi)}$ and $i \mathbb{D}_Z^\star \circledast \mathbb{X}_{(\Psi)}$ are respectively the scalar part and vector part of the quantum-field potential, $\mathbb{A}_{(\Psi)}$ . Apparently, in the tangent-frame $\Theta$ , the quantum-field potential $\mathbb{A}_{(\Psi)}$ contains not only the derivative of the quantum integrating function of field potential, $\mathbb{X}_{(\Psi)}$ , but also the spatial parameter of complex-sedenion curved space.

\subsection{\label{sec:level1}Quantum-field strength}

In the orthogonal and equal-length tangent-frame $\Pi$ of the complex-sedenion curved space, the quantum-field strength can be written as,
\begin{eqnarray}
\mathbb{F}_{(\Psi)} = \mathbb{D}_Z \circ \mathbb{A}_{(\Psi)}   ~,
\end{eqnarray}
where $\mathbb{F}_{(\Psi)} = \mathbb{Z}_F \circ \mathbb{F}$ , with $\mathbb{Z}_F$ being a dimensionless auxiliary quantity. $\mathbb{F}_{(\Psi)} = \mathbb{F}_{(\Psi)g} + k_{eg} \mathbb{F}_{(\Psi)e} + k_{wg} \mathbb{F}_{(\Psi)w} + k_{sg} \mathbb{F}_{(\Psi)s}$. $\mathbb{F}_{(\Psi)g}$ , $\mathbb{F}_{(\Psi)e}$ , $\mathbb{F}_{(\Psi)w}$ , and $\mathbb{F}_{(\Psi)s}$ are the components of quantum-field strength $\mathbb{F}_{(\Psi)}$ in four complex-quaternion spaces, $\mathbb{H}_g$, $\mathbb{H}_e$ , $\mathbb{H}_w$ , and $\mathbb{H}_s$ , respectively. $\mathbb{F}_{(\Psi)g} = i \textbf{\emph{I}}_{g0} F^{g0}_{(\Psi)} + \textbf{\emph{I}}_{gq} F^{gq}_{(\Psi)}$ . $\mathbb{F}_{(\Psi)e} = i \textbf{\emph{I}}_{e0} F^{e0}_{(\Psi)} + \textbf{\emph{I}}_{eq} F^{eq}_{(\Psi)}$. $\mathbb{F}_{(\Psi)w} = i \textbf{\emph{I}}_{w0} F^{w0}_{(\Psi)} + \textbf{\emph{I}}_{wq} F^{wq}_{(\Psi)}$. $\mathbb{F}_{(\Psi)s} = i \textbf{\emph{I}}_{s0} F^{s0}_{(\Psi)} + \textbf{\emph{I}}_{sq} F^{sq}_{(\Psi)}$. $F^{gj}_{(\Psi)}$ , $F^{ej}_{(\Psi)}$ , $F^{wj}_{(\Psi)}$ , and $F^{sj}_{(\Psi)}$ are complex numbers.

In the orthogonal and unequal-length tangent-frame $\Theta$ , the complex-sedenion quantum-field strength, $\mathbb{F}_{(\Psi)}$ , can be expanded in terms of the tangent-frame component $\textbf{\emph{e}}_\alpha$ ,
\begin{eqnarray}
\mathbb{F}_{(\Psi)} ( F^\alpha_{(\Psi)} ) = F^\alpha_{(\Psi)} \textbf{\emph{e}}_\alpha ~,
\end{eqnarray}
where $F_{(\Psi)}^0 = i f_{(\Psi)}^0$ , $F_{(\Psi)}^q = f_{(\Psi)}^q$ . $F_{(\Psi)}^4 = i f_{(\Psi)}^4$ , $F_{(\Psi)}^{q+4} = f_{(\Psi)}^{q+4}$ . $F_{(\Psi)}^8 = i f_{(\Psi)}^8$, $F_{(\Psi)}^{q+8} = f_{(\Psi)}^{q+8}$. $F_{(\Psi)}^{12} = i f_{(\Psi)}^{12}$ , $F_{(\Psi)}^{q+12} = f_{(\Psi)}^{q+12}$. And the components, $f_{(\Psi)}^j$ , $f_{(\Psi)}^{j+4}$ , $f_{(\Psi)}^{j+8}$ , and $f_{(\Psi)}^{j+12}$ , correspond to $F_{(\Psi)}^{gj}$ , $k_{eg} F_{(\Psi)}^{ej}$ , $k_{wg} F_{(\Psi)}^{wj}$ , and $k_{sg} F_{(\Psi)}^{sj}$ respectively.

In the tangent-frame $\Theta$ , making use of the property of the quantum-field potential, $\mathbb{A}_{(\Psi)}$ , the complex-sedenion quantum-field strength, $\mathbb{F}_{(\Psi)} = \mathbb{D}_Z \circ \mathbb{A}_{(\Psi)}$, can be separated into,
\begin{eqnarray}
\mathbb{F}_{(\Psi)} = \mathbb{D}_Z \circledcirc \mathbb{A}_{(\Psi)} + \mathbb{D}_Z \circledast \mathbb{A}_{(\Psi)}  ~,
\end{eqnarray}
where $\mathbb{D}_Z \circledcirc \mathbb{A}_{(\Psi)}$ and $\mathbb{D}_Z \circledast \mathbb{A}_{(\Psi)}$ are respectively the scalar part and vector part of the quantum-field strength, $\mathbb{F}_{(\Psi)}$ . Apparently, in the tangent-frame $\Theta$ , the quantum-field strength $\mathbb{F}_{(\Psi)}$ consists of not only the derivative of quantum-field potential, $\mathbb{A}_{(\Psi)}$ , but also the spatial parameter of complex-sedenion curved space.

\subsection{\label{sec:level1}Quantum-field source}

In the orthogonal and equal-length tangent-frame $\Pi$ of the complex-sedenion curved space, the quantum-field source can be written as,
\begin{eqnarray}
\mu \mathbb{S}_{(\Psi)} = - \mathbb{D}_Z^\ast \circ \mathbb{F}_{(\Psi)}   ~,
\end{eqnarray}
where $\mathbb{S}_{(\Psi)} = \mathbb{Z}_S \circ \mathbb{S}$ , with $\mathbb{Z}_S$ being a dimensionless auxiliary quantity. $\mu \mathbb{S}_{(\Psi)} = \mu_g \mathbb{S}_{(\Psi)g} + k_{eg} \mu_e \mathbb{S}_{(\Psi)e} + k_{wg} \mu_w \mathbb{S}_{(\Psi)w} + k_{sg} \mu_s \mathbb{S}_{(\Psi)s}$ . $\mathbb{S}_{(\Psi)g}$ , $\mathbb{S}_{(\Psi)e}$ , $\mathbb{S}_{(\Psi)w}$ , and $\mathbb{S}_{(\Psi)s}$ are the components of quantum-field source $\mathbb{S}_{(\Psi)}$ in four complex-quaternion spaces, $\mathbb{H}_g$, $\mathbb{H}_e$ , $\mathbb{H}_w$ , and $\mathbb{H}_s$ , respectively. $\mathbb{S}_{(\Psi)g} = i \textbf{\emph{I}}_{g0} S^{g0}_{(\Psi)} + \textbf{\emph{I}}_{gq} S^{gq}_{(\Psi)}$ . $\mathbb{S}_{(\Psi)e} = i \textbf{\emph{I}}_{e0} S^{e0}_{(\Psi)} + \textbf{\emph{I}}_{eq} S^{eq}_{(\Psi)}$. $\mathbb{S}_{(\Psi)w} = i \textbf{\emph{I}}_{w0} S^{w0}_{(\Psi)} + \textbf{\emph{I}}_{wq} S^{wq}_{(\Psi)}$ . $\mathbb{S}_{(\Psi)s} = i \textbf{\emph{I}}_{s0} S^{s0}_{(\Psi)} + \textbf{\emph{I}}_{sq} S^{sq}_{(\Psi)}$ . $S^{gj}_{(\Psi)}$, $S^{ej}_{(\Psi)}$ , $S^{wj}_{(\Psi)}$ , and $S^{sj}_{(\Psi)}$ are complex numbers. From the above, it is able to deduce the quantum-field equations and so forth, in the tangent-frame $\Pi$ of flat space $\mathbb{K}_Z$ . And the quantum-field equations can be degenerated into the Yang-Mills equations (see Ref.[42]).

In the orthogonal and unequal-length tangent-frame $\Theta$ , the complex-sedenion quantum-field source, $\mathbb{S}_{(\Psi)}$ , is expanded in terms of the tangent-frame component $\textbf{\emph{e}}_\alpha$,
\begin{eqnarray}
\mathbb{S}_{(\Psi)} ( S^\alpha_{(\Psi)} ) = S^\alpha_{(\Psi)} \textbf{\emph{e}}_\alpha ~,
\end{eqnarray}
where $S_{(\Psi)}^0 = i s_{(\Psi)}^0$ , $S_{(\Psi)}^q = s_{(\Psi)}^q$ . $S_{(\Psi)}^4 = i s_{(\Psi)}^4$ , $S_{(\Psi)}^{q+4} = s_{(\Psi)}^{q+4}$ . $S_{(\Psi)}^8 = i s_{(\Psi)}^8$, $S_{(\Psi)}^{q+8} = s_{(\Psi)}^{q+8}$. $S_{(\Psi)}^{12} = i s_{(\Psi)}^{12}$ , $S_{(\Psi)}^{q+12} = s_{(\Psi)}^{q+12}$. And the components, $s_{(\Psi)}^j$ , $s_{(\Psi)}^{j+4}$ , $s_{(\Psi)}^{j+8}$, and $s_{(\Psi)}^{j+12}$, correspond to $S_{(\Psi)}^{gj}$ , $k_{eg} S_{(\Psi)}^{ej}$ , $k_{wg} S_{(\Psi)}^{wj}$ , and $k_{sg} S_{(\Psi)}^{sj}$ respectively.

In the tangent-frame $\Theta$ , in virtue of the property of the quantum-field strength, $\mathbb{F}_{(\Psi)}$ , the complex-sedenion quantum-field source, $\mu \mathbb{S}_{(\Psi)} = - \mathbb{D}_Z^\ast \circ \mathbb{F}_{(\Psi)}$, can be separated into,
\begin{eqnarray}
\mu \mathbb{S}_{(\Psi)} = - \mathbb{D}_Z^\ast \circledcirc \mathbb{F}_{(\Psi)} - \mathbb{D}_Z^\ast \circledast \mathbb{F}_{(\Psi)}   ~,
\end{eqnarray}
where $- \mathbb{D}_Z^\ast \circledcirc \mathbb{F}_{(\Psi)}$ and $- \mathbb{D}_Z^\ast \circledast \mathbb{F}_{(\Psi)}$ are respectively the scalar part and vector part of the quantum-field source, $\mu \mathbb{S}_{(\Psi)}$. Obviously, in the tangent-frame $\Theta$, the quantum-field source $\mathbb{S}_{(\Psi)}$ comprises not only the derivative of quantum-field strength, $\mathbb{F}_{(\Psi)}$ , but also the spatial parameter of complex-sedenion curved space. From the above, it is able to deduce the quantum-field equations and so forth, in the tangent-frame $\Theta$ of the curved space $\mathbb{K}_Z$ . And the quantum-field equations can be degenerated into the Yang-Mills equations in the curved space.

\subsection{\label{sec:level1}Quantum linear momentum}

In the orthogonal and equal-length tangent-frame $\Pi$ of the complex-sedenion curved space, the quantum linear momentum can be written as,
\begin{eqnarray}
\mathbb{P}_{(\Psi)} = \mu \mathbb{S}_{(\Psi)} / \mu_g^g ~,
\end{eqnarray}
where $\mathbb{P}_{(\Psi)} = \mathbb{Z}_P \circ \mathbb{P}$ , with $\mathbb{Z}_P$ being a dimensionless auxiliary quantity. $\mathbb{P}_{(\Psi)} = \mathbb{P}_{(\Psi)g} + k_{eg} \mathbb{P}_{(\Psi)e} + k_{wg} \mathbb{P}_{(\Psi)w} + k_{sg} \mathbb{P}_{(\Psi)s}$. $\mathbb{P}_{(\Psi)g}$ , $\mathbb{P}_{(\Psi)e}$ , $\mathbb{P}_{(\Psi)w}$ , and $\mathbb{P}_{(\Psi)s}$ are the components of quantum linear momentum $\mathbb{P}_{(\Psi)}$ in four complex-quaternion spaces, $\mathbb{H}_g$ , $\mathbb{H}_e$ , $\mathbb{H}_w$ , and $\mathbb{H}_s$ , respectively. $\mathbb{P}_{(\Psi)g} = \mu_g \mathbb{S}_{(\Psi)g} / \mu_g^g $. $\mathbb{P}_{(\Psi)e} = \mu_e \mathbb{S}_{(\Psi)e} / \mu_g^g $. $\mathbb{P}_{(\Psi)w} = \mu_w \mathbb{S}_{(\Psi)w} / \mu_g^g $. $\mathbb{P}_{(\Psi)s} = \mu_s \mathbb{S}_{(\Psi)s} / \mu_g^g $ . $\mathbb{P}_{(\Psi)g} = i \textbf{\emph{I}}_{g0} P^{g0}_{(\Psi)} + \textbf{\emph{I}}_{gq} P^{gq}_{(\Psi)}$ . $\mathbb{P}_{(\Psi)e} = i \textbf{\emph{I}}_{e0} P^{e0}_{(\Psi)} + \textbf{\emph{I}}_{eq} P^{eq}_{(\Psi)}$ . $\mathbb{P}_{(\Psi)w} = i \textbf{\emph{I}}_{w0} P^{w0}_{(\Psi)} + \textbf{\emph{I}}_{wq} P^{wq}_{(\Psi)}$. $\mathbb{P}_{(\Psi)s} = i \textbf{\emph{I}}_{s0} P^{s0}_{(\Psi)} + \textbf{\emph{I}}_{sq} P^{sq}_{(\Psi)}$. $P^{gj}_{(\Psi)}$, $P^{ej}_{(\Psi)}$, $P^{wj}_{(\Psi)}$ , and $P^{sj}_{(\Psi)}$ are complex numbers.

In the orthogonal and unequal-length tangent-frame $\Theta$ , the complex-sedenion quantum linear momentum, $\mathbb{P}_{(\Psi)}$ , can be expanded in terms of the tangent-frame component $\textbf{\emph{e}}_\alpha$ ,
\begin{eqnarray}
\mathbb{P}_{(\Psi)} ( P^\alpha_{(\Psi)} ) = P^\alpha_{(\Psi)} \textbf{\emph{e}}_\alpha ~,
\end{eqnarray}
where $P_{(\Psi)}^0 = i p_{(\Psi)}^0$ , $P_{(\Psi)}^q = p_{(\Psi)}^q$ . $P_{(\Psi)}^4 = i p_{(\Psi)}^4$ , $P_{(\Psi)}^{q+4} = p_{(\Psi)}^{q+4}$ . $P_{(\Psi)}^8 = i p_{(\Psi)}^8$, $P_{(\Psi)}^{q+8} = p_{(\Psi)}^{q+8}$. $P_{(\Psi)}^{12} = i p_{(\Psi)}^{12}$ , $P_{(\Psi)}^{q+12} = p_{(\Psi)}^{q+12}$. And the components, $p_{(\Psi)}^j$ , $p_{(\Psi)}^{j+4}$ , $p_{(\Psi)}^{j+8}$ , and $p_{(\Psi)}^{j+12}$ , correspond to $P_{(\Psi)}^{gj}$ , $k_{eg} P_{(\Psi)}^{ej}$ , $k_{wg} P_{(\Psi)}^{wj}$ , and $k_{sg} P_{(\Psi)}^{sj}$ respectively.

\subsection{\label{sec:level1}Quantum angular momentum}

In the orthogonal and equal-length tangent-frame $\Pi$ of the complex-sedenion curved space, the quantum angular momentum can be written as,
\begin{eqnarray}
\mathbb{L}_{(\Psi)} = \mathbb{U}_{(\Psi)}^\star \circ \mathbb{P}_{(\Psi)}  ~,
\end{eqnarray}
where $\mathbb{L}_{(\Psi)} = \mathbb{Z}_L \circ \mathbb{L}$ , with $\mathbb{Z}_L$ being a dimensionless auxiliary quantity. $\mathbb{L}_{(\Psi)} = \mathbb{L}_{(\Psi)g} + k_{eg} \mathbb{L}_{(\Psi)e} + k_{wg} \mathbb{L}_{(\Psi)w} + k_{sg} \mathbb{L}_{(\Psi)s}$. $\mathbb{L}_{(\Psi)g}$ , $\mathbb{L}_{(\Psi)e}$ , $\mathbb{L}_{(\Psi)w}$ , and $\mathbb{L}_{(\Psi)s}$ are the components of quantum angular momentum $\mathbb{L}_{(\Psi)}$ in four complex-quaternion spaces, $\mathbb{H}_g$ , $\mathbb{H}_e$ , $\mathbb{H}_w$, and $\mathbb{H}_s$ , respectively. $\mathbb{L}_{(\Psi)g} = i \textbf{\emph{I}}_{g0} L^{g0}_{(\Psi)} + \textbf{\emph{I}}_{gq} L^{gq}_{(\Psi)}$ . $\mathbb{L}_{(\Psi)e} = i \textbf{\emph{I}}_{e0} L^{e0}_{(\Psi)} + \textbf{\emph{I}}_{eq} L^{eq}_{(\Psi)}$ . $\mathbb{L}_{(\Psi)w} = i \textbf{\emph{I}}_{w0} L^{w0}_{(\Psi)} + \textbf{\emph{I}}_{wq} L^{wq}_{(\Psi)}$ . $\mathbb{L}_{(\Psi)s} = i \textbf{\emph{I}}_{s0} L^{s0}_{(\Psi)} + \textbf{\emph{I}}_{sq} L^{sq}_{(\Psi)}$ . $L^{gj}_{(\Psi)}$ , $L^{ej}_{(\Psi)}$ , $L^{wj}_{(\Psi)}$, and $L^{sj}_{(\Psi)}$ are complex numbers.

In the orthogonal and unequal-length tangent-frame $\Theta$ , the complex-sedenion quantum angular momentum, $\mathbb{L}_{(\Psi)}$, can be expanded in terms of the tangent-frame component $\textbf{\emph{e}}_\alpha$ ,
\begin{eqnarray}
\mathbb{L}_{(\Psi)} ( L^\alpha_{(\Psi)} ) = L^\alpha_{(\Psi)} \textbf{\emph{e}}_\alpha ~,
\end{eqnarray}
where $L_{(\Psi)}^0 = i l_{(\Psi)}^0$ , $L_{(\Psi)}^q = l_{(\Psi)}^q$ . $L_{(\Psi)}^4 = i l_{(\Psi)}^4$ , $L_{(\Psi)}^{q+4} = l_{(\Psi)}^{q+4}$ . $L_{(\Psi)}^8 = i l_{(\Psi)}^8$, $L_{(\Psi)}^{q+8} = l_{(\Psi)}^{q+8}$. $L_{(\Psi)}^{12} = i l_{(\Psi)}^{12}$ , $L_{(\Psi)}^{q+12} = l_{(\Psi)}^{q+12}$. And the components, $l_{(\Psi)}^j$ , $l_{(\Psi)}^{j+4}$ , $l_{(\Psi)}^{j+8}$, and $l_{(\Psi)}^{j+12}$ , correspond to $L_{(\Psi)}^{gj}$ , $k_{eg} L_{(\Psi)}^{ej}$ , $k_{wg} L_{(\Psi)}^{wj}$ , and $k_{sg} L_{(\Psi)}^{sj}$ respectively.

In the tangent-frame $\Theta$ , in virtue of the properties of the quantum composite radius vector $\mathbb{U}_{(\Psi)}$ and quantum linear momentum $\mathbb{P}_{(\Psi)}$ , the complex-sedenion quantum angular momentum, $\mathbb{L}_{(\Psi)} = \mathbb{U}_{(\Psi)}^\star \circ \mathbb{P}_{(\Psi)}$ , can be separated into,
\begin{eqnarray}
\mathbb{L}_{(\Psi)} = \mathbb{U}_{(\Psi)}^\star \circledcirc \mathbb{P}_{(\Psi)} + \mathbb{U}_{(\Psi)}^\star \circledast \mathbb{P}_{(\Psi)}   ~,
\end{eqnarray}
where $\mathbb{U}_{(\Psi)}^\star \circledcirc \mathbb{P}_{(\Psi)}$ and $\mathbb{U}_{(\Psi)}^\star \circledast \mathbb{P}_{(\Psi)}$ are respectively the scalar part and vector part of the quantum angular momentum, $\mathbb{L}_{(\Psi)}$ . Obviously, in the tangent-frame $\Theta$ , the quantum angular momentum $\mathbb{L}_{(\Psi)}$ comprises not only the derivative of the quantum composite radius vector $\mathbb{U}_{(\Psi)}$ and quantum linear momentum $\mathbb{P}_{(\Psi)}$, but also the spatial parameter of complex-sedenion curved space.

\subsection{\label{sec:level1}Quantum torque}

In the orthogonal and equal-length tangent-frame $\Pi$ of the complex-sedenion curved space, the quantum torque can be written as,
\begin{eqnarray}
\mathbb{W}_{(\Psi)} = - v^0 \mathbb{D}_Z \circ \mathbb{L}_{(\Psi)}  ~,
\end{eqnarray}
where $\mathbb{W}_{(\Psi)} = \mathbb{Z}_W \circ \mathbb{W}$ , with $\mathbb{Z}_W$ being a dimensionless auxiliary quantity. $\mathbb{W}_{(\Psi)} = \mathbb{W}_{(\Psi)g} + k_{eg} \mathbb{W}_{(\Psi)e} + k_{wg} \mathbb{W}_{(\Psi)w} + k_{sg} \mathbb{W}_{(\Psi)s}$. $\mathbb{W}_{(\Psi)g}$ , $\mathbb{W}_{(\Psi)e}$ , $\mathbb{W}_{(\Psi)w}$ , and $\mathbb{W}_{(\Psi)s}$ are the components of quantum torque $\mathbb{W}_{(\Psi)}$ in four complex-quaternion spaces, $\mathbb{H}_g$ , $\mathbb{H}_e$ , $\mathbb{H}_w$, and $\mathbb{H}_s$ , respectively. $\mathbb{W}_{(\Psi)g} = i \textbf{\emph{I}}_{g0} W^{g0}_{(\Psi)} + \textbf{\emph{I}}_{gq} W^{gq}_{(\Psi)}$ . $\mathbb{W}_{(\Psi)e} = i \textbf{\emph{I}}_{e0} W^{e0}_{(\Psi)} + \textbf{\emph{I}}_{eq} W^{eq}_{(\Psi)}$. $\mathbb{W}_{(\Psi)w} = i \textbf{\emph{I}}_{w0} W^{w0}_{(\Psi)} + \textbf{\emph{I}}_{wq} W^{wq}_{(\Psi)}$ . $\mathbb{W}_{(\Psi)s} = i \textbf{\emph{I}}_{s0} W^{s0}_{(\Psi)} + \textbf{\emph{I}}_{sq} W^{sq}_{(\Psi)}$. $W^{gj}_{(\Psi)}$ , $W^{ej}_{(\Psi)}$, $W^{wj}_{(\Psi)}$ , and $W^{sj}_{(\Psi)}$ are complex numbers. When $\mathbb{W}_{(\Psi)} = 0$, from the definition of quantum torque, $\mathbb{W}_{(\Psi)} = - v^0 \mathbb{D}_Z \circ \mathbb{L}_{(\Psi)}$ , it is able to deduce the Dirac wave equation. Further the latter can be degenerated into the Schr\"{o}dinger wave equation.

In the orthogonal and unequal-length tangent-frame $\Theta$ , the complex-sedenion quantum torque, $\mathbb{W}_{(\Psi)}$, is expanded in terms of the tangent-frame component $\textbf{\emph{e}}_\alpha$,
\begin{eqnarray}
\mathbb{W}_{(\Psi)} ( W^\alpha_{(\Psi)} ) = W^\alpha_{(\Psi)} \textbf{\emph{e}}_\alpha ~,
\end{eqnarray}
where $W_{(\Psi)}^0 = i w_{(\Psi)}^0$ , $W_{(\Psi)}^q = w_{(\Psi)}^q$ . $W_{(\Psi)}^4 = i w_{(\Psi)}^4$ , $W_{(\Psi)}^{q+4} = w_{(\Psi)}^{q+4}$ . $W_{(\Psi)}^8 = i w_{(\Psi)}^8$, $W_{(\Psi)}^{q+8} = w_{(\Psi)}^{q+8}$ . $W_{(\Psi)}^{12} = i w_{(\Psi)}^{12}$, $W_{(\Psi)}^{q+12} = w_{(\Psi)}^{q+12}$. And the components, $w_{(\Psi)}^j$ , $w_{(\Psi)}^{j+4}$ , $w_{(\Psi)}^{j+8}$, and $w_{(\Psi)}^{j+12}$ , correspond to $W_{(\Psi)}^{gj}$ , $k_{eg} W_{(\Psi)}^{ej}$ , $k_{wg} W_{(\Psi)}^{wj}$ , and $k_{sg} W_{(\Psi)}^{sj}$ respectively.

In the tangent-frame $\Theta$ , in virtue of the property of the quantum angular momentum $\mathbb{L}_{(\Psi)}$ , the complex-sedenion quantum torque, $\mathbb{W}_{(\Psi)} = - v^0 \mathbb{D}_Z \circ \mathbb{L}_{(\Psi)}$, can be separated into,
\begin{eqnarray}
\mathbb{W}_{(\Psi)} = - v^0 \mathbb{D}_Z \circledcirc \mathbb{L}_{(\Psi)} - v^0 \mathbb{D}_Z \circledast \mathbb{L}_{(\Psi)}   ~,
\end{eqnarray}
where $- v^0 \mathbb{D}_Z \circledcirc \mathbb{L}_{(\Psi)}$ and $- v^0 \mathbb{D}_Z \circledast \mathbb{L}_{(\Psi)}$ are respectively the scalar part and vector part of the quantum torque, $\mathbb{W}_{(\Psi)}$. Apparently, in the tangent-frame $\Theta$ , the quantum torque $\mathbb{W}_{(\Psi)}$ contains not only the derivative of quantum angular momentum $\mathbb{L}_{(\Psi)}$ , but also the spatial parameter of complex-sedenion curved space.

\subsection{\label{sec:level1}Quantum force}

In the orthogonal and equal-length tangent-frame $\Pi$ of the complex-sedenion curved space, the quantum force can be written as,
\begin{eqnarray}
\mathbb{N}_{(\Psi)} = - \mathbb{D}_Z \circ \mathbb{W}_{(\Psi)}  ~,
\end{eqnarray}
where $\mathbb{N}_{(\Psi)} = \mathbb{Z}_N \circ \mathbb{N}$ , with $\mathbb{Z}_N$ being a dimensionless auxiliary quantity. $\mathbb{N}_{(\Psi)} = \mathbb{N}_{(\Psi)g} + k_{eg} \mathbb{N}_{(\Psi)e} + k_{wg} \mathbb{N}_{(\Psi)w} + k_{sg} \mathbb{N}_{(\Psi)s}$. $\mathbb{N}_{(\Psi)g}$ , $\mathbb{N}_{(\Psi)e}$ , $\mathbb{N}_{(\Psi)w}$ , and $\mathbb{N}_{(\Psi)s}$ are the components of quantum force $\mathbb{N}_{(\Psi)}$ in four complex-quaternion spaces, $\mathbb{H}_g$ , $\mathbb{H}_e$, $\mathbb{H}_w$, and $\mathbb{H}_s$ , respectively. $\mathbb{N}_{(\Psi)g} = i \textbf{\emph{I}}_{g0} N^{g0}_{(\Psi)} + \textbf{\emph{I}}_{gq} N^{gq}_{(\Psi)}$ . $\mathbb{N}_{(\Psi)e} = i \textbf{\emph{I}}_{e0} N^{e0}_{(\Psi)} + \textbf{\emph{I}}_{eq} N^{eq}_{(\Psi)}$. $\mathbb{N}_{(\Psi)w} = i \textbf{\emph{I}}_{w0} N^{w0}_{(\Psi)} + \textbf{\emph{I}}_{wq} N^{wq}_{(\Psi)}$ . $\mathbb{N}_{(\Psi)s} = i \textbf{\emph{I}}_{s0} N^{s0}_{(\Psi)} + \textbf{\emph{I}}_{sq} N^{sq}_{(\Psi)}$. $N^{gj}_{(\Psi)}$ , $N^{ej}_{(\Psi)}$, $N^{wj}_{(\Psi)}$ , and $N^{sj}_{(\Psi)}$ are complex numbers.

In the orthogonal and unequal-length tangent-frame $\Theta$ , the complex-sedenion quantum force, $\mathbb{N}_{(\Psi)}$ , is expanded in terms of the tangent-frame component $\textbf{\emph{e}}_\alpha$ ,
\begin{eqnarray}
\mathbb{N}_{(\Psi)} ( N^\alpha_{(\Psi)} ) = N^\alpha_{(\Psi)} \textbf{\emph{e}}_\alpha ~,
\end{eqnarray}
where $N_{(\Psi)}^0 = i n_{(\Psi)}^0$ , $N_{(\Psi)}^q = n_{(\Psi)}^q$ . $N_{(\Psi)}^4 = i n_{(\Psi)}^4$ , $N_{(\Psi)}^{q+4} = n_{(\Psi)}^{q+4}$ . $N_{(\Psi)}^8 = i n_{(\Psi)}^8$, $N_{(\Psi)}^{q+8} = n_{(\Psi)}^{q+8}$ . $N_{(\Psi)}^{12} = i n_{(\Psi)}^{12}$ , $N_{(\Psi)}^{q+12} = n_{(\Psi)}^{q+12}$. And the components, $n_{(\Psi)}^j$ , $n_{(\Psi)}^{j+4}$ , $n_{(\Psi)}^{j+8}$ , and $n_{(\Psi)}^{j+12}$ , correspond to $N_{(\Psi)}^{gj}$ , $k_{eg} N_{(\Psi)}^{ej}$ , $k_{wg} N_{(\Psi)}^{wj}$ , and $k_{sg} N_{(\Psi)}^{sj}$ respectively.

In the tangent-frame $\Theta$ , in virtue of the property of the quantum torque $\mathbb{W}_{(\Psi)}$, the complex-sedenion quantum force, $\mathbb{N}_{(\Psi)} = - \mathbb{D}_Z \circ \mathbb{W}_{(\Psi)}$ , is separated into,
\begin{eqnarray}
\mathbb{N}_{(\Psi)} = - \mathbb{D}_Z \circledcirc \mathbb{W}_{(\Psi)} - \mathbb{D}_Z \circledast \mathbb{W}_{(\Psi)}   ~,
\end{eqnarray}
where $- \mathbb{D}_Z \circledcirc \mathbb{W}_{(\Psi)}$ and $- \mathbb{D}_Z \circledast \mathbb{W}_{(\Psi)}$ are respectively the scalar part and vector part of the quantum force, $\mathbb{N}_{(\Psi)}$. Apparently, in the tangent-frame $\Theta$ , the quantum force $\mathbb{N}_{(\Psi)}$ contains not only the derivative of quantum torque $\mathbb{W}_{(\Psi)}$ , but also the spatial parameter of complex-sedenion curved space.

\begin{table}[h]
\tbl{The quantum-field equations, for the quantum mechanics on the microscopic scale, in the complex-sedenion curved quantum space $\mathbb{K}_Z$ .}
{\begin{tabular}{@{}ll@{}}
\toprule
quantum~physics~quantity     &  definition                                                                  \\
\colrule
quantum~field~potential      &  $\mathbb{A}_{(\Psi)} = i \mathbb{D}_Z^\star \circ \mathbb{X}_{(\Psi)}  $    \\
quantum~field~strength       &  $\mathbb{F}_{(\Psi)} = \mathbb{D}_Z \circ \mathbb{A}_{(\Psi)}  $            \\
quantum~field~source         &  $\mu \mathbb{S}_{(\Psi)} = - \mathbb{D}_Z^\ast \circ \mathbb{F}_{(\Psi)} $  \\
quantum~linear~momentum      &  $\mathbb{P}_{(\Psi)} = \mu \mathbb{S}_{(\Psi)} / \mu_g^g  $                 \\
quantum~angular~momentum~~~~~&  $\mathbb{L}_{(\Psi)} = \mathbb{U}^\star \circ \mathbb{P}_{(\Psi)}$          \\
quantum~torque               &  $\mathbb{W}_{(\Psi)} = - v^0 \mathbb{D}_Z \circ \mathbb{L}_{(\Psi)}  $      \\
quantum~force                &  $\mathbb{N}_{(\Psi)} = - \mathbb{D}_Z \circ \mathbb{W}_{(\Psi)} $           \\
\botrule
\end{tabular}}
\end{table}

\section{\label{sec:level1}Composite space}

In the complex-sedenion curved space $\mathbb{K}$ , it is able to express the contribution of the curved space on the four classical-fields. However it is incapable of establishing directly the relationship between the physical quantity and spatial parameter, neither exploring the reason to result in the warping of space. By contrast, in the complex-sedenion curved composite-space, it is able to deduce the formula between the physical quantity and spatial parameter, achieving a few inferences in accordance with that derived from the GR and complex-octonion curved composite-space. Apparently the research methods of the complex-sedenion curved space in the above can be extended into that of the curved composite-space described with the complex-sedenions.

\subsection{\label{sec:level1}Composite radius vector}

In the complex-sedenion flat space, the radius vector and the integrating function of field potential can be combined together to become the composite radius vector. The latter should be considered as a whole to take into account in the following context, enabling the energy expression to include the `electromagnetic potential energy' and `gravitational potential energy' and so forth. Further the composite radius vector, $\mathbb{U}$, can be regarded as the radius vector in one function space, which is called as the composite-space, $\mathbb{K}_U$ , temporarily.

In the flat composite-space, $\mathbb{K}_U$ , described with the complex-sedenions, it is appropriate to choose an orthogonal and equal-length tangent-frame as the coordinate system (Table 4). In the coordinate system, from Eqs.(1) and (12), the composite radius vector can be written as,
\begin{eqnarray}
&& \mathbb{U} ( H^\alpha ) = i H^0 \textbf{\emph{i}}_0 + H^q \textbf{\emph{i}}_q + i H^4 \textbf{\emph{i}}_4 + H^{4+q} \textbf{\emph{i}}_{4+q}
\nonumber
\\
&&
~~~~~~~~~~~~
+ i H^8 \textbf{\emph{i}}_8 + H^{8+q} \textbf{\emph{i}}_{8+q} + i H^{12} \textbf{\emph{i}}_{12} + H^{12+q} \textbf{\emph{i}}_{12+q}   ~ ,
\end{eqnarray}
where $H^\alpha$ being the coordinate value, with $H^\alpha = h^\alpha + k_{rx} x^\alpha $ .

In the curved composite-space, $\mathbb{K}_U$ , described with the complex-sedenions, it is able to choose an orthogonal and unequal-length tangent-frame as the coordinate system. In the coordinate system of the tangent space, from Eqs.(2) and (13), the composite radius vector, $\mathbb{U}$ , can be expanded in terms of the tangent-frame component $\textbf{\emph{E}}_\alpha$ ,
\begin{eqnarray}
&& \mathbb{U} ( C^\alpha ) = i C^0 \textbf{\emph{E}}_0 + C^q \textbf{\emph{E}}_q + i C^4 \textbf{\emph{E}}_4 + C^{4+q} \textbf{\emph{E}}_{4+q}
\nonumber
\\
&&
~~~~~~~~~~~~
+ i C^8 \textbf{\emph{E}}_8 + C^{8+q} \textbf{\emph{E}}_{8+q} + i C^{12} \textbf{\emph{E}}_{12} + C^{12+q} \textbf{\emph{E}}_{12+q}   ~ ,
\end{eqnarray}
where the coordinate values, $C^\alpha$ and $y^\alpha$ , both are real. $C^\alpha = c^\alpha + k_{rx} y^\alpha $ . $\textbf{\emph{E}}_\alpha$ is unequal-length, and $\textbf{\emph{E}}_\alpha$ corresponds to $\textbf{\emph{i}}_\alpha$ .

Further, the above can be rewritten as,
\begin{eqnarray}
\mathbb{U} ( U^\alpha ) = U^\alpha \textbf{\emph{E}}_\alpha ~,
\end{eqnarray}
where $U^0 = i C^0$ , $U^q = C^q$ . $U^4 = i C^4$ , $U^{q+4} = C^{q+4}$ . $U^8 = i C^8$ , $U^{q+8} = C^{q+8}$ . $U^{12} = i C^{12}$ , $U^{q+12} = C^{q+12}$ .
$\textbf{\emph{E}}_\alpha = \partial \mathbb{U} / \partial U^\alpha$ . $(\textbf{\emph{E}}_0)^2 > 0$, and $(\textbf{\emph{E}}_\xi)^2 < 0$.

From the viewpoint of function space, the composite-space, $\mathbb{K}_U$ , is one type of function space. The function space may possess multiple physical quantities, and each coordinate value may be a function. In other words, the coordinate value $U^\alpha$ of composite-space is one function, while the coordinate value $U^\alpha$ consists of the spatial parameter $c^\alpha$ and physical quantity $y^\alpha$ . When the composite-space $\mathbb{K}_U$ is distorted, it is found that there may be some other formula between the spatial parameter and physical quantity.

\subsection{\label{sec:level1}Curved composite-space}

In the complex-sedenion curved composite-space, $\mathbb{K}_U$ , the partial derivative of composite radius vector with respect to the coordinate value is utilized for the component of tangent-frame in the tangent space. Obviously, this is a new type of curved space, on the basis of the composite radius vector $\mathbb{U}$ . And the underlying space and tangent space both are the complex-sedenion composite-spaces. Making use of the property of complex-sedenion composite radius vector, it is capable of deducing some arguments in the curved composite-space, including the metric coefficient, connection coefficient, covariant derivative, and curvature tensor and so forth. After decomposing the curvature tensor, it is found some relations between the spatial parameter and physical quantity, in the complex-sedenion curved space $\mathbb{K}$ .

In the complex-sedenion curved composite-space $\mathbb{K}_U$ , the partial derivative, $\textbf{\emph{E}}_\alpha$, of composite radius vector with respect to the coordinate value, $U^\alpha$ , is chosen as the component of tangent-frame in the tangent space. These orthogonal and unequal-length components of tangent-frame constitute one coordinate system. Making use of the tangent-frame component and the norm of composite radius vector, it is able to define the metric of complex-sedenion curved composite-space $\mathbb{K}_U$ as follows,
\begin{eqnarray}
d S^2_{(R,X)} = g_{\overline{\alpha}\beta(R,X)} d \overline{U^\alpha} d U^\beta  ~ ,
\end{eqnarray}
where the metric coefficient, $g_{\overline{\alpha}\beta(R,X)} = \textbf{\emph{E}}_\alpha^\ast \circ \textbf{\emph{E}}_\beta$ , is sedenion-Hermitian. $\textbf{\emph{E}}_\alpha$ is the component of tangent-frame, with $\textbf{\emph{E}}_\alpha = \partial \mathbb{U} / \partial U^\alpha$ . $g_{\overline{\alpha}\beta(R,X)}$ is scalar, due to the orthogonal tangent-frames. $( U^\alpha )^\ast = \overline{U^\alpha}$ , and it indicates that the correlated tangent-frame component, $\textbf{\emph{E}}_\alpha$ , is sedenion conjugate.

In the complex-sedenion curved composite-space $\mathbb{K}_U$ , substituting the coordinate value $U^\alpha$ , tangent-frame component $\textbf{\emph{E}}_\alpha$ , and metric coefficient $g_{\overline{\alpha} \beta (R,X)}$ for the coordinate value $u^\alpha$ , tangent-frame component $\textbf{\emph{e}}_\alpha$ , and metric coefficient $g_{\overline{\alpha} \beta}$ in the complex-sedenion curved space $\mathbb{K}$ respectively, it is capable of inferring the connection coefficients of the curved composite-space $\mathbb{K}_U$ . And there are,
\begin{eqnarray}
&& \Gamma_{\overline{\lambda} , \beta \gamma (R,X)} = (1/2) ( \partial g_{\overline{\gamma} \lambda (R,X)} / \partial U^\beta
\nonumber
\\
&&
~~~~~~~~~~~~~~~~~~
+ \partial g_{\overline{\lambda} \beta (R,X)} / \partial U^\gamma - \partial g_{\overline{\gamma} \beta (R,X)} /\partial U^\lambda ) ~ ,
\\
&& \Gamma_{\overline{\lambda} , \overline{\beta} \gamma (R,X)} = (1/2) ( \partial g_{\overline{\gamma} \lambda (R,X)} / \partial \overline{U^\beta}
\nonumber
\\
&&
~~~~~~~~~~~~~~~~~~
+ \partial g_{\overline{\lambda} \beta (R,X)} / \partial \overline{U^\gamma} - \partial g_{\overline{\gamma} \beta (R,X)} /\partial \overline{U^\lambda} ) ~ ,
\end{eqnarray}
where $\Gamma_{\overline{\lambda} , \beta \gamma (R,X)}$ and $\Gamma_{\overline{\lambda} , \overline{\beta} \gamma (R,X)}$ are connection coefficients, and both of them are all scalar. $\Gamma_{\overline{\lambda} , \beta \gamma (R,X)} = g_{\overline{\lambda} \alpha (R,X)} \Gamma_{\beta \gamma (R,X)}^\alpha$. $ \Gamma_{\beta \gamma (R,X)}^\alpha = g^{\alpha \overline{\lambda}}_{(R,X)} \Gamma_{\overline{\lambda} , \beta \gamma (R,X)} $ . $\Gamma_{\beta \gamma (R,X)}^\alpha = \Gamma_{\gamma \beta (R,X)}^\alpha $. $ \Gamma_{\overline{\beta} \gamma (R,X)}^\alpha = g^{\alpha \overline{\lambda}}_{(R,X)} \Gamma_{\overline{\lambda} , \overline{\beta} \gamma (R,X)} $ . $\Gamma_{\overline{\lambda} , \overline{\beta} \gamma (R,X)} = g_{\overline{\lambda} \alpha (R,X)} \Gamma_{\overline{\beta} \gamma (R,X)}^\alpha$. $[ ( \Gamma_{\overline{\beta} \gamma (R,X)}^\alpha )^* ]^T = \Gamma_{\gamma \overline{\beta} (R,X)}^\alpha $. $\Gamma_{\overline{\beta} \gamma (R,X)}^\alpha = \Gamma_{\gamma \overline{\beta} (R,X)}^\alpha $ . $ g^{\alpha \overline{\lambda}}_{(R,X)} g_{\overline{\lambda} \beta (R,X)} = \delta^\alpha_\beta $ .

In the curved composite-space $\mathbb{K}_U$ described with the complex-sedenions, when the physical quantity, $\mathbb{Y} = Y^\beta \textbf{\emph{E}}_\beta $ , is transferred from a point $M_1$ to the next point $M_2$ , to meet the requirement of parallel translation, it means that the differential of quantity $\mathbb{Y}$ equals to zero. This condition of parallel translation, $d \mathbb{Y} = 0$, will infer,
\begin{equation}
d Y^\beta = - \Gamma_{\alpha \gamma (R,X)}^\beta Y^\alpha d U^\gamma  ~,
\end{equation}
with
\begin{equation}
\partial^2 \mathbb{U} / \partial U^\beta \partial U^\gamma = \Gamma_{\beta \gamma (R,X)}^\alpha \textbf{\emph{E}}_\alpha  ~ .
\end{equation}

In the curved composite-space $\mathbb{K}_U$ described with the complex-sedenions, for the first-rank contravariant tensor $Y^\beta$ of a point $M_2$ , the component of covariant derivative with respect to the coordinate $U^\gamma$ is written as,
\begin{eqnarray}
\nabla_\gamma Y^\beta = \partial ( \delta_\alpha^\beta Y^\alpha ) / \partial U^\gamma + \Gamma_{\alpha \gamma (R,X)}^\beta Y^\alpha   ~ ,
\\
\nabla_{\overline{\gamma}} Y^\beta = \partial ( \delta_\alpha^\beta Y^\alpha ) / \partial \overline{U^\gamma} + \Gamma_{\alpha \overline{\gamma} (R,X)}^\beta Y^\alpha   ~ ,
\end{eqnarray}
where $Y^\beta$ and $\Gamma_{\alpha \overline{\gamma} (R,X)}^\beta Y^\alpha$ both are scalar.

From the property of covariant derivative of tensor, the definition of curvature tensor is,
\begin{equation}
\nabla_{\overline{\alpha}} ( \nabla_\beta Y^\gamma ) - \nabla_\beta ( \nabla_{\overline{\alpha}} Y^\gamma ) = R_{\beta \overline{\alpha} \nu ~~(R,X)}^{~~~~~~\gamma} Y^\nu + T_{\beta \overline{\alpha} (R,X)}^\lambda ( \nabla_\lambda Y^\gamma )  ~   ,
\end{equation}
with
\begin{eqnarray}
&& R_{\beta \overline{\alpha} \nu ~~(R,X)}^{~~~~~~\gamma} = \partial \Gamma_{\nu \beta (R,X)}^\gamma / \partial \overline{U^\alpha} - \partial \Gamma_{\nu \overline{\alpha} (R,X)}^\gamma  / \partial U^\beta
\nonumber
\\
&&
~~~~~~~~~~~~~~~~~~~
+ \Gamma_{\lambda \overline{\alpha} (R,X)}^\gamma \Gamma_{\nu \beta (R,X)}^\lambda - \Gamma_{\lambda \beta (R,X)}^\gamma \Gamma_{\nu \overline{\alpha} (R,X)}^\lambda   ~~ ,
\end{eqnarray}
where $R_{\beta \overline{\alpha} \nu ~~(R,X)}^{~~~~~~\gamma}$ and $T_{\beta \overline{\alpha} (R,X)}^\lambda$ both are scalar. $R_{\beta \overline{\alpha} \nu ~~(R,X)}^{~~~~~~\gamma}$ is the curvature tensor, while $T_{\beta \overline{\alpha} (R,X)}^\lambda$ is the torsion tensor. $T_{\beta \overline{\alpha} (R,X)}^\lambda = \Gamma_{\overline{\alpha} \beta (R,X)}^\lambda - \Gamma_{\beta \overline{\alpha} (R,X)}^\lambda $. In the paper, we merely discuss the case, $T_{\beta \overline{\alpha} (R,X)}^\lambda = 0$, that is, $\Gamma_{\overline{\alpha} \beta (R,X)}^\lambda = \Gamma_{\beta \overline{\alpha} (R,X)}^\lambda $ .

In the curved composite-space, $\mathbb{K}_U$ , described with the complex-sedenions, it is found that the spatial parameter and physical quantity both make a contribution to these arguments, by means of the metric coefficient, connection coefficient, and curvature tensor and so forth. In case the coupling of spatial parameters and physical quantities of these arguments could be decoupled under some circumstances, it may be able to seek out some direct interrelations between the spatial parameter and physical quantity.

\subsection{\label{sec:level1}Coupling term}

From the preceding analysis, it is found that the radius vector and the integrating function of field potential can be linearly superposed together to become the composite radius vector, in the complex-sedenion curved composite-space $\mathbb{K}_U$ . However, as the derivative of composite radius vector, the interrelation between the spatial parameter with physical quantity in the following context will be complicated enough, including certain interconnections among the metric coefficient, connection coefficient, and curvature tensor. The coupling term (c.t. for short) between the spatial parameter and physical quantity is always extremely intricate, and it is very tough to decouple in general. Only in some special cases, it may be decoupled approximately. And it is able to allow us to attempt to grasp at the relations of spatial parameter and physical quantity.

In the metric equation, Eq.(61), the contributions, coming from the spatial parameter and physical quantity, to the metric coefficient are interrelated closely. Each of partial derivatives, $ \partial \mathbb{R} / \partial U^\gamma $ and $ \partial \mathbb{X} / \partial U^\gamma $ , will exert an influence on the metric coefficient. In a few comparatively simple circumstances, the metric tensor can be separated into,
\begin{equation}
g_{\overline{\alpha} \beta (R,X)} = g_{\overline{\alpha} \beta (R)} + g_{\overline{\alpha} \beta (X)} + c.t.( g_{\overline{\alpha} \beta} , \mathbb{R} , \mathbb{X} )  ~,
\end{equation}
where the partial derivative, $ \partial \mathbb{X} / \partial U^\alpha $ , possesses the dimension of field potential. In case the contribution of $\mathbb{X}$ could be neglected, $g_{\overline{\alpha} \beta (R,X)}$ will be reduced into $g_{\overline{\alpha} \beta (R)}$, which contains only the contribution of $ \partial \mathbb{R} / \partial U^\alpha $ . Meanwhile, if the contribution of $\mathbb{R}$ could be neglected, $g_{\overline{\alpha} \beta (R,X)}$ will be degenerated into $g_{\overline{\alpha} \beta (X)}$, which covers merely the contribution of $ \partial \mathbb{X} / \partial U^\alpha $ .

In the connection coefficient, Eq.(62), it is found that the contributions, coming from the spatial parameter and physical quantity, to the connection coefficient are mingled together. Both of partial derivatives, $\partial g_{\overline{\alpha} \beta (R)} / \partial U^\alpha$ and $\partial g_{\overline{\alpha} \beta (X)} / \partial U^\alpha$ , have the influences on the connection coefficient. In several comparatively simple cases, the connection coefficient can be divided into,
\begin{equation}
\Gamma_{\alpha \beta (R,X)}^\gamma = \Gamma_{\alpha \beta (R)}^\gamma + \Gamma_{\alpha \beta (X)}^\gamma + c.t.( \Gamma_{\alpha \beta}^\gamma , \mathbb{R} , \mathbb{X} ) ~ ,
\end{equation}
where the partial derivative, $(k_{rx}^{~~-1}) \partial g_{\overline{\alpha} \beta (X)} / \partial U^\gamma  $ , possesses the dimension of field strength. If the contribution of $g_{\overline{\alpha} \beta (X)}$ could be neglected, $\Gamma_{\alpha \beta (R,X)}^\gamma$ will be reduced into $\Gamma_{\alpha \beta (R)}^\gamma$ , which contains only the contribution of $ \partial g_{\overline{\alpha} \beta (R)} / \partial U^\gamma  $. In case the contribution of $g_{\overline{\alpha} \beta (R)}$ could be neglected, $\Gamma_{\alpha \beta (H,X)}^\gamma$ will be simplified into $\Gamma_{\alpha \beta (X)}^\gamma$ , which contains only the contribution of $ \partial g_{\overline{\alpha} \beta (X)} / \partial U^\gamma  $ . The decomposition process of the connection coefficient, Eq.(63), is similar to that of Eq.(62), under some comparatively simple circumstances.

The analysis of the curvature tensor, Eq.(69), shows that the contributions, coming from the spatial parameter and physical quantity, to the curvature tensor are mixed complicatedly. Four partial derivatives, $ \partial \Gamma_{\alpha \beta (R)}^\gamma / \partial U^\nu $ , $ \partial \Gamma_{\alpha \beta (X)}^\gamma / \partial U^\nu $, $ \partial \Gamma_{\overline{\alpha} \beta (R)}^\gamma / \partial U^\nu $, and $ \partial \Gamma_{\overline{\alpha} \beta (X)}^\gamma / \partial U^\nu $ , will directly impact the curvature tensor. Under some comparatively simple cases, the curvature tensor can be separated into,
\begin{equation}
R_{\beta \overline{\alpha} \nu ~~(R,X)}^{~~~~\gamma} = R_{\beta \overline{\alpha} \nu ~~(R)}^{~~~~\gamma} + R_{\beta \overline{\alpha} \nu ~~(X)}^{~~~~\gamma} + c.t.( R_{\beta \overline{\alpha} \nu }^{~~~~\gamma} , \mathbb{R} , \mathbb{X} )   ~   ,
\end{equation}
where two partial derivatives, $ (k_{rx}^{~~-1}) \partial \Gamma_{\alpha \beta (X)}^\gamma / \partial U^\nu $ and $ (k_{rx}^{~~-1}) \partial \Gamma_{\overline{\alpha} \beta (X)}^\gamma / \partial U^\nu $, both possess the dimension of field source. When the contributions of two terms, $\Gamma_{\alpha \beta (X)}^\gamma$ and $\Gamma_{\overline{\alpha} \beta (X)}^\gamma$ , could be neglected, $R_{\beta \overline{\alpha} \nu ~~(R,X)}^{~~~~\gamma}$ will be reduced into $R_{\beta \overline{\alpha} \nu ~~(R)}^{~~~~\gamma}$ , which contains the contributions of $ \partial \Gamma_{\alpha \beta (R)}^\gamma / \partial U^\nu $ and $ \partial \Gamma_{\overline{\alpha} \beta (R)}^\gamma / \partial U^\nu $. If the contribution of two terms, $\Gamma_{\alpha \beta (R)}^\gamma$ and $\Gamma_{\overline{\alpha} \beta (R)}^\gamma$ , could be neglected, $R_{\beta \overline{\alpha} \nu ~~(R,X)}^{~~~~\gamma}$ will be reduced into $R_{\beta \overline{\alpha} \nu ~~(X)}^{~~~~\gamma}$ , which contains the contribution of $ \partial \Gamma_{\alpha \beta (X)}^\gamma / \partial U^\nu $ and $ \partial \Gamma_{\overline{\alpha} \beta (X)}^\gamma / \partial U^\nu $ .

In the complex-sedenion curved composite-space $\mathbb{K}_U$ , in virtue of the precise experimental measurements, it is able to determine whether the composite-space is distorted. Apparently, it is not reasonable to presume whether the composite-space is distorted, before finishing some required experiment measurements. In other words, there may be one flat composite-space under some circumstances, that is, $R_{\beta \overline{\alpha} \nu ~~(R,X)}^{~~~~\gamma} = 0$. Even if neither of ingredients, $R_{\beta \overline{\alpha} \nu ~~(R)}^{~~~~\gamma}$ and $R_{\beta \overline{\alpha} \nu ~~(X)}^{~~~~\gamma}$, is equal to zero, the curvature tensor $R_{\beta \overline{\alpha} \nu ~~(R,X)}^{~~~~\gamma}$ is still possible to be zero. Undoubtedly there must be several relations between these two ingredients, $R_{\beta \overline{\alpha} \nu ~~(R)}^{~~~~\gamma}$ and $R_{\beta \overline{\alpha} \nu ~~(X)}^{~~~~\gamma}$ .

When $R_{\beta \overline{\alpha} \nu ~~(R,X)}^{~~~~\gamma} = 0$ , in the complex-sedenion flat composite-space $\mathbb{K}_U$, it is able to deduce a simple equation among a few ingredients of curvature tensor as follows,
\begin{equation}
0 = R_{\beta \overline{\alpha} \nu ~~(R)}^{~~~~\gamma} + R_{\beta \overline{\alpha} \nu ~~(X)}^{~~~~\gamma} + c.t.( R_{\beta \overline{\alpha} \nu }^{~~~~\gamma} , \mathbb{R} , \mathbb{X} )   ~   ,
\end{equation}
further, in case the coupling term, $c.t.( R_{\beta \overline{\alpha} \nu }^{~~~~\gamma} , \mathbb{R} , \mathbb{X} )$ , is equal to zero, the above can be reduced into one simpler equation,
\begin{equation}
R_{\beta \overline{\alpha} \nu ~~(R)}^{~~~~\gamma} = - R_{\beta \overline{\alpha} \nu ~~(X)}^{~~~~\gamma}    ~   ,
\end{equation}
where the left side is the contribution of spatial parameters, and the right side is the contribution of physical quantities. The term, $(k_{rx}^{~~-1}) R_{\beta \overline{\alpha} \nu ~~(X)}^{~~~~\gamma}$ , possesses the dimension of field source, and are much more complicated than that of field source.

The above consists with the academic thought in the GR and complex-octonion curved composite-space, that is, the existence of field dominates the bending of space. And the above intuitively expands this academic thought from another point of view. Certainly the GR, complex-octonion composite-space, and the paper succeed to the Cartesian academic thought of `the space is the extension of substance'.

After achieving the spatial parameters, it is found that these parameters will exert an influence on certain operators, impacting some field equations of the four classical-fields. On the basis of composite radius vector $\mathbb{U}$ , some `physical quantities' (such as, $g_{\overline{\alpha} \beta (X)}$ , $\Gamma_{\alpha \beta (X)}^\gamma$ , and $R_{\beta \overline{\alpha} \nu ~~(X)}^{~~~~\gamma}$ ) may make a contribution towards the spatial parameters, including
$g_{\overline{\alpha} \beta (R)}$ , $\Gamma_{\alpha \beta (R)}^\gamma$ , and $R_{\beta \overline{\alpha} \nu ~~(R)}^{~~~~\gamma}$ . There is, $\mathbb{R} + k_{rx} \mathbb{X} \approx \mathbb{R}$ , in general, so these spatial parameters will approximate to that in Section 2. Next, these spatial parameters have the influences on some operators in the complex-sedenion curved space $\mathbb{K}$ , including the divergence, gradient, and curl. Finally, these operators are capable of impacting directly some field equations of the four classical-fields in the curved space $\mathbb{K}$ (see Section 4), even in the curved quantum-space $\mathbb{K}_Z$ (see Section 5). By all appearances, a majority of arguments in the composite-space $\mathbb{K}_U$ are different from that in the fundamental-space $\mathbb{K}$ , or quantum-space $\mathbb{K}_Z$ . Suddenly, it is found that the study of curved space $\mathbb{K}$ is so closely interrelated with the four classical-fields described with the complex-sedenions, by means of the composite-space $\mathbb{K}_U$ .

The research method in the curved composite-space $\mathbb{K}_U$ , based on the composite radius vector, can be extended into the curved product-space, based on the quantum composite radius vector, $\mathbb{U}_{(\Psi)}$ . In this curved product-space, some physical quantities are possible to dominate the spatial parameters of curved quantum-space, $\mathbb{K}_Z$, impacting some field equations of four quantum-fields from different standpoints.

\begin{table}[ht]
\tbl{Contrast of some tangent-frames in the complex-sedenion flat and curved spaces.}
{\begin{tabular}{@{}llll@{}}
\toprule
complex-sedenion~space~            &  flat~space~                   &  curved~space~                        &   decomposition                            \\
\colrule
$\mathbb{K}$ , fundamental-space   &  $\textbf{\emph{i}}_\alpha$    &  $\textbf{\emph{e}}_\alpha$           &                                            \\
$\mathbb{K}_U$ , composite-space   &  $\textbf{\emph{i}}_\alpha$    &  $\textbf{\emph{E}}_\alpha$           &   $\textbf{\emph{E}}_{\alpha(R)}$ , or $\textbf{\emph{E}}_{\alpha(X)}$      \\
$\mathbb{K}_Z$ , quantum-space     &  $\textbf{\emph{i}}_\alpha$    &  $\textbf{\emph{e}}_\alpha$           &                                            \\
$\mathbb{K}_M$ , product-space     &  $\textbf{\emph{i}}_\alpha$    &  $\textbf{\emph{E}}_{\alpha(Z,R,X)}$  &   $\textbf{\emph{E}}_{\alpha(Z,R)}$ , or $\textbf{\emph{E}}_{\alpha(Z,X)}$    \\
\botrule
\end{tabular}}
\end{table}

\section{\label{sec:level1}Product space}

In the complex-sedenion curved quantum-space $\mathbb{K}_Z$ , it is able to describe the contribution of the curved quantum-space on the four quantum-fields. But it is incapable of establishing directly the interrelation between the physical quantity and spatial parameter, neither researching the reason to lead to distorting the quantum-space $\mathbb{K}_Z$. By contrast, in the complex-sedenion curved product-space, it is able to infer the equation between the physical quantity and spatial parameter, achieving a few inferences in accordance with that derived from the complex-octonion curved composite-space and GR. Obviously the research methods of the complex-sedenion curved quantum-space in the above can be extended into that of the curved product-space described with the complex-sedenions.

\subsection{\label{sec:level1}Quantum composite radius vector}

In the complex-sedenion flat space, the quantum composite radius vector, $\mathbb{M} = \mathbb{Z}_U \circ \mathbb{U}$, is the sedenion product of the composite radius vector, $\mathbb{U}$ , and auxiliary quantity, $\mathbb{Z}_U$. This quantum composite radius vector, $\mathbb{M}$ or $\mathbb{U}_{(\Psi)}$ , can be considered as the radius vector of one function space, which is called as the product-space, $\mathbb{K}_M$, temporarily.

In the complex-sedenion flat product-space, $\mathbb{K}_M$ , it is able to choose an orthogonal and equal-length tangent-frame as the coordinate system. In the coordinate system, the quantum composite radius vector can be expanded in terms of the tangent-frame component $\textbf{\emph{i}}_\alpha$ ,
\begin{equation}
\mathbb{M} ( m^\alpha ) = m^\alpha \textbf{\emph{i}}_\alpha   ~   ,
\end{equation}
where the scalar $m^\alpha$ is the coordinate value in the flat product-space.

Similarly, in the complex-sedenion curved product-space, $\mathbb{K}_M$ , it is appropriate to choose an orthogonal and unequal-length tangent-frame as the coordinate system. In the coordinate system of the tangent space, the quantum composite radius vector, $\mathbb{M}$ , can be expanded in terms of the tangent-frame component $\textbf{\emph{E}}_{\alpha(Z,R,X)}$ ,
\begin{equation}
\mathbb{M} ( M^\alpha ) = M^\alpha \textbf{\emph{E}}_{\alpha(Z,R,X)}   ~   ,
\end{equation}
where the scalar $M^\alpha$ is the coordinate value in the curved product-space. $\textbf{\emph{E}}_{\alpha(Z,R,X)}$ and $M^\alpha$ correspond to $\textbf{\emph{i}}_\alpha$ and $m^\alpha$ respectively. $\textbf{\emph{E}}_{\alpha(Z,R,X)} =  \partial \mathbb{M} / \partial M^\alpha$ , and $\textbf{\emph{E}}_{\alpha(Z,R,X)}$ is unequal-length. $(\textbf{\emph{E}}_{0(Z,R,X)})^2 > 0$, and $(\textbf{\emph{E}}_{\xi(Z,R,X)})^2 < 0$.

From the viewpoint of function space, the product-space, $\mathbb{K}_M$ , is one type of function space. The function space possesses multiple arguments, and each coordinate value is a function. In other words, the coordinate value $M^\alpha$ of product-space is one function, and contains a few spatial parameters and physical quantities. When the product-space is distorted, there may be some other formulae between the spatial parameter and physical quantity.

\subsection{\label{sec:level1}Curved product-space}

In the complex-sedenion curved product-space, $\mathbb{K}_M$ , the partial derivative of quantum composite radius vector, $\mathbb{M}$, with respect to the coordinate value is chosen for the component of tangent-frame in the tangent space. Obviously, it is a new type of curved space, on the basis of the quantum composite radius vector $\mathbb{M}$ . And the underlying space and tangent space both are the complex-sedenion product-spaces. In virtue of the property of complex-sedenion quantum composite radius vector, it is capable of inferring some arguments in the curved product-space, including the metric coefficient, connection coefficient, covariant derivative, and curvature tensor and so forth. Decomposing the curvature tensor, we can seek out some relations between the spatial parameter and physical quantity, in the complex-sedenion curved quantum-space $\mathbb{K}_Z$ . It means that it is able to find the contribution of some other physical quantities on the curved quantum-space, from other points of view in the curved product-space, $\mathbb{K}_M$ .

In the tangent space of the complex-sedenion curved product-space $\mathbb{K}_M$ , the coordinate value is $M^\alpha$ , and the component of tangent-frame is $\textbf{\emph{E}}_{\alpha(Z,R,X)}$. These orthogonal and unequal-length components of tangent-frame compose one coordinate system. Making use of the tangent-frame component and the norm of quantum composite radius vector, it is able to define the metric of complex-sedenion curved product-space $\mathbb{K}_M$ as follows,
\begin{equation}
d S^2_{(Z,R,X)} = g_{\overline{\alpha} \beta (Z,R,X)} d \overline{M^\alpha} d M^\beta  ~ ,
\end{equation}
where the metric coefficient, $g_{\overline{\alpha} \beta (Z,R,X)} = \textbf{\emph{E}}_{\alpha (Z,R,X)}^\ast \circ \textbf{\emph{E}}_{\beta (Z,R,X)} $ , is sedenion-Hermitian. $\textbf{\emph{E}}_{\alpha(Z,R,X)}$  is the component of tangent-frame, with $\textbf{\emph{E}}_{\beta (Z,R,X)} = \partial \mathbb{M} / \partial M^\beta $ . $g_{\overline{\alpha} \beta (Z,R,X)}$ is scalar, due to the orthogonal tangent-frames. $( M^\alpha )^\ast = \overline{M^\alpha} $, and it indicates that the correlated tangent-frame component, $\textbf{\emph{E}}_{\alpha(Z,R,X)}$, is sedenion conjugate.

In the complex-sedenion curved product-space $\mathbb{K}_M$ , substituting the coordinate value $M^\alpha$ , tangent-frame component $\textbf{\emph{E}}_{\alpha(Z,R,X)}$ , and metric coefficient $g_{\overline{\alpha} \beta (Z,R,X)}$ for the coordinate value $U^\alpha$ , tangent-frame component $\textbf{\emph{E}}_{\alpha}$ , and metric coefficient $g_{\overline{\alpha} \beta (R,X)}$ in the complex-sedenion curved composite space $\mathbb{K}_U$ respectively, it is capable of inferring the connection coefficients of the curved product-space $\mathbb{K}_M$ . And there are,
\begin{eqnarray}
&& \Gamma_{\overline{\lambda} , \beta \gamma  (Z,R,X)} = (1/2) ( \partial g_{\overline{\gamma} \lambda (Z,R,X)} / \partial M^\beta
\nonumber
\\
&&
~~~~~~~~~~~~~~~~~~~
+ \partial g_{\overline{\lambda} \beta (Z,R,X)} / \partial M^\gamma - \partial g_{\overline{\gamma} \beta (Z,R,X)} /\partial M^\lambda ) ~ ,
\\
&& \Gamma_{\overline{\lambda} , \overline{\beta} \gamma  (Z,R,X)} = (1/2) ( \partial g_{\overline{\gamma} \lambda (Z,R,X)} / \partial \overline{M^\beta}
\nonumber
\\
&&
~~~~~~~~~~~~~~~~~~~
+ \partial g_{\overline{\lambda} \beta (Z,R,X)} / \partial \overline{M^\gamma} - \partial g_{\overline{\gamma} \beta (Z,R,X)} /\partial \overline{M^\lambda} ) ~ ,
\end{eqnarray}
where $\Gamma_{\overline{\lambda} , \beta \gamma (Z,R,X)}$ and $\Gamma_{\overline{\lambda} , \overline{\beta} \gamma (Z,R,X)}$ both are connection coefficients, and are scalar. $\Gamma_{\overline{\lambda} , \beta \gamma  (Z,R,X)} = g_{\overline{\lambda} \alpha (Z,R,X)} \Gamma_{\beta \gamma (Z,R,X)}^\alpha$. $\Gamma_{\overline{\lambda} , \overline{\beta} \gamma  (Z,R,X)} = g_{\overline{\lambda} \alpha (Z,R,X)} \Gamma_{\overline{\beta} \gamma (Z,R,X)}^\alpha$. $ \Gamma_{\beta \gamma (Z,R,X)}^\alpha = g^{\alpha \overline{\lambda}}_{ (Z,R,X)} \Gamma_{\overline{\lambda} , \beta \gamma  (Z,R,X)} $. $[ ( \Gamma_{\overline{\beta} \gamma (Z,R,X)}^\alpha )^* ]^T = \Gamma_{\gamma \overline{\beta} (Z,R,X)}^\alpha $. $ \Gamma_{\overline{\beta} \gamma (Z,R,X)}^\alpha =  g^{\alpha \overline{\lambda}}_{ (Z,R,X)} \Gamma_{\overline{\lambda} , \overline{\beta} \gamma  (Z,R,X)} $. $\Gamma_{\beta \gamma (Z,R,X)}^\alpha = \Gamma_{\gamma \beta (Z,R,X)}^\alpha $. $\Gamma_{\overline{\beta} \gamma (Z,R,X)}^\alpha = \Gamma_{\gamma \overline{\beta} (Z,R,X)}^\alpha $. $ g^{\alpha \overline{\lambda}}_{ (Z,R,X)}  g_{\overline{\lambda} \beta (Z,R,X)} = \delta^\alpha_\beta $.

In the curved product-space $\mathbb{K}_M$ , described with the complex-sedenions, when the physical quantity, $\mathbb{Y} = Y^\beta \textbf{\emph{E}}_{\beta (Z,R,X)}$ , is transferred from a point $M_1$ to the next point $M_2$ ,  to meet the requirement of parallel translation, it means that the differential of quantity $\mathbb{Y}$ equals to zero. This condition of parallel translation, $d \mathbb{Y} = 0$, will infer,
\begin{equation}
d Y^\beta = - \Gamma_{\alpha \gamma (Z,R,X)}^\beta Y^\alpha d M^\gamma  ~,
\end{equation}
with
\begin{equation}
\partial^2 \mathbb{M} / \partial M^\beta \partial M^\gamma = \Gamma_{\beta \gamma (Z,R,X)}^\alpha \textbf{\emph{E}}_{\alpha (Z,R,X)}  ~ .
\end{equation}

In the curved product-space, $\mathbb{K}_M$ , described with the complex-sedenions, for the first-rank contravariant tensor $Y^\beta$ of a point $M_2$ , the component of covariant derivative with respect to the coordinate $M^\gamma$ is written as,
\begin{eqnarray}
\nabla_\gamma Y^\beta = \partial ( \delta_\alpha^\beta Y^\alpha ) / \partial M^\gamma + \Gamma_{\alpha \gamma (Z,R,X)}^\beta Y^\alpha   ~ ,
\\
\nabla_{\overline{\gamma}} Y^\beta = \partial ( \delta_\alpha^\beta Y^\alpha ) / \partial \overline{M^\gamma} + \Gamma_{\alpha \overline{\gamma} (Z,R,X)}^\beta Y^\alpha   ~ ,
\end{eqnarray}
where $Y^\beta$ and $\Gamma_{\alpha \overline{\gamma} (Z,R,X)}^\beta$ both are scalar.

From the tensor property of covariant derivative, the definition of curvature tensor is,
\begin{equation}
\nabla_{\overline{\alpha}} ( \nabla_\beta Y^\gamma ) - \nabla_\beta ( \nabla_{\overline{\alpha}} Y^\gamma ) = R_{\beta \overline{\alpha} \nu ~(Z,R,X)}^{~~~~\gamma} Y^\nu + T_{\beta \overline{\alpha} (Z,R,X)}^\lambda ( \nabla_\lambda Y^\gamma )  ~   ,
\end{equation}
with
\begin{eqnarray}
&& R_{\beta \overline{\alpha} \nu ~(Z,R,X)}^{~~~~\gamma} = \partial \Gamma_{\nu \beta (Z,R,X)}^\gamma / \partial \overline{M^\alpha} - \partial \Gamma_{\nu \overline{\alpha} (Z,R,X)}^\gamma  / \partial M^\beta
\nonumber
\\
&&
~~~~~~~~~~~~~~~~~~~
+ \Gamma_{\lambda \overline{\alpha} (Z,R,X)}^\gamma \Gamma_{\nu \beta (Z,R,X)}^\lambda - \Gamma_{\lambda \beta (Z,R,X)}^\gamma \Gamma_{\nu \overline{\alpha} (Z,R,X)}^\lambda   ~ ,
\\
&& T_{\beta \overline{\alpha} (Z,R,X)}^\lambda = \Gamma_{\overline{\alpha} \beta (Z,R,X)}^\lambda - \Gamma_{\beta \overline{\alpha} (Z,R,X)}^\lambda  ~,
\end{eqnarray}
where $R_{\beta \overline{\alpha} \nu ~(Z,R,X)}^{~~~~\gamma}$ and $T_{\beta \overline{\alpha} (Z,R,X)}^\lambda$ both are scalar. $R_{\beta \overline{\alpha} \nu ~(Z,R,X)}^{~~~~\gamma}$ is the curvature tensor, while $T_{\beta \overline{\alpha} (Z,R,X)}^\lambda$ is the torsion tensor. In the paper, we only discuss the case, $T_{\beta \overline{\alpha} (Z,R,X)}^\lambda = 0$, that is, $\Gamma_{\overline{\alpha} \beta (Z,R,X)}^\lambda = \Gamma_{\beta \overline{\alpha} (Z,R,X)}^\lambda $ .

Making use of the properties associated with the metric coefficient, connection coefficient, and curvature tensor and so forth, it is found that the spatial parameter and physical quantity both make a contribution towards these arguments, in the complex-sedenion curved product-space, $\mathbb{K}_M$ . When the coupling of spatial parameters and physical quantities of these arguments could be decoupled under some circumstances, it may be able to find out some direct interrelations between the spatial parameter and physical quantity. But it may be more complicated than the situation of the curved composite-space, $\mathbb{K}_U$ .

\subsection{\label{sec:level1}Decoupling}

From the preceding equations, in the complex-sedenion curved product-space $\mathbb{K}_M$ , it is found that the spatial parameter, $\mathbb{Z}_U \circ \mathbb{R}$ , and physical quantity, $\mathbb{Z}_U \circ \mathbb{X}$ , can be superposed linearly also, which are similar to two arguments, $\mathbb{R}$ and $\mathbb{X}$ , respectively in the Section 4. As the derivative of these two arguments, the interrelation between the spatial parameter with physical quantity will be comparatively complicated in the following context, including certain interconnections among the metric coefficient, connection coefficient, and curvature tensor. The coupling term between the spatial parameter and physical quantity is always too intricate to decouple in general. Only in some particular cases, it may be decoupled approximately, attempting to grasp at the relations of spatial parameter and physical quantity.

From the metric equation, Eq.(77), it is found that the contributions of two arguments, $\mathbb{Z}_U \circ \mathbb{R}$ and $\mathbb{Z}_U \circ \mathbb{X}$ , to the metric coefficient are interrelated closely. Each of partial derivatives, $ \partial (\mathbb{Z}_U \circ \mathbb{R}) / \partial M^\alpha $ and $ \partial (\mathbb{Z}_U \circ \mathbb{X}) / \partial M^\alpha $ , will exert an influence on the metric coefficient. In a few comparatively simple circumstances, the metric tensor can be separated into,
\begin{equation}
g_{\overline{\alpha} \beta (Z,R,X)} = g_{\overline{\alpha} \beta (Z,R)} + g_{\overline{\alpha} \beta (Z,X)} + c.t.( g_{\overline{\alpha} \beta} , \mathbb{Z}_U , \mathbb{R} , \mathbb{X} )   ~   ,
\end{equation}
where the partial derivative, $ \partial (\mathbb{Z}_U \circ \mathbb{X}) / \partial M^\alpha $ , possesses the dimension of quantum-field potential. In case the contribution of $\mathbb{Z}_U \circ \mathbb{X}$ could be neglected, $g_{\overline{\alpha} \beta (Z,R,X)}$ will be reduced into $g_{\overline{\alpha} \beta (Z,R)}$ , which contains only the contribution of $ \partial (\mathbb{Z}_U \circ \mathbb{R}) / \partial M^\alpha $. Meanwhile, if the contribution of $\mathbb{Z}_U \circ \mathbb{R}$ could be neglected, $g_{\overline{\alpha} \beta (Z,R,X)}$ will be degenerated into $g_{\overline{\alpha} \beta (Z,X)}$ , which covers merely the contribution of $ \partial (\mathbb{Z}_U \circ \mathbb{X}) / \partial M^\alpha $.

In the connection coefficient, Eq.(78), it is found that the contributions of two arguments, $g_{\overline{\alpha} \beta (Z,R)}$ and $g_{\overline{\alpha} \beta (Z,X)}$ , to the connection coefficient are mingled together. Each of partial derivatives, $ \partial g_{\overline{\alpha} \beta (Z,R)} / \partial M^\alpha $ and $ \partial g_{\overline{\alpha} \beta (Z,X)} / \partial M^\alpha $, has an influence on the connection coefficient. In several comparatively simple cases, the connection coefficient can be divided into,
\begin{equation}
\Gamma_{\alpha \beta (Z,R,X)}^\gamma = \Gamma_{\alpha \beta (Z,R)}^\gamma + \Gamma_{\alpha \beta (Z,X)}^\gamma + c.t.( \Gamma_{\alpha \beta}^\gamma , \mathbb{Z}_U , \mathbb{R} , \mathbb{X} )   ~   ,
\end{equation}
where the partial derivative, $(k_{rx}^{~~-1}) \partial g_{\overline{\alpha} \beta (Z,X)} / \partial M^\alpha $ , possesses the dimension of quantum-field strength. If the contribution of $g_{\overline{\alpha} \beta (Z,X)}$ could be neglected, $\Gamma_{\alpha \beta (Z,R,X)}^\gamma$ will be reduced into $\Gamma_{\alpha \beta (Z,R)}^\gamma$ , which contains only the contribution of $ \partial g_{\overline{\alpha} \beta (Z,R)} / \partial M^\alpha $. In case the contribution of $g_{\overline{\alpha} \beta (Z,R)}$ could be neglected, $\Gamma_{\alpha \beta (Z,R,X)}^\gamma$ will be simplified into $\Gamma_{\alpha \beta (Z,X)}^\gamma$, which contains only the contribution of $ \partial g_{\overline{\alpha} \beta (Z,X)} / \partial M^\alpha $ . The decomposition process of the connection coefficient, Eq.(79), is similar to that of Eq.(78), under some comparatively simple circumstances.

The analysis of the curvature tensor, Eq.(85), shows that the contributions, coming from the spatial parameter and physical quantity, to the curvature tensor are mixed complicatedly. And each of four partial derivatives, $ \partial \Gamma_{\alpha \beta (Z,R)}^\gamma / \partial M^\nu $, $ \partial \Gamma_{\alpha \beta (Z,X)}^\gamma / \partial M^\nu $, $ \partial \Gamma_{\overline{\alpha} \beta (Z,R)}^\gamma / \partial M^\nu $, and $ \partial \Gamma_{\overline{\alpha} \beta (Z,X)}^\gamma / \partial M^\nu $ ,
will directly impact the curvature tensor. Under some comparatively simple cases, the curvature tensor can be separated into,
\begin{equation}
R_{\beta \overline{\alpha} \nu ~(Z,R,X)}^{~~~~\gamma} = R_{\beta \overline{\alpha} \nu ~(Z,R)}^{~~~~\gamma} + R_{\beta \overline{\alpha} \nu ~(Z,X)}^{~~~~\gamma} + c.t.( R_{\beta \overline{\alpha} \nu }^{~~~~\gamma} , \mathbb{Z}_U , \mathbb{R} , \mathbb{X} )   ~   ,
\end{equation}
where the partial derivatives, $(k_{rx}^{~~-1}) \partial \Gamma_{\alpha \beta (Z,X)}^\gamma / \partial M^\nu $ and $(k_{rx}^{~~-1}) \partial \Gamma_{\overline{\alpha} \beta (Z,X)}^\gamma / \partial M^\nu $, will possess the dimension of quantum-field source. When the contributions of two terms, $\Gamma_{\alpha \beta (Z,X)}^\gamma$ and $\Gamma_{\overline{\alpha} \beta (Z,X)}^\gamma$ , could be neglected, $R_{\beta \overline{\alpha} \nu ~(Z,R,X)}^{~~~~\gamma}$ will be reduced into $R_{\beta \overline{\alpha} \nu ~(Z,R)}^{~~~~\gamma}$, which contains the contributions of $ \partial \Gamma_{\alpha \beta (Z,R)}^\gamma / \partial M^\nu $ and $ \partial \Gamma_{\overline{\alpha} \beta (Z,R)}^\gamma / \partial M^\nu $ . If the contribution of two terms, $\Gamma_{\alpha \beta (Z,R)}^\gamma$ and $\Gamma_{\overline{\alpha} \beta (Z,R)}^\gamma$ , could be neglected, $R_{\beta \overline{\alpha} \nu ~(Z,R,X)}^{~~~~\gamma}$ will be reduced into $R_{\beta \overline{\alpha} \nu ~(Z,X)}^{~~~~\gamma}$, which contains the contribution of $ \partial \Gamma_{\alpha \beta (Z,X)}^\gamma / \partial M^\nu $ and $ \partial \Gamma_{\overline{\alpha} \beta (Z,X)}^\gamma / \partial M^\nu $ .

In the complex-sedenion curved product-space $\mathbb{K}_M$ , in virtue of the precise experimental measurements, it is able to verify whether the product-space is distorted. In other words, there may be the flat product-space under some circumstances, that is, $R_{\beta \overline{\alpha} \nu ~(Z,R,X)}^{~~~~\gamma} = 0$. Even if neither of ingredients, $R_{\beta \overline{\alpha} \nu ~(Z,R)}^{~~~~\gamma}$ and $R_{\beta \overline{\alpha} \nu ~(Z,X)}^{~~~~\gamma}$ , is equal to zero, the curvature tensor $R_{\beta \overline{\alpha} \nu ~(Z,R,X)}^{~~~~\gamma}$ may be zero. Undoubtedly there must be a few relations between two ingredients, $R_{\beta \overline{\alpha} \nu ~(Z,R)}^{~~~~\gamma}$ and $R_{\beta \overline{\alpha} \nu ~(Z,X)}^{~~~~\gamma}$ .

When $R_{\beta \overline{\alpha} \nu ~(Z,R,X)}^{~~~~\gamma} = 0$ , in the complex-sedenion flat product-space $\mathbb{K}_M$, it is able to deduce a simple equation among a few ingredients of curvature tensor as follows,
\begin{equation}
0 = R_{\beta \overline{\alpha} \nu ~(Z,R)}^{~~~~\gamma} + R_{\beta \overline{\alpha} \nu ~(Z,X)}^{~~~~\gamma} + c.t.( R_{\beta \overline{\alpha} \nu }^{~~~~\gamma} , \mathbb{Z}_U , \mathbb{R} , \mathbb{X} )   ~   ,
\end{equation}
further, if there is, $c.t.( R_{\beta \overline{\alpha} \nu }^{~~~~\gamma} , \mathbb{Z}_U , \mathbb{R} , \mathbb{X} ) = 0$ , the above can be reduced into,
\begin{equation}
R_{\beta \overline{\alpha} \nu ~(Z,R)}^{~~~~\gamma} = - R_{\beta \overline{\alpha} \nu ~(Z,X)}^{~~~~\gamma}   ~   ,
\end{equation}
where the left side is the contribution of spatial parameter, $ \mathbb{Z}_U \circ \mathbb{R} $ , and the right side is the contribution of physical quantity, $ \mathbb{Z}_U \circ \mathbb{X} $ . The term, $(k_{rx}^{~~-1}) R_{\beta \overline{\alpha} \nu ~(Z,X)}^{~~~~\gamma}$, possesses the dimension of quantum-field source, and are much more complicated than that of field source.

The above is also accordant with the Cartesian academic thought of `the space is the extension of substance'. In other words, the existence of quantum-field dominates the bending of space also, in the complex-sedenion curved product-space $\mathbb{K}_M$. Apparently, Eq.(91) expands this academic thought from a new point of view. It means that some other physical quantities, besides the physical quantities in the Section 5, are still able to impact the bending degree of curved space.

In the complex-sedenion curved product-space $\mathbb{K}_M$ , when the contributions of some physical quantities, $g_{\overline{\alpha} \beta (Z,X)}$, $\Gamma_{\alpha \beta (Z,X)}^\gamma$ , $\Gamma_{\overline{\alpha} \beta (Z,X)}^\gamma$ , and $R_{\beta \overline{\alpha} \nu ~(Z,X)}^{~~~~\gamma}$ , are comparatively huge, they will exert an evident influence on the spatial parameters, $g_{\overline{\alpha} \beta (Z,R)}$ , $\Gamma_{\alpha \beta (Z,R)}^\gamma$ , $\Gamma_{\overline{\alpha} \beta (Z,R)}^\gamma$ , and $R_{\beta \overline{\alpha} \nu ~(Z,R)}^{~~~~\gamma}$ , of the curved space. Subsequently, the spatial parameters will make a contribution to several operators in the curved quantum-space $\mathbb{K}_Z$ , including the divergence, gradient, and curl. Finally these operators are able to impact directly some field equations for the four kinds of quantum-fields in the curved quantum-space $\mathbb{K}_Z$ . Obviously, most of physical quantities in the product-space $\mathbb{K}_M$ are different from that in the fundamental-space $\mathbb{K}$ , quantum-space $\mathbb{K}_Z$ , or composite-space $\mathbb{K}_U$ .

Making a comparison and analysis of the preceding studies, it is found that there are a few discrepancies between the curved product-space $\mathbb{K}_M$ and composite-space $\mathbb{K}_U$. (a) Tangent-frame. In the curved composite-space $\mathbb{K}_U$ , the tangent-frame is relevant to the composite radius vector $\mathbb{U}$ . In the curved product-space $\mathbb{K}_M$, the tangent-frame is associated with the quantum composite radius vector $\mathbb{M}$ . (b) Physical quantity. In the curved composite-space $\mathbb{K}_U$ , the physical quantity is associated with the integrating function, $\mathbb{X}$ , of field potential, impacting the spatial parameter of the curved composite-space $\mathbb{K}_U$ . In the curved product-space $\mathbb{K}_M$ , the physical quantity is related to the component, $ \mathbb{Z}_U \circ \mathbb{X} $, affecting the spatial parameter of the curved product-space $\mathbb{K}_M$ . (c) Spatial parameter. In the curved composite-space $\mathbb{K}_U$ , the spatial parameters are correlative with the radius vector $\mathbb{R}$ , including the arguments, $g_{\overline{\alpha} \beta (R)}$ , $\Gamma_{\alpha \beta (R)}^\gamma$ , $\Gamma_{\overline{\alpha} \beta (R)}^\gamma$, and $R_{\beta \overline{\alpha} \nu ~(R)}^{~~~~\gamma}$ . In the curved product-space $\mathbb{K}_M$ , the spatial parameters are relevant to the component, $ \mathbb{Z}_U \circ \mathbb{R} $, including the arguments, $g_{\overline{\alpha} \beta (Z,R)}$, $\Gamma_{\alpha \beta (Z,R)}^\gamma$ , $\Gamma_{\overline{\alpha} \beta (Z,R)}^\gamma$ , and $R_{\beta \overline{\alpha} \nu ~(Z,R)}^{~~~~\gamma}$ .

To be brief, in virtue of the complex-sedenion curved product-space $\mathbb{K}_M$, it is able to catch sight of several relations between the spatial parameters and physical quantities. Next the spatial parameters will make a contribution towards some operators of the curved quantum-space $\mathbb{K}_Z$ . Further, these operators may directly exert an influence on several field equations for the four types of quantum-fields, in the curved quantum-space $\mathbb{K}_Z$ .

\begin{table}[ht]
\tbl{The classical-fields and quantum-fields in the complex-sedenion space.}
{\begin{tabular}{@{}llll@{}}
\toprule
interaction              &   classical/quantum field                   &   fundamental/adjoint field         \\
\colrule
gravitational            &   classical-field                           &   fundamental classical-field       \\
                         &                                             &   adjoint classical-field           \\
                         &   quantum-field                             &   fundamental quantum-field         \\
                         &                                             &   adjoint quantum-field             \\
electromagnetic          &   classical-field                           &   fundamental classical-field       \\
                         &                                             &   adjoint classical-field           \\
                         &   quantum-field                             &   fundamental quantum-field         \\
                         &                                             &   adjoint quantum-field             \\
W-nuclear                &   classical-field                           &   fundamental classical-field       \\
                         &                                             &   adjoint classical-field           \\
                         &   quantum-field                             &   fundamental quantum-field         \\
                         &                                             &   adjoint quantum-field             \\
strong-nuclear           &   classical-field                           &   fundamental classical-field       \\
                         &                                             &   adjoint classical-field           \\
                         &   quantum-field                             &   fundamental quantum-field         \\
                         &                                             &   adjoint quantum-field             \\
\botrule
\end{tabular}}
\end{table}

\section{\label{sec:level1}Approximations}

By means of the similar deduction methods for the approximate theory in the complex-octonion curved composite-space $\mathbb{O}_U$ (see Ref.[41]), it is able to achieve respectively two approximate theories in the complex-sedenion curved composite-space $\mathbb{K}_U$ and product-space $\mathbb{K}_M$ . The approximate theory in the curved composite-space $\mathbb{K}_U$ can be applied to explore several physical phenomena, which are associated with the classical mechanics on the macroscopic scale, in the curved fundamental-space $\mathbb{K}$ . Meanwhile the approximate theory in the curved product-space $\mathbb{K}_M$ can be utilized to study some physical phenomena, which are related with the quantum mechanics on the microscopic scale, in the curved quantum-space $\mathbb{K}_Z$ (Table 5).

\subsection{\label{sec:level1}Composite-space}

In the complex-sedenion curved composite-space $\mathbb{K}_U$ based on the composite radius vector, an approximate theory will be degenerated from the preceding discussions associated with the curved composite-space $\mathbb{K}_U$ , under certain approximate circumstances. In the classical mechanics on the macroscopic scale, some inferences of this approximate theory correlate with that of the GR, especially in the case that the energy-momentum tensor is equal to zero.

For the sake of achieving the approximate theory, it is necessary to meet the requirement of two approximate conditions. (a) The bending degree of composite-space is extremely tiny, and even it is flat, allowing us to separate relevant arguments of curved composite-space $\mathbb{K}_U$ into two components, spatial parameter and physical quantity. (b) The coupling term between the spatial parameter and physical quantity equals to zero approximately, that is, they could be decoupled.

In the curved composite-space described with the complex-sedenions, the tangent-frame, $\textbf{\emph{E}}_\alpha = \partial ( \mathbb{R} + k_{rx} \mathbb{X} ) / \partial U^\alpha $, can be separated into two components, that is, $\textbf{\emph{E}}_\alpha = \textbf{\emph{E}}_{\alpha(R)} + \textbf{\emph{E}}_{\alpha(X)}$ . Herein the component, $\textbf{\emph{E}}_{\alpha(R)} = \partial \mathbb{R} / \partial U^\alpha$ , is related with the spatial parameter, while the component,
$\textbf{\emph{E}}_{\alpha(X)} = \partial ( k_{rx} \mathbb{X}) / \partial U^\alpha$, is associated with the physical quantity.

Therefore, the metric coefficient,
\begin{eqnarray}
g_{\overline{\alpha} \beta (R,X)} = ( \textbf{\emph{E}}_{\alpha (R)} + \textbf{\emph{E}}_{\alpha (X)} )^\ast \circ ( \textbf{\emph{E}}_{\beta (R)} + \textbf{\emph{E}}_{\beta (X)} ) ~ ,
\end{eqnarray}
in Eq.(61) can be expanded in a Taylor series,
\begin{eqnarray}
g_{\overline{\alpha} \beta (R,X)} \approx 1 + A_{\overline{\alpha} \beta}  ~ ,
\end{eqnarray}
where $A_{\overline{\alpha} \beta} = \textbf{\emph{E}}_{\alpha (R)}^\ast \circ \textbf{\emph{E}}_{\beta (X)} + \textbf{\emph{E}}_{\alpha (X)}^\ast \circ \textbf{\emph{E}}_{\beta (R)} $ . The minor term, $\textbf{\emph{E}}_{\alpha (X)}^\ast \circ \textbf{\emph{E}}_{\beta (X)}$, can be neglected. The norm of the component of tangent-frame, $\textbf{\emph{E}}_{\alpha (R)}$, associated with the spatial parameter is close to 1. That is, $\parallel \textbf{\emph{E}}_{\alpha (R)} \parallel \approx 1$ , and $\parallel \textbf{\emph{E}}_{\beta (R)} \parallel \approx 1$ . $ \textbf{\emph{E}}_{\alpha (R)}^\ast \circ  \textbf{\emph{E}}_{\beta (R)} \approx 1 $ . The term $( A_{\overline{\alpha} \beta} / k_{rx} )$ possesses the dimension of field potential, but is not the field potential. And it is merely considered as a term to be equivalent to the field potential. In other words, there are certain complicated interrelations between the term $( A_{\overline{\alpha} \beta} / k_{rx} )$ with the field potential.

Starting from the above metric coefficient, it is able to infer the connection coefficient, curvature tensor, and geodetic line and so forth of the curved composite-space $\mathbb{K}_U$ , under the approximate conditions. Further it is capable of derived several inferences from the curved composite-space $\mathbb{K}_U$ , which are consistent with that in the GR, especially in some vacuum solutions when the energy-momentum tensor is zero. However this approximate method, in the curved composite-space $\mathbb{K}_U$ , can merely explain a small quantity of physical phenomena. In contrast, if we contemplate the contributions of the arguments (such as, metric coefficient, connection coefficient, and curvature tensor) on the divergence, gradient, and curl in the curved space $\mathbb{K}$ , under certain conditions, it will be able to explain plenty of physical phenomena in the curved space $\mathbb{K}$ , from the field equations of four classical-fields in the curved space $\mathbb{K}$ .

In the GR, the metric coefficient is written as, $g_{jk} = 1 + h_{jk}$ . The term $h_{jk}$ is regarded as the gravitational potential in the Newtonian mechanics. In the paper, the minor term $A_{\overline{\alpha} \beta}$ is equivalent to the minor term $h_{jk}$ . In other words, the viewpoint of $A_{\overline{\alpha} \beta}$ in the composite-space $\mathbb{K}_U$ happens to coincide with that of $h_{jk}$ in the GR, and even both of them give mutual support to each other to a certain extent. Meanwhile, it is found that the physical quantity $R_{\beta \overline{\alpha} \nu ~(X)}^{~~~~\gamma}$ in the paper corresponds to the energy-momentum tensor in the GR, although there are a few discrepancies between the paper and GR. As a result, some inferences in the complex-sedenion composite-space $\mathbb{K}_U$ can be reduced into that in the complex-octonion composite-space $\mathbb{O}_U$ and the GR.

By all appearances, the approximate method in the curved composite-space, $\mathbb{K}_U$, can be extended into the curved product-space, $\mathbb{K}_M$ .

\subsection{\label{sec:level1}Product-space}

In the curved product-space $\mathbb{K}_M$ based on the quantum composite radius vector, an approximate theory will be reduced from the preceding discussions associated with the curved product-space $\mathbb{K}_M$ , under certain approximate circumstances. Some inferences of this approximate theory are similar to that of GR.

In order to acquire the approximate theory of the curved product-space, it is necessary to meet the demand of two approximate conditions. (a) The bending degree of product-space is quite tiny, and even it is flat, separating the arguments of curved product-space into two ingredients, spatial parameter and physical quantity. (b) The coupling term between the spatial parameter and physical quantity is zero approximately, that is, they can be decoupled.

In the complex-sedenion curved product-space $\mathbb{K}_M$ , the tangent-frame, $\textbf{\emph{E}}_{\alpha(Z,R,X)} = \partial \{ \mathbb{U}_Z \circ ( \mathbb{R} + k_{rx} \mathbb{X} )\} / \partial M^\alpha$ , can be separated into two components,
\begin{eqnarray}
\textbf{\emph{E}}_{\alpha(Z,R,X)} = \textbf{\emph{E}}_{\alpha(Z,R)} + \textbf{\emph{E}}_{\alpha(Z,X)}  ~ ,
\end{eqnarray}
where $\textbf{\emph{E}}_{\alpha(Z,R)} = \partial ( \mathbb{U}_Z \circ \mathbb{R} ) / \partial M^\alpha$ , which is related with the spatial parameter. $\textbf{\emph{E}}_{\alpha(Z,X)} = \partial ( k_{rx} \mathbb{U}_Z \circ \mathbb{X} ) / \partial M^\alpha$ , which is associated with the physical quantity.

Consequently, the metric coefficient,
\begin{eqnarray}
g_{\overline{\alpha} \beta (Z,R,X)} = ( \textbf{\emph{E}}_{\alpha(Z,R)} + \textbf{\emph{E}}_{\alpha(Z,X)} )^\ast \circ ( \textbf{\emph{E}}_{\alpha(Z,R)} + \textbf{\emph{E}}_{\alpha(Z,X)} )     ~ ,
\end{eqnarray}
in Eq.(77) can be expanded in a Taylor series,
\begin{eqnarray}
g_{\overline{\alpha} \beta (Z,R,X)} \approx 1 + A_{\overline{\alpha} \beta (Z)}  ~ ,
\end{eqnarray}
where $A_{\overline{\alpha} \beta (Z)} = \textbf{\emph{E}}_{\alpha (Z,R)}^\ast \circ \textbf{\emph{E}}_{\beta (Z,X)} + \textbf{\emph{E}}_{\alpha (Z,X)}^\ast \circ \textbf{\emph{E}}_{\beta (Z,R)} $ . The minor term, $\textbf{\emph{E}}_{\alpha (Z,X)}^\ast \circ \textbf{\emph{E}}_{\beta (Z,X)}$, can be neglected. The norm of the component of tangent-frame, $\textbf{\emph{E}}_{\alpha (Z,R)}$, associated with the spatial parameter is close to 1. That is, $\parallel \textbf{\emph{E}}_{\alpha (Z,R)} \parallel \approx 1$ , and $\parallel \textbf{\emph{E}}_{\beta (Z,R)} \parallel \approx 1$. $ \textbf{\emph{E}}_{\alpha (Z,R)}^\ast \circ  \textbf{\emph{E}}_{\beta (Z,R)} \approx 1 $ . Obviously, Eq.(96) in the curved product-space $\mathbb{K}_M$ is similar to Eq.(93) in the curved composite-space $\mathbb{K}_U$ .

According to the definition of quantum-field potential, the term $( A_{\overline{\alpha} \beta (Z)} / k_{rx} )$ is seized of the dimension of quantum-field potential. However this term is not the quantum-field potential, and is merely regarded as one term to be equivalent to the quantum-field potential. In other words, there are several complicated interrelations between the term $( A_{\overline{\alpha} \beta (Z)} / k_{rx} )$ with the quantum-field potential.

From the above metric coefficient, it is able to infer the connection coefficient, curvature tensor, and geodetic line and so forth in the curved product-space $\mathbb{K}_M$, under the approximate conditions. Further it is capable of inferring a few conclusions from the curved product-space $\mathbb{K}_M$ , which are similar to that in the GR. However this approximate method can merely explain a small quantity of physical phenomena in the curved product-space $\mathbb{K}_M$ . In contrast to the above, if we consider the influences of the arguments (such as, metric coefficient, connection coefficient, and curvature tensor) on the divergence, gradient, and curl in the curved quantum-space $\mathbb{K}_Z$ , under the approximate conditions, it is capable of explaining plenty of physical phenomena in the curved space $\mathbb{K}_Z$ , from some field equations of four quantum-fields in the curved space $\mathbb{K}_Z$ .

It is similar to the case of the complex-sedenion curved space $\mathbb{K}$ , the complex-sedenion curved quantum-space $\mathbb{K}_Z$ can be degenerated into the complex-quaternion curved quantum-space, and even the four space-time. In the curved quantum-space $\mathbb{K}_Z$, the metric coefficient can be reduced into, $g_{jk (Z,R,X)} = 1 + h_{jk (Z)}$ . And the term $h_{jk (Z)}$ is regarded as the quantum-field potential in the complex-quaternion quantum-space. This inference is similar to that derived from the GR, under the approximate circumstances. In other words, the viewpoint of $A_{\overline{\alpha} \beta (Z)}$ in the quantum-space $\mathbb{K}_Z$ coincides with that of $h_{jk}$ in the GR. And the curved quantum-space $\mathbb{K}_Z$ inherits and extends the Cartesian academic thought of `the space is the extension of substance' apparently.

\section{\label{sec:level1}Conclusions}

One complex-quaternion space may describe a type of field, especially the field equations of fundamental interaction. Generally, four perpendicular complex-quaternion spaces can combine together to become one complex-sedenion space, exploring the physical properties of four fundamental interactions. On the one hand, in the paper, the four fields comprise the gravitational field, electromagnetic field, and strong-nuclear field, except for the weak-nuclear field. The complex-sedenion electromagnetic field can be degenerated into the electroweak field, while the weak-nuclear field can be regarded as the adjoint-field of the complex-sedenion electromagnetic field. So the weak-nuclear field cannot be considered as an independent field any more. On the other hand, according to the multiplicative closure of sedenions, there must be four types of fields in the complex-sedenion space simultaneously. That is, the complex-sedenion four fields consist of the gravitational field, electromagnetic field, strong-nuclear field, and a new unknown field (or W-nuclear field).

One approximate theory can be reduced from the above curved composite-space $\mathbb{K}_U$, in certain approximate circumstances. In the curved composite-space $\mathbb{K}_U$, some inferences, derived from the approximate theory, correlate with that from GR. Apparently, the curved composite-space $\mathbb{K}_U$ and GR both inherit the Cartesian academic thought of `the space is the extension of substance'. Next, in the curved composite-space $\mathbb{K}_U$ , it is able to establish some interrelations between the spatial parameter and physical quantity. Some physical quantities make a contribution to the spatial parameter of curved space. The spatial parameter has an influence on a few operators (such as, divergence, gradient, and curl) in the curved space $\mathbb{K}$ , impacting the field equations of four classical-fields. Further, in the curved space $\mathbb{K}$ , it is able to deduce the physical quantities for the four classical-fields, which are associated with the classical mechanics on the macroscopic scale, including the complex-sedenion field potential, field strength, field source, linear momentum, angular momentum, torque, and force and so forth.

In virtue of the auxiliary quantity, from the complex-sedenion fundamental-space $\mathbb{K}$ , it is able to deduce the complex-sedenion quantum-space $\mathbb{K}_Z$ and product-space $\mathbb{K}_M$ , achieving another approximate theory. Firstly, under certain circumstances, one approximate theory can be reduced from the complex-sedenion curved product-space $\mathbb{K}_M$. In the curved product-space $\mathbb{K}_M$ , some inferences of the approximate theory are similar to that of GR. Apparently, this conclusion expands the Cartesian academic thought of `the space is the extension of substance'. Secondly, in the curved product-space $\mathbb{K}_M$ , it is able to establish some relations among the spatial parameters and physical quantities. Some physical quantities make a contribution to the spatial parameter of the curved quantum-space $\mathbb{K}_Z$. The spatial parameter exerts an influence on several operators (such as, divergence, gradient, and curl) in the curved space $\mathbb{K}_Z$ , impacting the field equations of four quantum-fields. Thirdly, in the curved space $\mathbb{K}_Z$, making use of the complex-sedenion exponential form and wavefunction, it is capable of deducing the physical quantities for the four quantum-fields, which are associated with the quantum mechanics on the microscopic scale, including the Dirac wave equation and Yang-Mills equation.

It should be noted that the paper discussed merely some simple cases about not only the relations among spatial parameters and physical quantities, but also the influences of the curved spaces on the physical quantities of four fundamental interactions, in the complex-sedenion curved composite-space $\mathbb{K}_U$ and product-space $\mathbb{K}_M$. However it clearly states that the physical quantities dominate the spatial parameters of the curved composite-space $\mathbb{K}_U$ or product-space $\mathbb{K}_M$. Later, these spatial parameters make a contribution towards the divergence, gradient, and curl and others, exerting an influence on the field equations in the curved fundamental-space $\mathbb{K}$ or quantum-space $\mathbb{K}_Z$. Under certain circumstances, the paper can deduce several inferences in accordance with that derived from the GR. In the following study, it is going to apply the complex-sedenion curved composite-space, $\mathbb{K}_U$ , to explore the influence of the physical quantities of four classical-fields, in the classical mechanics, on the equilibrium equations and so forth. And we plan to utilize the complex-sedenion curved product-space, $\mathbb{K}_M$ , to research the influence of the physical quantities of four quantum-fields, in the quantum mechanics, on the topological structures and so on. And it may intend to analyze the discrepancies among some tangent-spaces of different complex-sedenion curved spaces, and their influences on some operators and field equations associated with the four fundamental interactions.

\section*{Acknowledgements}

The author is indebted to the anonymous referees for their valuable comments on the previous manuscripts. This project was supported partially by the National Natural Science Foundation of China under grant number 60677039.

\appendix

\section{Curvature tensor}

By means of the similar methods in the complex-octonion curved space, it is able to deduce the metric coefficient, connection coefficient, and curvature tensor and so forth, in the complex-sedenion curved space $\mathbb{K}$ . In the curved space described with the complex-sedenions, there is a relation between the metric coefficient and connection coefficient. The definition of metric coefficient must meet the requirement of this relation, inferring the connection coefficient and curvature tensor from the metric coefficient.

From the condition of parallel translation, $d \mathbb{Y} = 0$ , there is,
\begin{eqnarray}
 ( d Y^\beta ) \textbf{\emph{e}}_\beta + Y^\beta ( d \textbf{\emph{e}}_\beta ) = 0  ~ ,
\end{eqnarray}
with
\begin{equation}
\partial^2 \mathbb{R} / \partial u^\beta \partial u^\gamma = \Gamma_{\beta \gamma }^\alpha \textbf{\emph{e}}_\alpha    ~   ,
\end{equation}
where $\textbf{\emph{e}}_\alpha = \partial \mathbb{R} / \partial u^\alpha$ . $d \textbf{\emph{e}}_\beta = \{ \partial^2 \mathbb{R} / ( \partial u^\beta \partial u^\gamma ) \} d u^\gamma$ . $\Gamma_{\beta \gamma }^\alpha$ is the connection coefficient. $\alpha, \beta, \gamma, \lambda, \nu = 0, 1, 2, 3, 4, 5, 6, 7, 8, 9, 10, 11, 12, 13, 14, 15$ . The paper continues to use the contraction of tensor.

From the above and, $d ( Y^\beta Y_\beta ) = 0$, it is able to obtain,
\begin{eqnarray}
&& d Y^\beta = - \Gamma_{\alpha \gamma}^\beta Y^\alpha d u^\gamma ~ ,
\\
&& d Y_\beta = \Gamma_{\beta \gamma }^\alpha Y_\alpha d u^\gamma ~ .
\end{eqnarray}

Multiplying the component $\textbf{\emph{e}}_\lambda^\ast$ from the left of Eq.(A.2) produces,
\begin{equation}
( \partial \mathbb{R}^\ast / \partial u^\lambda ) \circ ( \partial^2 \mathbb{R} / \partial u^\beta \partial u^\gamma ) = g_{\overline{\lambda} \alpha} \Gamma_{\beta \gamma }^\alpha   ~   ,
\end{equation}
while multiplying the component $\textbf{\emph{e}}_\lambda$ from the right of the conjugate of Eq.(A.2) yields,
\begin{equation}
( \partial^2 \mathbb{R}^\ast / \partial u^\beta \partial u^\gamma ) \circ ( \partial \mathbb{R} / \partial u^\lambda ) = \overline{\Gamma_{\beta \gamma }^\alpha}  g_{\overline{\alpha} \lambda} ~   ,
\end{equation}
where $\partial^2 \mathbb{R}^\ast / \partial u^\beta \partial u^\gamma = \overline{\Gamma_{\beta \gamma }^\alpha} \textbf{\emph{e}}_\alpha^\ast $ . $\overline{\Gamma_{\beta \gamma }^\alpha}$ is a connection coefficient. $\textbf{\emph{e}}_\alpha^\ast = \partial \mathbb{R}^\ast / \partial u^\alpha$.

In the orthogonal and unequal-length tangent-frame, from the last two equations, it is found that the partial derivative of the metric tensor, $g_{\overline{\lambda} \gamma} = \textbf{\emph{e}}_\lambda^\ast \circ \textbf{\emph{e}}_\gamma$, with respect to the coordinate value, $u^\beta$ , is able to generate,
\begin{equation}
g_{\overline{\lambda} \alpha} \Gamma_{\beta \gamma }^\alpha + \overline{\Gamma_{\beta \lambda }^\alpha}  g_{\overline{\alpha} \gamma} = \partial g_{\overline{\lambda} \gamma} / \partial u^\beta  ~ ,
\end{equation}
similarly there are,
\begin{eqnarray}
g_{\overline{\beta} \alpha} \Gamma_{\gamma \lambda }^\alpha + \overline{\Gamma_{\gamma \beta }^\alpha}  g_{\overline{\alpha} \lambda} = \partial g_{\overline{\beta} \lambda} / \partial u^\gamma  ~ ,
\\
g_{\overline{\gamma} \alpha} \Gamma_{\lambda \beta }^\alpha + \overline{\Gamma_{\lambda \gamma }^\alpha}  g_{\overline{\alpha} \beta} = \partial g_{\overline{\gamma} \beta} / \partial u^\lambda  ~ ,
\end{eqnarray}
further, from the last three equations, there is,
\begin{equation}
\Gamma_{ \overline{\lambda} , \beta \gamma } = (1/2) ( \partial g_{ \overline{\gamma} \lambda } / \partial u^\beta + \partial g_{ \overline{\lambda} \beta } / \partial u^\gamma - \partial g_{ \overline{\gamma} \beta } / \partial u^\lambda )    ~  ,
\end{equation}
where $ [ ( g_{\overline{\lambda} \alpha} \Gamma_{\beta \gamma }^\alpha )^\ast ]^T = \overline{\Gamma_{\gamma \beta }^\alpha}  g_{\overline{\alpha} \lambda} $ ,
and $ [ ( \overline{\Gamma_{\gamma \beta }^\alpha}  g_{\overline{\alpha} \lambda} )^\ast ]^T = g_{\overline{\lambda} \alpha} \Gamma_{\beta \gamma }^\alpha $ .

On the other hand, from the condition of parallel translation, $d \mathbb{Y} = 0$ , there is,
\begin{equation}
\partial^2 \mathbb{R} / \partial \overline{u^\beta} \partial u^\gamma = \Gamma_{\overline{\beta} \gamma }^\alpha \textbf{\emph{e}}_\alpha    ~   ,
\end{equation}
similarly, there are,
\begin{eqnarray}
&& ( \partial \mathbb{R}^\ast / \partial u^\lambda ) \circ ( \partial^2 \mathbb{R} / \partial \overline{u^\beta} \partial u^\gamma ) = g_{\overline{\lambda} \alpha} \Gamma_{\overline{\beta} \gamma }^\alpha   ~   ,
\\
&& ( \partial^2 \mathbb{R}^\ast / \partial \overline{u^\beta} \partial u^\gamma ) \circ ( \partial \mathbb{R} / \partial u^\lambda ) = \overline{\Gamma_{\overline{\beta} \gamma }^\alpha}  g_{\overline{\alpha} \lambda} ~   ,
\end{eqnarray}
where $\partial^2 \mathbb{R}^\ast / \partial \overline{u^\beta} \partial u^\gamma = \overline{\Gamma_{\overline{\beta} \gamma }^\alpha} \textbf{\emph{e}}_\alpha^\ast $ . $\Gamma_{\overline{\beta} \gamma }^\alpha$ and $\overline{\Gamma_{\overline{\beta} \gamma }^\alpha}$ are coefficients.

From the last two equations, it is found that the partial derivative of the metric tensor, $g_{\overline{\lambda} \gamma} = \textbf{\emph{e}}_\lambda^\ast \circ \textbf{\emph{e}}_\gamma$ , with respect to the coordinate value, $\overline{u^\beta}$ , is able to produce,
\begin{equation}
g_{\overline{\lambda} \alpha} \Gamma_{\overline{\beta} \gamma }^\alpha + \overline{\Gamma_{\overline{\beta} \lambda }^\alpha}  g_{\overline{\alpha} \gamma} = \partial g_{\overline{\lambda} \gamma} / \partial \overline{u^\beta}  ~ ,
\end{equation}
similarly, there are,
\begin{eqnarray}
g_{\overline{\beta} \alpha} \Gamma_{\overline{\gamma} \lambda }^\alpha + \overline{\Gamma_{\overline{\gamma} \beta }^\alpha}  g_{\overline{\alpha} \lambda} = \partial g_{\overline{\beta} \lambda} / \partial \overline{u^\gamma}  ~ ,
\\
g_{\overline{\gamma} \alpha} \Gamma_{\overline{\lambda} \beta }^\alpha + \overline{\Gamma_{\overline{\lambda} \gamma }^\alpha}  g_{\overline{\alpha} \beta} = \partial g_{\overline{\gamma} \beta} / \partial \overline{u^\lambda}  ~ ,
\end{eqnarray}
further, from the last three equations, there is,
\begin{equation}
\Gamma_{ \overline{\lambda} , \overline{\beta} \gamma } = (1/2) ( \partial g_{ \overline{\gamma} \lambda } / \partial \overline{u^\beta} + \partial g_{ \overline{\lambda} \beta } / \partial \overline{u^\gamma} - \partial g_{ \overline{\gamma} \beta } / \partial \overline{u^\lambda} )    ~  ,
\end{equation}
where $ [ ( g_{\overline{\lambda} \alpha} \Gamma_{\overline{\beta} \gamma }^\alpha )^\ast ]^T = \overline{\Gamma_{\overline{\gamma} \beta }^\alpha}  g_{\overline{\alpha} \lambda} $ , and $ [ ( \overline{\Gamma_{\overline{\gamma} \beta }^\alpha}  g_{\overline{\alpha} \lambda} )^\ast ]^T = g_{\overline{\lambda} \alpha} \Gamma_{\overline{\beta} \gamma }^\alpha $ .

From the above, it is capable of inferring the curvature tensor and torsion tensor. In the curved space described with the complex-sedenions, for a tensor $Y^\gamma$ with contravariant rank 1, the covariant derivatives are,
\begin{equation}
\nabla_\beta Y^\gamma = \partial Y^\gamma / \partial u^\beta + \Gamma _{\lambda \beta}^\gamma Y^\lambda ~ , ~~~~~
\nabla_{\overline{\alpha}} Y^\gamma = \partial Y^\gamma / \partial \overline{u^\alpha} + \Gamma _{\lambda \overline{\alpha}}^\gamma Y^\lambda  ~,
\end{equation}
meanwhile, for a mixed tensor, $Z_\nu^\gamma$ , with contravariant rank 1 and covariant rank 1, the covariant derivative can be written as,
\begin{eqnarray}
\nabla_\beta Z_\nu^\gamma = \partial Z_\nu^\gamma / \partial u^\beta - \Gamma_{\beta \nu}^\lambda Z_\lambda^\gamma + \Gamma_{\beta \lambda}^\gamma Z_\nu^\lambda  ~ ,
\\
\nabla_{\overline{\alpha}} Z_\nu^\gamma = \partial Z_\nu^\gamma / \partial \overline{u^\alpha} - \Gamma_{\overline{\alpha} \nu}^\lambda Z_\lambda^\gamma + \Gamma_{\overline{\alpha} \lambda}^\gamma Z_\nu^\lambda  ~ ,
\end{eqnarray}
where $Y^\gamma$ and $Z_\nu^\gamma$ both are scalar.

Apparently, the above equations deduce,
\begin{eqnarray}
\nabla_{\overline{\alpha}} ( \nabla_\beta Y^\gamma )  -  \nabla_\beta ( \nabla_{\overline{\alpha}} Y^\gamma ) = R_{\beta \overline{\alpha} \nu}^{~~~~\gamma} Y^\nu + T_{\beta \overline{\alpha}}^\lambda (\nabla_\lambda Y^\gamma)  ~  ,
\end{eqnarray}
and
\begin{eqnarray}
&& R_{\beta \overline{\alpha} \nu}^{~~~~\gamma} = \partial \Gamma_{\nu \beta}^\gamma / \partial \overline{u^\alpha} - \partial \Gamma_{\nu \overline{\alpha}}^\gamma /\partial u^\beta + \Gamma_{\lambda \overline{\alpha}}^\gamma \Gamma_{\nu \beta}^\lambda - \Gamma_{\lambda \beta}^\gamma \Gamma_{\nu \overline{\alpha}}^\lambda  ~ ,
\\
&& T_{\beta \overline{\alpha}}^\lambda = \Gamma_{\overline{\alpha} \beta }^\lambda - \Gamma_{\beta \overline{\alpha}}^\lambda  ~ ,
\end{eqnarray}
where $T_{\beta \overline{\alpha}}^\lambda$ is the torsion tensor, and $R_{\beta \overline{\alpha} \nu}^{~~~~\gamma}$ is the curvature tensor. In the paper, there is, $ T_{\beta \overline{\alpha}}^\lambda = 0 $ .

The discovery process of complex-sedenion spaces may be similar to that of electromagnetic spectrum. In the electromagnetic spectrum, the visible light spectrum was first found, while the discovery of the invisible light spectrum was much later. Until H. R. Hertz discovered the electromagnetic waves, it began to recognize the existence of the invisible light spectrum. Similarly, in the complex-sedenion spaces, the scholars recognize the influence of complex-quaternion space $\mathbb{H}_g$ for the gravitational field nowadays. And they will find the impact of other complex-quaternion spaces, $\mathbb{H}_e$ , $\mathbb{H}_w$ , and $\mathbb{H}_s$ , for other fields someday.


\begin{thebibliography}{0}


\bibitem{GUT1}
      K. Dimopoulos,
      Correlated curvature perturbations and magnetogenesis from the GUT gauge bosons,
      {\it Astropart. Phys.}
      {\bf 42} (2013), 86--89.

\bibitem{GUT2}
      D. J. Cirilo-Lombardo,
      Unified Field Theoretical Models from Generalized Affine Geometries,
      {\it Int. J. Theor. Phys.}
      {\bf 49} (6) (2010), 1288--1301.

\bibitem{GUT3}
      T.-J. Li,
      GUT breaking on $M^4¡Á\times T^2/(Z_2\times Z_2)$,
      {\it Phys. Lett. B}
      {\bf 520} (3-4) (2001), 377--384.

\bibitem{GUT4}
      C. Gerhardt,
      Combining gravity with the forces of the standard model on a cosmological scale,
      {\it Classical Quant. Grav.}
      {\bf 27} (15) (2010), 155008.

\bibitem{GUT5}
      C. H. Albright, S. M. Barr,
      Construction of a minimal Higgs SO(10) supersymmetric grand unified model,
      {\it Phys. Rev. D}
      {\bf 62} (9) (2000), 093008.

\bibitem{GUT6}
      D. Staicova, M. Stoilov,
      Cosmological aspects of a unified dark energy and dust dark matter mode,
      {\it Mod. Phys. Lett. A}
      {\bf 32} (1) (2017), 1750006.

\bibitem{GUT7}
      M. I. Wanas, N. L. Youssef, A. M. Sid-Ahmed,
      Teleparallel Lagrange Geometry and a Unified Field Theory,
      {\it Classical Quant. Grav.}
      {\bf 27} (4) (2010), 045005.

\bibitem{GUT8}
      K. Bamba, G. G. L. Nashed, W. El Hanafy, Sh. K. Ibraheem,
      Bounce inflation in $f(T)$ Cosmology: A unified inflaton-quintessence field,
      {\it Phys. Rev. D}
      {\bf 94} (8) (2016), 083513.

\bibitem{GUT9}
      J. J. van der Bij,
      Cosmotopological relation for a unified field theory,
      {\it Phys. Rev. D}
      {\bf 76} (12) (2007), 121702.

\bibitem{GUT10}
      P. Dave, E. L. Green,
      Unified Field Theory From Enlarged Transformation Group: The Consistent Hamiltonian,
      {\it Int. J. Theor. Phys.}
      {\bf 42} (8) (2003), 1849--1873.

\bibitem{GUT11}
      Z. Berezhiani, A. Rossi,
      Predictive grand unified textures for quark and neutrino masses and mixings,
      {\it Nucl. Phys. B}
      {\bf 594} (1-2) (2001), 113--168.

\bibitem{GUT12}
      Pushpa, P. S. Bisht, T.-J. Li, O. P. S. Negi,
      Quaternion Octonion Reformulation of Grand Unified Theories,
      {\it Int. J. Theor. Phys.}
      {\bf 51} (10) (2012), 3228--3235.

\bibitem{GUT13}
      H. Tanaka, I. S. Sogami,
      SO(10) Grand Unified Theory in Generalized Covariant Derivative Formalism,
      {\it Prog. Theor. Phys.}
      {\bf 103} (3) (2000), 621--634.

\bibitem{GUT14}
      L. J. Hall, Y. Nomura,
      Grand Unification and Intermediate Scale Supersymmetry,
      {\it J. High Energy Phys.}
      {\bf 2014} (2014), 129.

\bibitem{GUT15}
      C. S. Aulakh,
      Bajc-Melfo vacua enable Yukawon ultraminimal grand unified theories,
      {\it Phys. Rev. D}
      {\bf 91} (5) (2015), 055012.

\bibitem{GUT16}
      Y.-Z. You, C.-K. Xu,
      Interacting Topological Insulator and Emergent Grand Unified Theory,
      {\it Phys. Rev. B}
      {\bf 91} (12) (2015), 125147.

\bibitem{GUT17}
      E. Allys,
      Bosonic condensates in realistic supersymmetric GUT cosmic strings,
      {\it J. Cosmol. Astropart. Phys.}
      {\bf 2016} (4) (2016), 009.

\bibitem{GUT18}
      N. Yamatsu,
      A Supersymmetric Grand Unified Model with Noncompact Horizontal Symmetry,
      {\it Prog. Theor. Exp. Phys.}
      {\bf 2013} (12) (2013), 123B01.

\bibitem{GUT19}
      K. Kojima, K. Takenaga, T. Yamashita,
      Grand Gauge-Higgs Unification,
      {\it Phys. Rev. D}
      {\bf 84} (5) (2011), 051701.

\bibitem{GUT20}
      T. Kobayashi, S. Raby, R.-J. Zhang,
      Constructing 5d orbifold grand unified theories from heterotic strings,
      {\it Phys. Lett. B}
      {\bf 593} (1-4) (2004), 262--270.

\bibitem{GUT21}
      B. Dutta, Y. Mimura, R. N. Mohapatra,
      An SO(10) grand unified theory of flavor,
      {\it J. High Energy Phys.}
      {\bf 2010} (2010), 34.

\bibitem{GUT22}
      K. Bhattacharya, U. Sarkar, C. R. Das, G. Rajasekaran,
      Seesaw fermion masses in an SO(10) grand unified theory,
      {\it Phys. Rev. D}
      {\bf 74} (1) (2006), 015003.

\bibitem{GUT23}
      D. A. Steer, T. Vachaspati,
      Domain walls and fermion scattering in Grand Unified models,
      {\it Phys. Rev. D}
      {\bf 73} (10) (2006), 105021.

\bibitem{GUT24}
      J. Rocher, M. Sakellariadou,
      Constraints on Supersymmetric Grand Unified Theories from Cosmology,
      {\it J. Cosmol. Astropart. Phys.}
      {\bf 2005} (3) (2005), 004.

\bibitem{weng1}
      Z.-H. Weng,
      Forces in the complex octonion curved space,
      {\it Int. J. Geom. Methods Mod. Phys.}
      {\bf 13} (6) (2016), 1650076.

\bibitem{quaternion2}
      V. Majernik,
      The energy density of the quaternionic field as dark energy in the universe,
      {\it Gen. Relat. Gravit.}
      {\bf 36} (2004), 2139--2149.

\bibitem{quaternion5}
      A. J. Davies,
      Quaternionic Dirac equation,
      {\it Phys. Rev. D}
      {\bf 41} (1990), 2628--2630.

\bibitem{quaternion6}
      M. Tanisli, M. E. Kansu, S. Demir.
      Reformulation of electromagnetic and gravito-electromagnetic equations for Lorentz system with octonion algebra,
      {\it Gen. Relat. Gravit.}
      {\bf 46} (2014), 1739.

\bibitem{quaternion7}
      C. Castro.
      On octonionic gravity, exceptional Jordan strings and nonassociative ternary gauge field theories,
      {\it Int. J. Geom. Methods Mod. Phys.}
      {\bf 9} (3) (2012), 1250021.

\bibitem{quaternion8}
      S. V. Ludkowski, W. Sprossig.
      Spectral Theory of Super-Differential Operators of Quaternion and Octonion Variables,
      {\it Adv. Appl. Clifford Al.}
      {\bf 21} (1) (2011), 165--191.

\bibitem{quaternion11}
      J. Koplinger,
      Gravity and electromagnetism on conic sedenions,
      {\it Appl. Math. Comput.}
      {\bf 188} (2007), 948--953.

\bibitem{quaternion12}
      V. L. Mironov and S. V. Mironov,
      Sedeonic generalization of relativistic quantum mechanics,
      {\it Int. J. Mod. Phys. A}
      {\bf 24} (32) (2009), 6237--6254.

\bibitem{quaternion14}
      B. C. Chanyal,
      Dual octonion electrodynamics with the massive field of dyons,
      {\it J. Math. Phys.}
      {\bf 57} (2016), 033503.

\bibitem{quaternion13}
      S. Demir and M. Tanisli,
      Sedenionic Formulation for Generalized Fields of Dyons,
      {\it Int. J. Theor. Phys.}
      {\bf 51} (2012), 1239--1252.

\bibitem{quaternion4}
      A. S. Rawat, O. P. S. Negi,
      Quaternion gravi-electromagnetism,
      {\it Int. J. Theor. Phys.}
      {\bf 51} (2012), 738--745.

\bibitem{quaternion3}
      S. M. Grusky, K. V. Khmelnytskaya, V. V. Kravchenko,
      On a quaternionic Maxwell equation for the time-dependent electromagnetic field in a chiral medium,
      {\it J. Phys. A}
      {\bf 37} (2004), 4641--4647.

\bibitem{quaternion9}
      C. G. Tsagas.
      Electromagnetic fields in curved spacetimes,
      {\it Classical Quant. Grav.}
      {\bf 22} (2) (2005), 393--407.

\bibitem{weng2}
      Z.-H. Weng,
      Some properties of dark matter field in the complex octonion space,
      {\it Int. J. Mod. Phys. A}
      {\bf 30} (35) (2015), 1550212.

\bibitem{quaternion1}
      S. P. Brumby, B. E. Hanlon, G. C. Joshi,
      Implications of quaternionic dark matter,
      {\it Phys. lett. B}
      {\bf 401} (1997), 247--253.

\bibitem{quaternion10}
      S. Furui.
      Axial anomaly and the triality symmetry of octonion,
      {\it Few-Body Syst.}
      {\bf 54} (11) (2013), 2097--2111.

\bibitem{weng3}
      Z.-H. Weng,
      Physical quantities and spatial parameters in the complex octonion curved space,
      {\it Gen. Relat. Gravit.}
      {\bf 48} (12) (2016), 153.

\bibitem{weng4}
      Z.-H. Weng,
      Color confinement of non-Abelian gauge fields in the complex sedenion space,
      {\it Adv. Math. Phys.}
      {\bf 2017} (2017), 9876464.



\end{thebibliography}
\end{document}